\documentclass[acmsmall]{acmart}

\AtBeginDocument{%
  }

\usepackage{geometry}

\usepackage{multirow}
\usepackage{graphicx}
\usepackage{array}
\newcolumntype{C}[1]{>{\centering\arraybackslash}m{#1}}

\usepackage{tabularx}
\usepackage{booktabs}
\usepackage{hyperref}
\usepackage{subcaption}
\usepackage{arydshln} % for \hdashline (table)

\usepackage[table]{xcolor} 
\usepackage{hhline}
\usepackage{enumitem}

\usepackage{makecell}      % For \makecell command (optional, for line breaks in cells)
\usepackage{threeparttable} % For \begin{tablenotes} (optional, if you want notes)

\usepackage{tcolorbox}
\tcbuselibrary{skins, breakable}

\newcommand{\link}[1]{%
   \texttt{\textcolor{linkblue}{#1}}%
}

\usepackage{algorithm} % algorithm counter
\usepackage[indLines]{algpseudocodex} % pseudocode + vertical indent guides

\newtcolorbox{algobox}[1]{
  enhanced,
  breakable,
  colback=white,
  colframe=black!10,
  boxrule=0.8pt,
  arc=0pt,
  outer arc=0pt,
  left=1mm,right=1mm,top=1mm,bottom=1mm,
  before skip=8pt, after skip=8pt,
  title={#1},
  fonttitle=\bfseries,
  coltitle=black
}

\newenvironment{breakablealgorithm}[2]{%
  \refstepcounter{algorithm}%
  \begin{algobox}{Algorithm \thealgorithm. #1}
  \label{#2}
  \begin{algorithmic}[1]
}{%
  \end{algorithmic}
  \end{algobox}
}

\definecolor{linkblue}{HTML}{3377FF}
\definecolor{subfigblue}{HTML}{E8F0FE}
\definecolor{originalcolor}{HTML}{FEDACC}
\definecolor{originalframe}{HTML}{FBA481}
\definecolor{simplifiedcolor}{HTML}{D9F2EC}
\definecolor{simplifiedframe}{HTML}{A0DDD1}
\definecolor{gtcolor}{HTML}{E9E1FF}
\definecolor{gtframe}{HTML}{C2A7ED}
\definecolor{rulecolor}{HTML}{E8F0FE}
\definecolor{ruleframe}{HTML}{A7C2ED}
\definecolor{learningcolor}{HTML}{FDF4E1}
\definecolor{learningframe}{HTML}{F8DAA1}

\newtcbox{\originalbox}{on line, colback=originalcolor!40, colframe=originalframe, 
  boxsep=0pt, left=1.3pt, right=1.3pt, top=1pt, bottom=1pt, arc=0.1pt, boxrule=0.5pt, fontupper=\sffamily}
\newtcbox{\simplifiedbox}{on line, colback=simplifiedcolor!40, colframe=simplifiedframe,
  boxsep=0pt, left=1.3pt, right=1.3pt, top=1pt, bottom=1pt, arc=0.1pt, boxrule=0.5pt, fontupper=\sffamily}
\newtcbox{\gtbox}{on line, colback=gtcolor!40, colframe=gtframe,
boxsep=0pt, left=1.3pt, right=1.3pt, top=1pt, bottom=1pt, arc=0.1pt, boxrule=0.5pt, fontupper=\sffamily}
\newtcbox{\rbbox}{on line, colback=rulecolor!40, colframe=ruleframe,
boxsep=0pt, left=1.3pt, right=1.3pt, top=1pt, bottom=1pt, arc=0.1pt, boxrule=0.5pt, fontupper=\sffamily}
\newtcbox{\lbbox}{on line, colback=learningcolor!40, colframe=learningframe,
boxsep=0pt, left=1.3pt, right=1.3pt, top=1pt, bottom=1pt, arc=0.1pt, boxrule=0.5pt, fontupper=\sffamily}
\newtcbox{\subfigbox}{on line,
  colback=subfigblue, colframe=black,
  boxsep=0pt, left=1.5pt, right=1.5pt, top=1.5pt, bottom=1.5pt,
  arc=0pt, boxrule=0.5pt,
  height=2.5ex, width=2.5ex,
  valign=center, halign=center, tcbox raise base, fontupper=\sffamily}

\begin{document}

    \title[M\textsc{ake it} S\textsc{imple,} M\textsc{ake it} D\textsc{ance}: Dance Motion Simplification to Support Novices' Dance Learning]{M\textsc{ake it} S\textsc{imple,} M\textsc{ake it} D\textsc{ance}: Dance Motion Simplification to Support Novices' Dance Learning}


\author{Hyunyoung Han}
\orcid{0009-0002-4681-5021}
\authornote{This work was conducted while the author was at the Graduate School of Culture Technology, KAIST.}
\affiliation{%
  \institution{School of Electrical Engineering, KAIST}
  \city{Daejeon}
  \country{Republic of Korea}}
\email{hyhan@kaist.ac.kr}

\author{Murad Eynizada}
\orcid{0009-0004-0228-7903}
\affiliation{%
  \institution{Graduate School of Culture Technology, KAIST}
  \city{Daejeon}
  \country{Republic of Korea}}
\email{eynizadamurad@kaist.ac.kr}

\author{Son Xuan Nghiem}
\orcid{0009-0001-8659-0552}
\affiliation{%
  \institution{School of Computing, KAIST}
  \city{Daejeon}
  \country{Republic of Korea}}
\email{cashew@kaist.ac.kr}

\author{Sang Ho Yoon}
\orcid{0000-0002-3780-5350}
\affiliation{%
  \institution{Graduate School of Culture Technology, KAIST}
  \city{Daejeon}
  \country{Republic of Korea}}
\email{sangho@kaist.ac.kr}

\renewcommand{\shortauthors}{Han et al.}

\begin{abstract}
Online dance tutorials have gained widespread popularity. However, many novices encounter difficulties when dance motion complexity exceeds their skill level, potentially leading to discouragement. This study explores dance motion simplification to address this challenge. We surveyed 30 novices to identify challenging movements, then conducted focus groups with 30 professional choreographers across 10 genres to explore simplification strategies and collect paired original-simplified dance datasets. We identified five complexity factors and developed automated simplification methods using both rule-based and learning-based approaches. We validated our approach through three evaluations. Technical evaluation confirmed our complexity measures and algorithms. 20 professional choreographers assessed motion naturalness, simplification adequacy, and style preservation. 18 novices evaluated learning effectiveness through workload, self-efficacy, objective performance, and perceived difficulty. This work contributes to dance education technology by proposing methods that help make choreography more approachable for beginners while preserving essential characteristics.
\end{abstract}



%   \includegraphics[width=\textwidth]{Contents/figure/teaser.pdf}
%   \caption{We present a dance motion simplification approach developed through extensive user research. First, we identified complexity factors via workshops with 30 learners and 30 choreographers. Next, we created both rule-based and learning-based simplification methods. Finally, we validated our approach through technical metrics, expert assessment (n=20), and a learner study (n=18).}
%   \Description{teaser}


\maketitle

\section{Introduction}\label{sec:introduction}
\begin{figure}[t]
  \includegraphics[width=\textwidth]{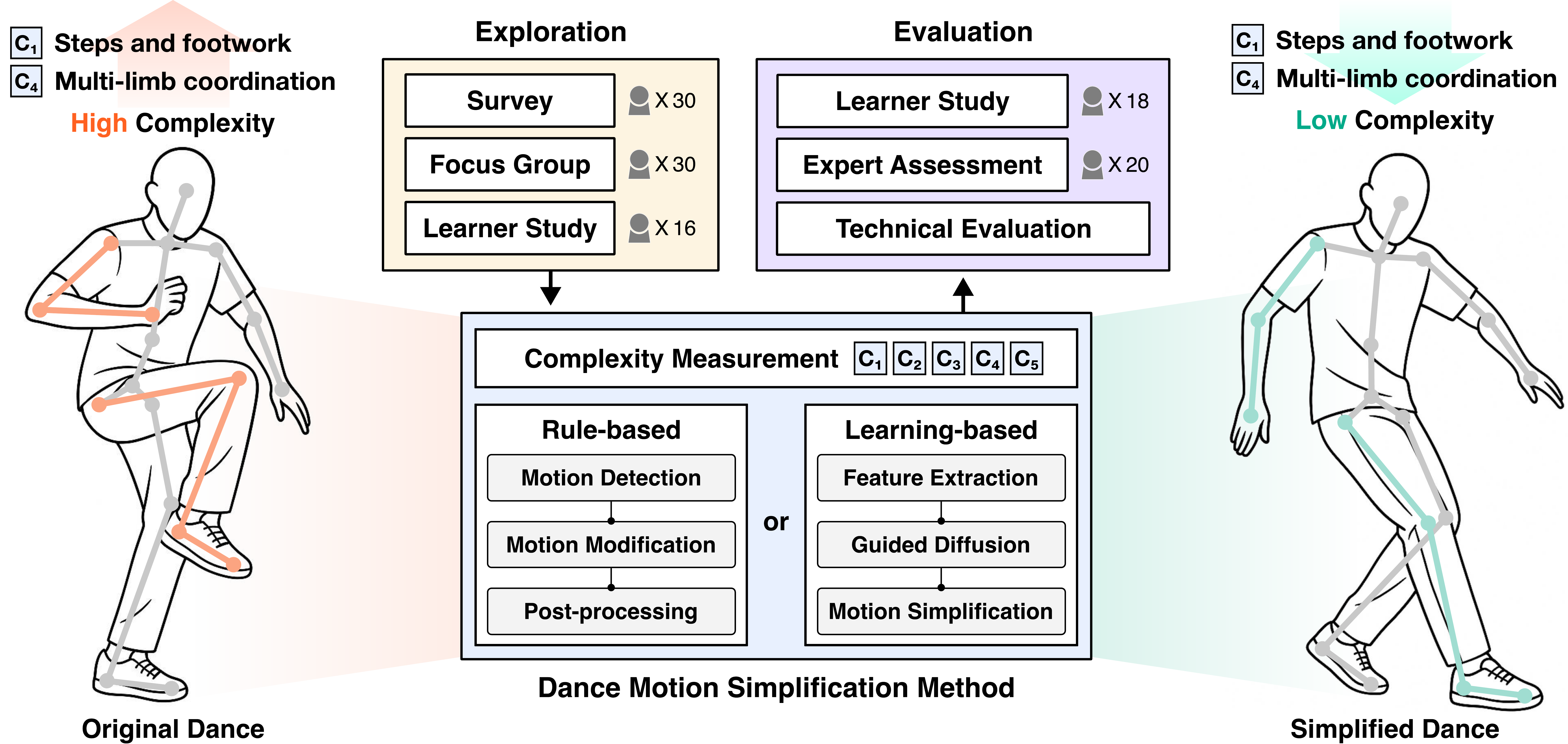}
  \caption{We present a dance motion simplification approach developed through extensive user research. First, we identified movement complexity factors via workshops with 30 learners and 30 choreographers. Next, we created both rule-based and learning-based simplification methods. Finally, we validated our approach through technical measures, expert assessment ($n=20$), and a learner study ($n=18$).}
  \Description{teaser}
  \label{fig:teaser}
\end{figure}

The rise of video-based social media platforms\footnote{For example, YouTube~(\url{https://youtube.com}), TikTok~(\url{https://www.tiktok.com}).} has transformed how people learn dance. These platforms provide millions of dance tutorials and videos, allowing people to learn dance movements free from traditional barriers such as time, location, or cost~\cite{warburton2024tiktok}. However, this increased accessibility reveals a fundamental limitation: while learners can pause, replay, and practice at their own pace, online videos cannot provide the personalized feedback and progressive skill development that characterize effective motor learning~\cite{kim2025cluster}. Most dance content presents movements at full complexity, creating a substantial gap between novices' current abilities and the skill level required. This gap often leads to frustration and discontinuation of practice~\cite{krasnowmotor}.

To address this challenge, researchers have explored various technological solutions. Several systems focus on helping learners better understand dance motion. Rivière et al.~\cite{riviere2019capturing} observed that choreographers naturally decompose dance motions when learning and developed interfaces to support this process. Endo et al.~\cite{endo2024automatic} created automatic video segmentation systems to help novices parse complex sequences. Others have experimented with alternative representations, such as verbal descriptions~\cite{blanchet2023integrating, hassan2024designing} or emojis~\cite{blanchet2023learnthatdance, blanchet2025enhancing} to make movements more intuitive. While these innovations effectively support comprehension, they do not resolve the fundamental issue: the movements themselves remain physically demanding. Learners may understand what they need to do, but the original dance still requires coordination, balance, and muscle control that exceed their current capabilities~\cite{limanskaya2021coordination, malkogeorgos2013physiological}.

We take a complementary approach that directly addresses execution difficulty. In practice, dance instructors routinely teach easier variations first, selectively reducing difficult elements while preserving recognizable anchors such as rhythm structure, salient poses, and stylistic intent.\footnote{An example is provided for easier understanding. \href{https://www.youtube.com/watch?v=CN4fffh7gmk\&list=RDCN4fffh7gmk\&start_radio=1}{\link{Original dance of Dynamite~(BTS, 2020)}}, and \href{https://www.youtube.com/watch?v=U-me_w_x3j4\&list=RDU-me_w_x3j4\&start_radio=1}{\link{its simplified version of the same dance}}.} Despite the observed effectiveness of this pedagogical strategy, current human-computer interaction~(HCI) research lacks a systematic and computational account of 1) which movement characteristics reliably trigger novice breakdowns across genres, 2) how such characteristics can be operationalized into interpretable and actionable criteria, and 3) how those criteria can be used to transform motions toward novice-friendly variants while preserving essential choreographic characteristics. Addressing these gaps leads us to three central research questions~(RQs).

\begin{itemize}[leftmargin=*]
    \item RQ1. What movement characteristics do novices find challenging across different dance genres?
    \item RQ2. Can we establish universal criteria for identifying and simplifying difficult movements based on both learner and instructor perspectives? 
    \item RQ3. Is it possible to automate the dance simplification process while preserving the essential characteristics of the original dance motion?
\end{itemize}

\begin{figure}[t]
    \centering
    \includegraphics[width=\linewidth]{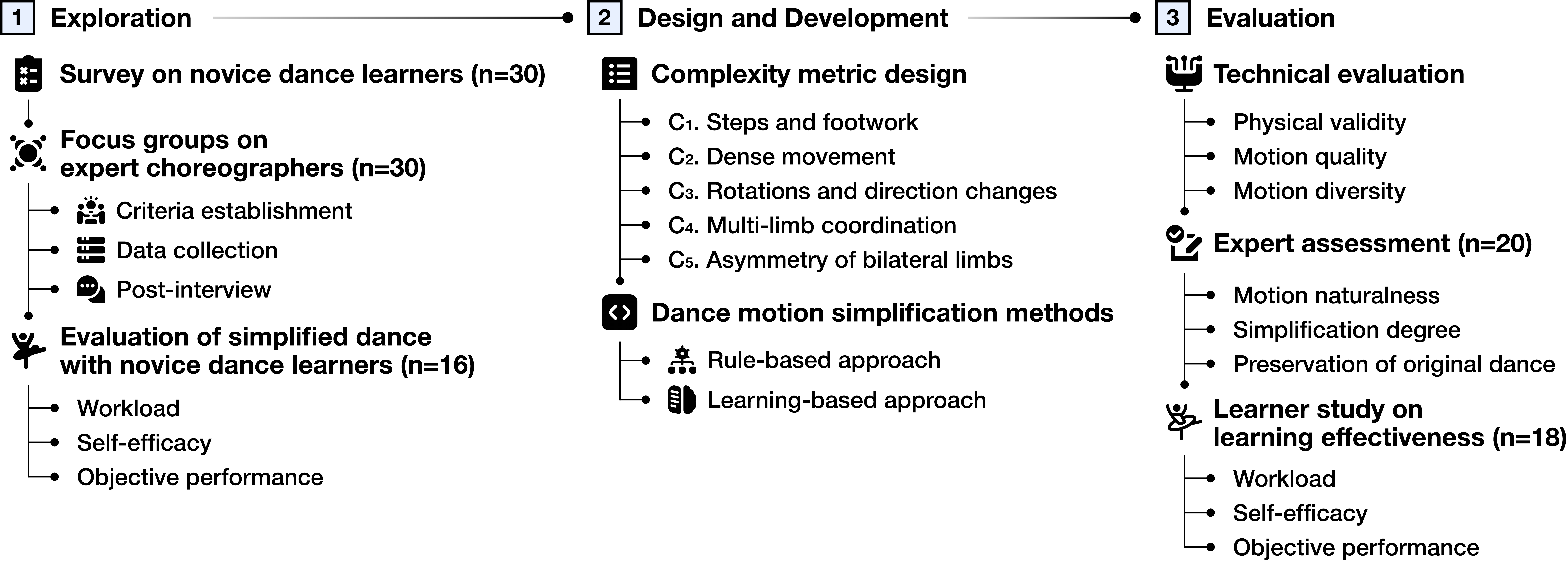}
    \caption{Overview of the entire process of our study. \subfigbox{1} Exploration is presented in \S\ref{sec:workshop}, \subfigbox{2} Design and Development are covered across \S\ref{sec:computational-metrics} and \S\ref{sec:methodology}, and \subfigbox{3} Evaluation is detailed in \S\ref{sec:evalution}.}
    \label{fig:overview-entire-study}
    \vspace{-0.5cm}
\end{figure}

To answer these questions, we conducted a comprehensive study on dance motion simplification. First, we surveyed 30 novice learners to identify movement patterns they found most challenging. Next, we organized focus groups with 30 professional choreographers across 10 dance genres to develop systematic simplification criteria and collect paired dance motion data of original and simplified sequences. Based on these empirical findings, we developed automated dance motion simplification methods---both rule-based and learning-based---that transform complex dance sequences into novice-friendly versions. Finally, we validated our methods through technical evaluation, expert assessment with 20 choreographers, and a learner study with 18 novices. Figure~\ref{fig:overview-entire-study} provides an overview of our study.

Our work makes four contributions:

\begin{itemize}[leftmargin=*]
    \item We establish empirically grounded, cross-genre criteria for identifying difficult movements and for systematically simplifying dance motions, based on consultation with both novice learners and professional choreographers.
    \item We create the first paired dataset of \texttt{\{original, simplified\}} dance sequences spanning multiple genres, which we make publicly available.\footnote{A preview of the dataset is available at \href{https://zenodo.org/records/17105212?token=eyJhbGciOiJIUzUxMiJ9.eyJpZCI6IjJmMzZiODUzLTBmZGItNDMyOS04NDRkLWRjZWQ1Zjg1NzhhMCIsImRhdGEiOnt9LCJyYW5kb20iOiI0MTcwM2E4M2VmZjE3ZDI2NmE2N2RlNzUyNGU3YzY4YyJ9.YhWo24pGYYVNMOTWVrNP1dcodyQ5C8ZaH_0-5R1TqfAOzf_EOcFKkDi9PLmoccVJTuee2XtaDv60vvNIMrGPBg}{\link{this link}}. The full dataset and code will be released after the review process.}
    \item We develop automated dance motion simplification methods using both rule-based and learning-based pipelines, enabling controllable complexity reduction toward a target level.
    \item We contribute a reproducible, multi-level evaluation framework (technical, expert, learner) and evidence-based implications for designing and deploying dance motion simplification for learning.
\end{itemize}
\section{Background and Related Work}\label{sec:RW}
We first introduce the motor learning theories that ground our approach, then review research on enhancing dance learning experience, dance motion complexity, and computational methods for motion manipulation.

\subsection{Theoretical Foundations: Motor Learning and Progressive Practice}\label{subsec:theory}

Dance learning is a form of whole-body motor skill acquisition that requires learners to coordinate many degrees of freedom under temporal constraints~\cite{chang2020whole}, maintain balance~\cite{berardi2004teaching}, and continuously correct errors through proprioceptive feedback~\cite{simmons2005sensory}. Novice progress depends not only on what learners understand about a movement, but also on whether the movement's execution demands fall within their current capability range~\cite{guadagnoli2004challenge}. We ground our work in motor learning theories that explain how complex skills are acquired through structured manipulation of task demands~\cite{guadagnoli2004challenge, lee2025motor}.

A central principle is that practice is most effective when task difficulty matches the learner's skill level. The challenge point framework~\cite{guadagnoli2004challenge} argues that learning benefits from an optimal level of challenge: tasks that are too easy provide insufficient learning-relevant information, while tasks that are too difficult overwhelm the learner and impede improvement. For complex skills with multiple interacting components, progressive part practice provides a practical scaffolding strategy: learners practice reduced or partial versions of the target skill before integrating components toward full performance~\cite{lee2025motor}. Crucially, effective scaffolding preserves essential characteristics that support transfer to the full skill, while temporarily attenuating elements that generate excessive control demands for novices~\cite{lee2025motor, wulf2002principles}. This perspective frames simplification as a deliberate rebalancing of execution demands that keeps practice within a productive learning zone while maintaining pathways for progression.

\subsection{Enhancing Dance Learning Experience}
Dance learning involves cognitive resources like memory, attention, and multi-limb coordination~\cite{merom2016dancing, low2016we}. Research in human-computer interaction has sought to support this process along two complementary directions: helping learners \textit{understand} movements and reducing the \textit{execution} demands of movements themselves.
On the understanding side, researchers have developed tools that make dance knowledge more accessible. Rivière et al.~\cite{riviere2019capturing} studied how choreographers decompose motions and translated these strategies into supportive interfaces, while Endo et al.~\cite{endo2024automatic} introduced automatic dance video segmentation to help novices parse complex choreography into manageable units. Other work has explored supplementary representations—verbal descriptions~\cite{hassan2024designing, blanchet2023integrating, heiland2012effects} and emoji-based visual metaphors~\cite{blanchet2023learnthatdance, blanchet2025enhancing}—to make abstract movement qualities more concrete. These approaches build on and automate existing pedagogical methods, yet they fundamentally assist comprehension rather than altering what the learner must physically perform.

On the execution side, attempts to simplify dance motions have been made in both offline and online instruction settings~\cite{wang2023pilot}. As noted in \S\ref{subsec:theory}, teaching simplified versions of movements to beginners is effective for both skill acquisition and retention~\cite{guadagnoli2004challenge}. However, current simplification practices rely on manual effort by instructors, and systematic, automated approaches remain largely unexplored. This study bridges this gap by introducing and evaluating rule-based and learning-based methods to automate dance simplification and reduce movement complexity.

\subsection{Understanding Dance Motion Complexity}
Realizing the scaffolding approach outlined in \S\ref{subsec:theory} requires a reliable way to measure movement difficulty. Prior work has approached this from both qualitative and quantitative perspectives. On the qualitative side, studies have examined what makes movements challenging within specific dance genres, such as jazz~\cite{poon2000learning} and hip-hop~\cite{sato2014key}. Earlier research also proposed criteria based on the number of articulated joints involved or kinematic parameters such as velocity variability~\cite{yang2010evaluating}, though these were typically limited to isolated body parts rather than whole-body movement.

On the quantitative side, Yang et al.~\cite{yang2010evaluating} proposed a computational measure of whole-body motion complexity based on two kinematic properties: \textit{un-correlation}, which quantifies how independently active joint dimensions move by averaging pairwise correlation coefficients across all joint pairs, and \textit{non-smoothness}, which captures motion jerkiness by measuring the number of interpolating frames required for cubic spline interpolation of each joint's temporal trajectory to fall within a distortion threshold. The overall complexity is computed as a weighted sum of these two factors. This measure was subsequently applied by Yang et al.\cite{yang2013generating} to automatically generate easy-to-complex learning paths for a dance lesson generation system.

However, there remain opportunities to extend these foundational approaches. The qualitative criteria were primarily derived from analyst observation or biomechanical heuristics, and may not fully capture the perspectives of actual stakeholders—namely, learners who experience difficulty firsthand and choreographers who routinely adapt movements for different skill levels. Similarly, while Yang et al.'s quantitative measures effectively characterize kinematic properties of isolated joints, factors such as inter-limb coordination demands or balance challenges—which our formative study also identifies as key sources of difficulty (\S\ref{sec:workshop})—have yet to be incorporated. This may partly explain why these measures do not always generalize across diverse dance genres with distinct movement vocabularies. Motivated by these observations, we conduct a multi-stage workshop with both novice learners and expert choreographers to establish empirically-grounded complexity measures, develop dance motion simplification strategies, and create the paired dataset of \texttt{\{original, simplified\}} dance sequences.

\subsection{Computational Methods to Motion Manipulation}

Despite advances in motion editing and generation, research specifically targeting dance motion simplification for teaching is lacking. Existing rule- and learning-based methods focus on motion generation~\cite{holden2016deep, zhang2024motiondiffuse, dai2024motionlcm, karunratanakul2023guided, abe2004momentum}, style transfer~\cite{hsu2005style, jang2022motion}, or retargeting~\cite{aberman2020skeleton, tak2005physically}, but overlook systematic complexity reduction for dance education.

Early rule-based approaches established foundational motion manipulation techniques. Bruderlin and Williams~\cite{bruderlin1995motion} treated joint trajectories as signals, applying multi-resolution filtering to smooth or exaggerate motion at different temporal scales. Witkin and Popovic~\cite{witkin1995motion} proposed motion warping, which uses sparse keyframe constraints to stretch or redirect a motion's path without disrupting its fine details. Gleicher~\cite{gleicher1997motion} formulated motion editing as a spacetime optimization problem, enabling edits such as fixing a hand's endpoint while automatically adjusting the rest of the body to remain natural. Chi et al.~\cite{chi2000emote} introduced EMOTE, a system that separates what a movement does from how it is performed, allowing stylistic qualities such as timing and intensity to be adjusted independently of the spatial trajectory. These methods offer powerful low-level control but do not represent learner-centered notions of difficulty or explicitly reason about which aspects of complexity should be reduced.

Recent learning-based models capture higher-level relationships between music and movement. FineDance~\cite{li2023finedance} contributes a a diffusion-based generator that produces fine-grained full-body and hand motions aligned to music. EDGE~\cite{tseng2023edge} employs a transformer-based diffusion model with music features to generate physically plausible dance while supporting editable operations such as joint-wise conditioning and in-betweening. POPDG~\cite{luo2024popdg} introduces a space-augmentation module that strengthens spatial relationships across joints, and an alignment module that improves music–motion synchronization. Notably, POPDG has been adopted as the state-of-the-art baseline in subsequent dance generation research~\cite{zhang2025danceeditor, zhao2025freedance}, owing to its superior performance in quality, diversity, and semantic understanding. These capabilities make it an ideal foundation for complexity-aware dance generation.

To develop an effective approach to dance motion simplification, we conducted user studies with novice learners and choreographers to understand how they perceive movement difficulty and what strategies they employ to simplify complex motions. Informed by these findings, we designed two complexity-aware modules applicable to both rule- and learning-based pipelines: a difficulty assessment module that quantifies movement complexity, and a simplification guidance module that reduces complexity to a target level while preserving musical anchors and essential movement characteristics. Together, these modules enable automated simplification that directly supports novice dance learning.
\section{Workshop on Dance Motion Simplification}\label{sec:workshop}
We conducted a multi-stage workshop to investigate the necessity, process, and effects of dance motion simplification. We focused on identifying challenging movement characteristics across genres~(RQ1) and establishing simplification criteria from both learner and instructor perspectives~(RQ2).
Our approach consisted of three stages. First, we surveyed novice learners during actual dance classes to identify challenging movements and assess their perceived need for simplification~(\S\ref{subsec:pre-survey}). Then, we organized focus groups~\cite{kuniavsky2003observing} with professional choreographers with teaching experience, where participants discussed challenging dance movements and simplification strategies, and collected paired original-simplified dance sequences~(\S\ref{subsec:focus-group} and \S\ref{subsec:results-and-findings-from-FG}). Finally, we evaluated learning effects by having novice participants learn both original and simplified versions of identical dance sequences, then comparing their performance outcomes~(\S\ref{subsec:eval-of-simplified-choreography}).

\subsection{Survey on Novice Dance Learners}\label{subsec:pre-survey}
To understand challenges in dance learning and the need for simplification, we surveyed 30 novice learners (26 female, mean age 28.03, $\sigma=7.26$) with less than 6 months of learning experience~\cite{senecal2020partnerdance} ($\mu=2.17$ months, $\sigma=1.70$). Survey was conducted via Google Forms\footnote{\url{https://workspace.google.com/products/forms/}}. Participants' demographics and survey questions are available in Appendices~\ref{appendix:demographics} and ~\ref{appendix:survey-choreo-learners}. We conducted a qualitative content analysis~\cite{forman2007qualitative} of the open-ended responses, coding each answer into categories of difficult movement types and reasons for difficulty. We then aggregated category frequencies to obtain an overview of the most apparent challenges reported by novice learners. 

When asked about difficult movements encountered during practice, participants most frequently cited coordination of multiple body parts (15 of 30, 50\%), particularly upper and lower body coordination (40\%). As one participant explained, \textit{``mixing hands with legs''} caused confusion: \textit{``my brain gets confused''} (P4). Other challenging areas included steps such as slides and step sequences (23.33\%) and upper body movements such as chest isolations (20\%).

Participants attributed these difficulties primarily to physical unfamiliarity (23.33\%)---movements their bodies were \textit{``not used to''} (P5)---and encountering movement patterns they had never performed before (20\%). Cognitive load from memorizing sequences and fast tempos were also mentioned, though less frequently (6.67\% each).

These challenges carried emotional weight: half of the participants reported negative emotions such as frustration (P1: \textit{``I get frustrated that I can't keep up''}) and self-consciousness when struggling. However, 40\% described positive feelings upon successful execution, with one noting they felt \textit{``relieved and not defeated''} even when performance was imperfect (P4). Participants rated simplification as helpful for both learning ($\mu=4.90$, $\sigma=1.65$) and confidence ($\mu=5.00$, $\sigma=1.76$) on a 7-point Likert scale.

\subsection{Focus Group Setup}\label{subsec:focus-group}
Following the preliminary survey on the need for dance motion simplification, we sought to establish expert consensus on dance motion simplification criteria. We conducted 10 focus groups with 3 participants each (total 30 choreographers; 13 female; mean age 26.87, $\sigma=5.72$) with an average choreographic experience of 9.87 years ($\sigma=5.79$) and teaching experience of 6.80 years ($\sigma=5.12$). Participants were recruited through snowball sampling, targeting choreographers with expertise in 10 genres from the AIST dance video dataset~\cite{aist-dance-db}~(ballet jazz, break, house, krump, LA-style hiphop, lock, middle hiphop, pop, street jazz, and waack) as shown in Table~\ref{tab:choreographers-info}.

\begin{table*}[tb]
\centering
\small\sffamily
\renewcommand{\arraystretch}{1.25}
\setlength{\tabcolsep}{4pt}
\caption{Demographics and experience information of focus group participants}
\label{tab:choreographers-info}
\resizebox{\linewidth}{!}{%
  \begin{tabular}{ccccccc}
    \hline
    \textbf{Group} & \textbf{Genre} & \textbf{Alias} & \textbf{Gender} & \textbf{Choreographic Exp.\,(yrs)} & \textbf{Teaching Exp.\,(yrs)} & \textbf{Teaching Freq. for Novices} \\
    \hline
    \arrayrulecolor{black!10}
    \multirow{3}{*}{1} 
    & \multirow{3}{*}{Ballet Jazz} 
      & P1  & Female & 15 & 7  & $\geq$4 times/week \\ \cline{3-7}
      & & P2  & Female & 7  & 3  & 4 times/week      \\ \cline{3-7}
      & & P3  & Female & 5  & 2  & 1 time/week       \\
    \arrayrulecolor{black!30}
    \hline
    \arrayrulecolor{black!10}
    \multirow{3}{*}{2}
    & \multirow{3}{*}{Break}      
      & P4  & Male   & 16 & 16 & $\geq$3 times/week \\ \cline{3-7}
      & & P5  & Male   & 13 & 11 & 2 times/week      \\ \cline{3-7}
      & & P6  & Male   & 14 & 6  & $\geq$1 time/week  \\
    \arrayrulecolor{black!30}
    \hline
    \arrayrulecolor{black!10}
    \multirow{3}{*}{3}
    & \multirow{3}{*}{House}     
      & P7  & Female & 8  & 5  & $\geq$4 times/week \\ \cline{3-7}
      & & P8  & Female & 10 & 4  & $\geq$2 times/week \\ \cline{3-7}
      & & P9  & Male   & 7  & 5  & 1--2 times/week    \\
    \arrayrulecolor{black!30}
    \hline
    \arrayrulecolor{black!10}
    \multirow{3}{*}{4}
    & \multirow{3}{*}{Krump}      
      & P10 & Male   & 5  & 3  & 2 times/week      \\ \cline{3-7}
      & & P11 & Male   & 4  & 4  & 1 time/week       \\ \cline{3-7}
      & & P12 & Male   & 13 & 13 & 4 times/month     \\
    \arrayrulecolor{black!30}
    \hline
    \arrayrulecolor{black!10}
    \multirow{3}{*}{5}
    & \multirow{3}{*}{LA-style Hiphop}
      & P13 & Male   & 30 & 25 & 2 times/week      \\ \cline{3-7}
      & & P14 & Male   & 14 & 7  & 5 times/week      \\ \cline{3-7}
      & & P15 & Male   & 4  & 3  & 2 times/week      \\
    \arrayrulecolor{black!30}
    \hline
    \arrayrulecolor{black!10}
    \multirow{3}{*}{6}
    & \multirow{3}{*}{Lock}
      & P16 & Male   & 10 & 10 & 4 times/month     \\ \cline{3-7}
      & & P17 & Male   & 10 & 10 & $\geq$1 time/week \\ \cline{3-7}
      & & P18 & Male   & 15 & 10 & 3 times/week      \\
    \arrayrulecolor{black!30}
    \hline
    \arrayrulecolor{black!10}
    \multirow{3}{*}{7}
    & \multirow{3}{*}{Middle Hiphop}
      & P19 & Male   & 5  & 2  & 4 times/week      \\ \cline{3-7}
      & & P20 & Female & 3  & 3  & 1--2 times/week   \\ \cline{3-7}
      & & P21 & Female & 4  & 3  & 1--2 times/week   \\
    \arrayrulecolor{black!30}
    \hline
    \arrayrulecolor{black!10}
    \multirow{3}{*}{8}
    & \multirow{3}{*}{Pop}     
      & P22 & Male   & 7  & 1  & None              \\ \cline{3-7}
      & & P23 & Male   & 7  & 2  & 1 time/week       \\ \cline{3-7}
      & & P24 & Male   & 11 & 11 & 2 times/week      \\
    \arrayrulecolor{black!30}
    \hline
    \arrayrulecolor{black!10}
    \multirow{3}{*}{9}
    & \multirow{3}{*}{Street Jazz}
      & P25 & Female & 20 & 11 & 1 time/week       \\ \cline{3-7}
      & & P26 & Female & 11 & 6  & $\geq$4 times/week \\ \cline{3-7}
      & & P27 & Female & 10 & 7  & $\geq$2 times/week \\
    \arrayrulecolor{black!30}
    \hline
    \arrayrulecolor{black!10}
    \multirow{3}{*}{10}
    & \multirow{3}{*}{Waack}
      & P28 & Female & 9  & 7  & 1 time/week       \\ \cline{3-7}
      & & P29 & Female & 4  & 3  & $\geq$5 times/week \\ \cline{3-7}
      & & P30 & Female & 5  & 4  & $\geq$8 times/week \\
    \arrayrulecolor{black}
    \hline
  \end{tabular}%
}
\vspace{-0.5cm}
\end{table*}

\begin{figure}[t]
    \centering
    \includegraphics[width=\linewidth]{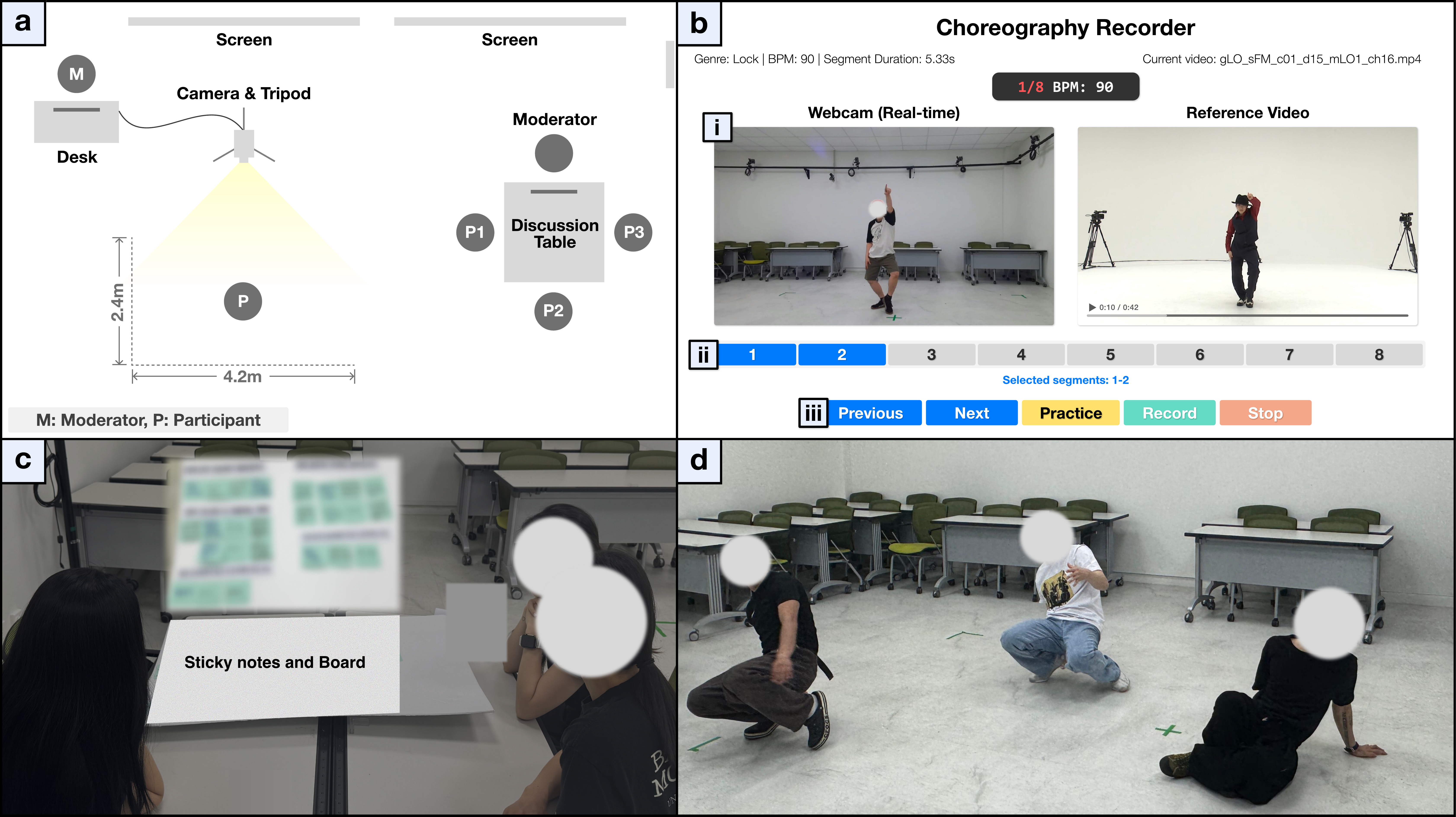}
    \caption{Overview of the focus group session and data collection: \subfigbox{a} the environmental setting, \subfigbox{b} the dance motion collecting tool including functions of \subfigbox{i} real-time display of the user alongside a reference video, \subfigbox{ii} selecting specific segments, and \subfigbox{3} recording, practice, and video selection, \subfigbox{c} an ongoing discussion among participants during the session, and \subfigbox{iii} the process of acquiring data on dance movements.}
    \label{fig:focus-group}
    \vspace{-10pt}
\end{figure}

We conducted focus groups in the environment shown in Figure~\ref{fig:focus-group}-\subfigbox{a}. Choreographers engaged in collaborative discussions at a designated table, with dance data collection performed in a 4.2~m~(width)~$\times$~2.4~m~(length) area to accommodate all proposed movements.
Each focus group session lasted approximately three hours, structured as follows: \emph{introduction}~(10-mins), \emph{interview for establishing simplification criteria}~(\emph{criteria establishment}; 30-mins), \emph{dance motion simplification data collection}~(\emph{data collection}; 120-mins), and \textit{post-interview and wrap-up}~(10-mins). Table~\ref{tab:question-for-focus-group} presents the questions used in each session along with their objectives. The \emph{introduction} phase included an explanation of the focus group process and consent form completion. During the \textit{criteria establishment} phase, we conducted semi-structured group interviews based on pre-survey results about movements that novices find difficult~(\S\ref{subsec:pre-survey}) and potential simplification approaches, with all three choreographers participating in collaborative discussions. The \emph{data collection} phase involved capturing both original and simplified dance motion. Participants explained their simplification process (what movements to simplify, why, and how) for `advanced dance' sequences, which include dynamic and individualized movements~\cite{baker2024computational} from the AIST dance video dataset.  
We developed a tool that records video and provides a real-time mirror view, along with a reference screen featuring the original AIST dance videos~(see Figure~\ref{fig:focus-group}-\subfigbox{b}). The tool segments videos into 8-count units using beats per minute~(BPM)~\cite{sofras2019dance}, allowing choreographers to select sections they find challenging for novices.
We employed think-aloud protocols~\cite{blandford2013semi} to capture choreographers' decision-making processes during dance simplification. We moderated sessions to ensure collaborative discussion among the three participants. Specifically, we encouraged participants to reach consensus on which movements to simplify and how to simplify them, rather than allowing individual simplification approaches to dominate. Focus groups were conducted under the institutional review board~(IRB) approval, and each participant received \$40 compensation per hour.

\begin{table*}[tb]
\centering
\setlength{\tabcolsep}{5pt}
\caption{Questions and objectives for each session in focus groups. Q1, Q2 were used in the \emph{criteria establishment} session, Q3--Q5 in the \emph{data collection} session, and Q6, Q7 in the \emph{post-interview} session.}
\label{tab:question-for-focus-group}
\resizebox{\linewidth}{!}{%
\small\sffamily
\begin{tabular}{p{0.49\linewidth} p{0.49\linewidth}}
    \hline
    \textbf{Questions} & \textbf{Objectives} \\
    \hline\arrayrulecolor{black!20}
    Q1. What movements do novices typically find difficult when learning dance?
    & Draw on movements novices find difficult (\S\ref{subsec:pre-survey}) to identify instructors' perspectives on challenges for novices. \\
    Q1.1. Why do you think they find these specific points challenging?
    & Understand underlying cognitive and physical factors contributing to movement difficulty. \\
    Q1.2. How do you simplify these difficult movements in your classes?
    & Document existing pedagogical strategies and practical simplification techniques. \\
    Q2. Do you have any strategies you routinely use to simplify dance motion?
    & Elicit systematic approaches and heuristics employed by expert instructors \\
    \hline
    Q3. What criteria did you use to simplify each section?
    & Validate and refine simplification criteria through applied examples. \\
    Q4. What considerations guided you when modifying movements?
    & Capture decision-making processes during simplification. \\
    Q5. Which movements did you retain, remove, or simplify?
    & Document specific simplification choices and rationale. \\
    \hline
    Q6. Of today's simplification discussions, which strategy or criterion do you consider most important?
    & Synthesize key insights and prioritize design implications. \\
    Q7. If simplification strategies were automated, what system features would be required?
    & Elicit system requirements and design considerations for computational implementation. \\
    \arrayrulecolor{black}
    \hline
\end{tabular}%
}
% \vspace{-0.5cm}
\end{table*}

\subsection{Results and Findings from Focus Groups}\label{subsec:results-and-findings-from-FG}
\begin{figure*}[t!]
    \centering
    \includegraphics[width=\linewidth]{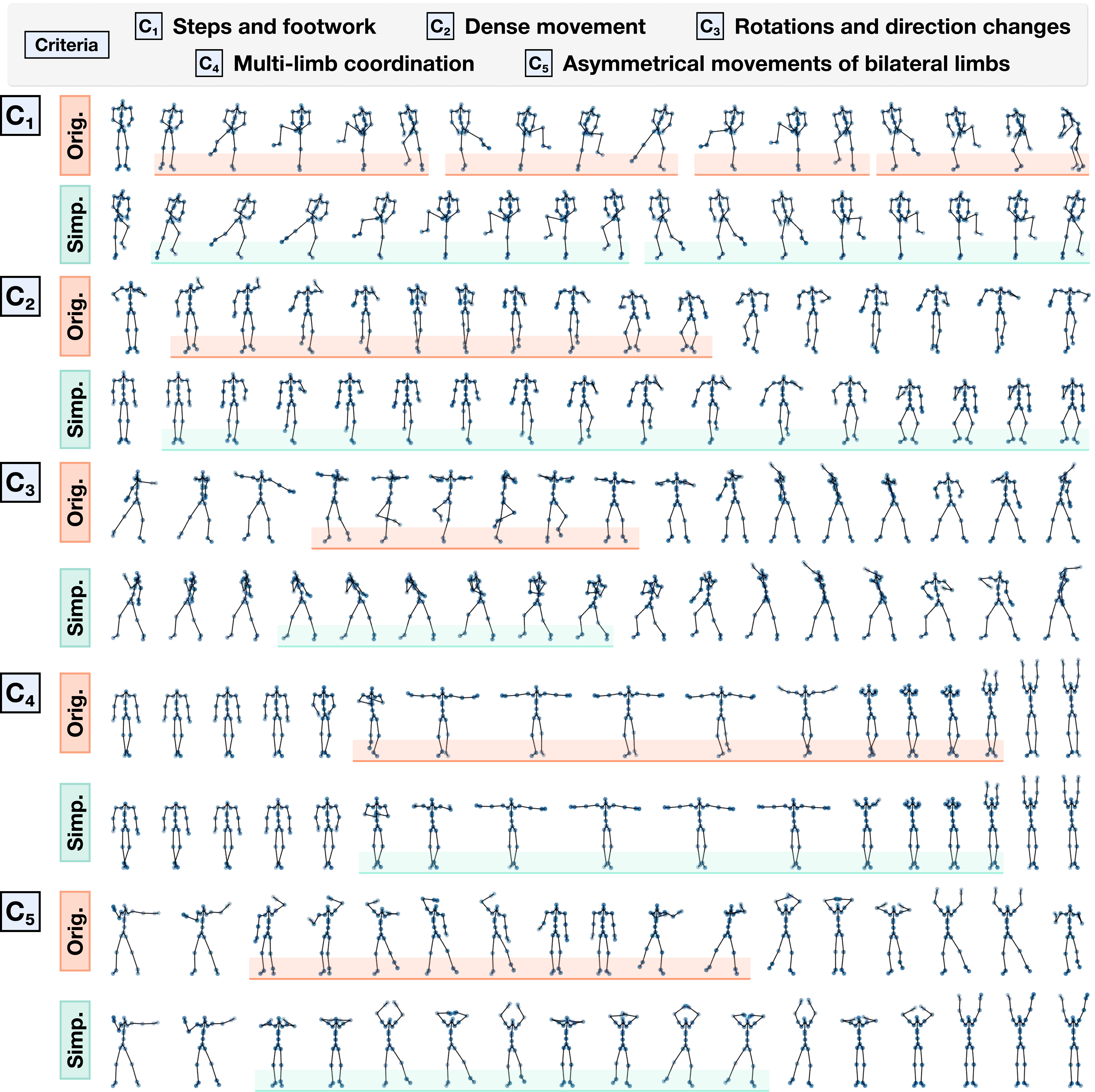}
    \caption{Original and simplified motion sequences for each simplification criterion. \subfigbox{$C_1$} Steps and footwork reducing repetitive stepping movements from two repetitions to one; \subfigbox{$C_2$} Dense movement slowing down rapidly executed movements; \subfigbox{$C_3$} Rotations and directional changes eliminating rotational movements; \subfigbox{$C_4$} Multi-limb coordination simplifying upper and lower body coordination by isolating upper body movement only; \subfigbox{$C_5$} Asymmetry of bilateral limbs transforming asymmetric bilateral arm movements into symmetric patterns. Colored boxes show key frames where simplification is most apparent. Motion flows left to right.}
    \label{fig:simplification-example}
\end{figure*}

We analyzed the focus group discussions to understand choreographers' simplification practices and to establish consensus criteria for dance motion simplification. 
Using qualitative content analysis~\cite{forman2007qualitative} centered on learner breakdowns and choreographers' simplification strategies, we coded all segments in which choreographers described 1) when novice learners tend to struggle or experience breakdowns, and 2) how they adapt choreography to support learning. 
By iteratively clustering related codes, we derived five primary categories~($\boldsymbol{C_\#}$)~of challenging movements, each associated with distinct simplification approaches. 
Figure~\ref{fig:simplification-example} shows the original and simplified motions illustrating the simplification process for each category. 
Below, we describe each category in terms of the learner challenges reported by choreographers and the strategies they commonly employ to simplify those movements.

\textbf{$\boldsymbol{C_1}.$ Steps and footwork.} Choreographers identified unfamiliar leg movements as particularly challenging due to limited usage in daily life~(P4, P17, P20). P17 explained that \textit{``many people don't know how to use their legs properly for the dance moves,''} emphasizing how the unfamiliarity of specific leg movements creates fundamental learning barriers. This difficulty was also described in a prior study~\cite{Hristova2015source}. To address this challenge, they proposed reducing the motion speed and unifying directional patterns: \textit{``Slow down the speed and make the directional patterns of steps consistent.''}

\textbf{$\boldsymbol{C_2}.$ Dense movement.} Multiple movements in limited beats cause information overload for learners (P26), increasing cognitive and motor complexity~\cite{CHAN2009316}. Choreographers address this by 1) slowing tempo through converting off-beat to on-beat~(P3, P5, P14, P19, P21--P25, P28--P30). 2) reducing repetitions~(P1, P3, P9, P16, P17, P26), and 3) substituting complex sequences with single poses, as P1 described: \textit{``Perform just one movement and then hold a different pose.''}

\textbf{$\boldsymbol{C_3}.$ Rotations and direction changes.} Rotations are challenging due to limited muscle memory, affecting balance and spatial orientation (P2, P4, P5, P12, P15, P20, P25, P28, P30). Many struggle to maintain balance while rotating (P2). Mirror dependency can cause spatial confusion (P7, P8, P12, P14, P19, P20, P29, P30), as P20 explained: \textit{``learners get confused when their mirrored image looks different from another direction,''} consistent with prior work~\cite{Hristova2015source}. To simplify, choreographers may remove rotations entirely (P1--P3, P5, P7, P11, P26), have learners face forward (P3), slow rotations (P4, P5, P11, P13, P29), or modify movements while keeping the rotation (P13, P18, P21, P25, P29).

\textbf{$\boldsymbol{C_4}.$ Multi-limb coordination.} Coordinating multiple body parts is cognitively demanding and often leaves learners feeling disconnected. P3 noted, \textit{``They feel like their body is malfunctioning. They move like robots.''} This challenge is greater with different rhythms in upper and lower body (P11), and is echoed in prior research~\cite{krasnow2015motor}. P12 proposed a difficulty hierarchy considering visual proximity and daily usage patterns: arms (closest to the line of sight), lower body (far from sight and requires constant weight shifting), and core (rarely performs torso and pelvis movements in daily life). This hierarchy informs simplification strategies, with the most common approach being isolation of upper or lower body movements (P3, P4, P6, P10, P12, P14, P16--P18, P21, P25--P27, P30). Additional strategies include synchronizing movement counts across body parts (P9, P11, P28, P30) and ensuring arms and legs move in the same direction to reduce cognitive load (P29, P30).

\textbf{$\boldsymbol{C_5}.$ Asymmetry of bilateral limbs.} Moving left and right sides of body differently creates major cognitive challenges for learners (P3, P8, P13, P20, P26, P27, P29). P20 articulated this difficulty through an analogy: \textit{``Using left and right body simultaneously in different directions is inevitably cognitively difficult. It's like trying to draw a square with your left hand while drawing a triangle with your right hand.''} This aligns with Yang et al.~\cite{yang2010evaluating}. This difficulty, also seen in drumming (P27), relates to upper-lower body coordination issues, a phenomenon also discussed before~\cite{wang2025mr}. 
To simplify, choreographers often make left and right movements symmetrical (P4, P9, P12--P14, P19, P25, P26, P29), as P12 and P13 explained: \textit{``Make left and right movements symmetrical.''} This reduces cognitive load and helps learners build skills before tackling asymmetry.

\subsection{Dataset Configuration}\label{subsec:dataset-configuration}
\begin{figure}[ht!]
    \centering
    \includegraphics[width=\linewidth]{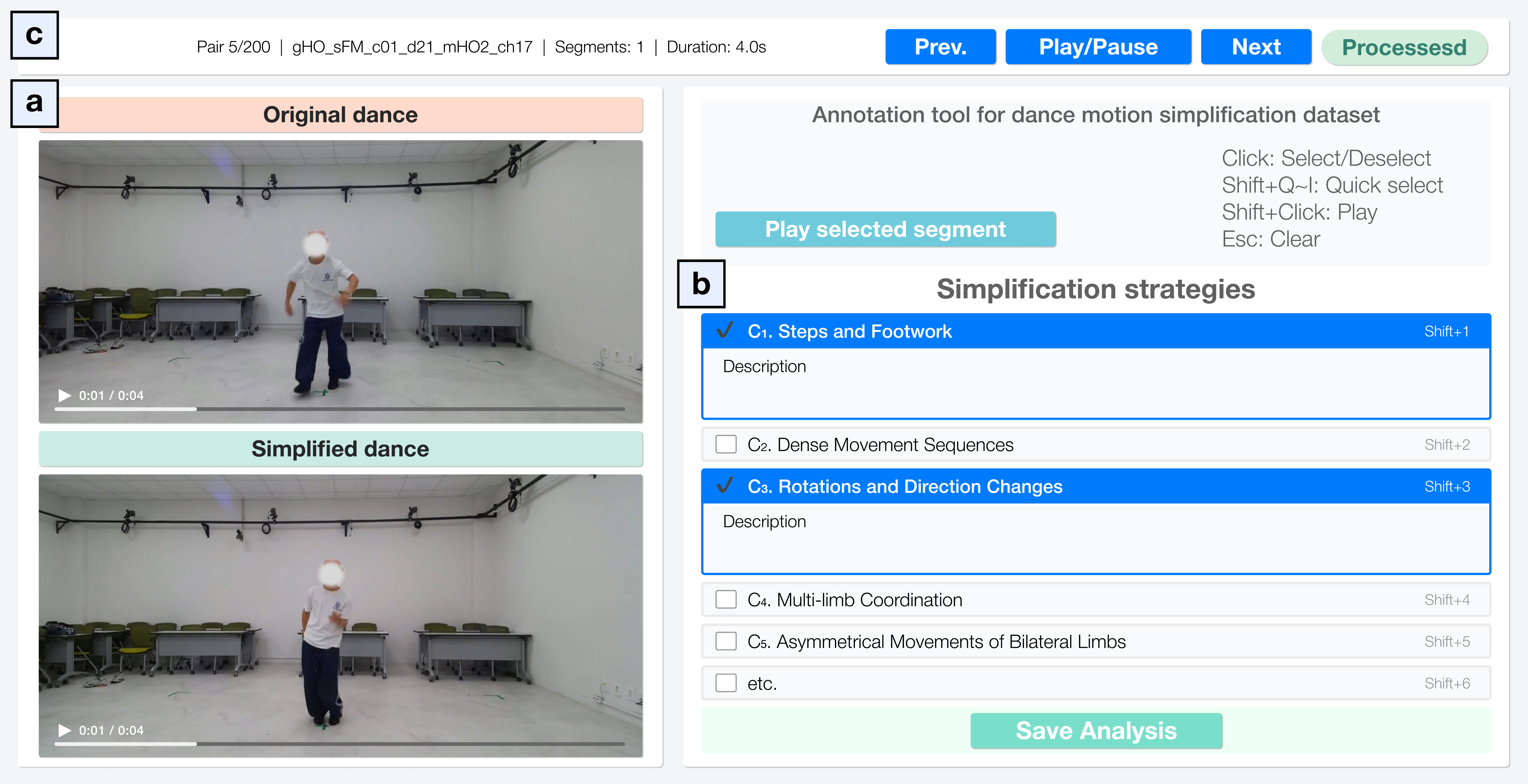}
    \caption{The custom annotation tool used to construct the dance motion simplification dataset. The interface allows for \subfigbox{a} inspecting original and simplified dance pairs, \subfigbox{b} labeling complexity criteria ($C_1$--$C_5$), and \subfigbox{c} segmenting, playing, and analyzing motion clips.} 
    \label{fig:annotation-tool}
    % \vspace{-10pt}
\end{figure}

We constructed a dataset consisting of both the original and simplified dance captured from \S\ref{subsec:focus-group}, extending prior collections~\cite{aist-dance-db, li2021learn, luo2024popdg}. Using `advanced dance' from the AIST dance video dataset~\cite{aist-dance-db}, we define the original performance as `original dance' and the corresponding simplified version as `simplified dance'. Our dataset encompasses 10 dance genres: ballet jazz, break, house, krump, LA-style hip-hop, lock, middle hip-hop, pop, street jazz, and waack. For annotation, we created a custom tool (Figure~\ref{fig:annotation-tool}) to select which of five simplification criteria from \S\ref{subsec:results-and-findings-from-FG} were applied and to describe the process. During the annotation process, once original and simplified videos were loaded, we selected segments where simplification occurred and labeled them with one or more simplification strategies. Strategies were chosen from six options ($C_1$--$C_5$ and others), and a text description of the chosen strategies was entered and saved for each segment. An annotation followed choreographers' rationale for each criterion. The dataset contains 548 pairs (82.9 minutes total), with details by criterion in Table~\ref{tab:dance-dataset}.

To enrich the dataset during model training and testing (detailed in \S\ref{subsec:complexity-eval} and \S\ref{sec:methodology}), we augmented the data in multiple ways based on prior work~\cite{jang2022motion}: 1) horizontal flipping to introduce lateral variability, 2) temporal scaling ($\{0.8, 0.9, 1.1, 1.2\}$) to simulate differences in execution speed, and 3) cropping 60–90\% of the sequence to approximate partial observations. Augmentation increased the amount of training data by eight-fold. Data were split 8:1:1 for train, validation, and test sets.

\begin{table}[ht!]
    \centering
    \small
    \sffamily
    \renewcommand{\arraystretch}{1.15}
    \caption{Dataset statistics of dance pairs by complexity criteria.}
    % \vspace{-0.3cm}
    \label{tab:dance-dataset}
    \begin{tabular}{lccc}
    \hline
    \textbf{Category} & \textbf{\# of Dance Pairs (Files)} & \textbf{Avg. Length [sec.] ($\sigma$)} & \textbf{Total Length [min.]} \\
    \hline\arrayrulecolor{black!10}
    $C_1$ & 121 (242 files) & 4.53 (0.68) & 18.1 \\
    \hline
    $C_2$ & 147 (294 files) & 4.60 (0.71) & 21.4 \\
    \hline
    $C_3$ & 84 (168 files)  & 4.76 (0.79) & 12.7 \\
    \hline
    $C_4$ & 140 (280 files) & 4.51 (0.71) & 21.5 \\
    \hline
    $C_5$ & 22 (44 files)   & 4.36 (0.58) & 3.2 \\
    \hline
    Others & 34 (68 files)  & 4.63 (0.96) & 5.0 \\
    \arrayrulecolor{black!30}
    \hline
    \textbf{Overall} & \textbf{548 (1,096 files)} & \textbf{4.54 (0.73)} & \textbf{82.9} \\
    \arrayrulecolor{black}
    \hline
    \end{tabular}
    % \vspace{-0.5cm}
\end{table}

\subsection{Evaluation of Simplified Dance with Novices}\label{subsec:eval-of-simplified-choreography}

\subsubsection{Experiment design and apparatus}
We evaluated how simplified dance affects learning efficacy using dance data from the prior phase. We employed a within-subjects repeated-measures design with two conditions (\originalbox{original} vs.\ \simplifiedbox{simplified}) crossed with five criteria ($C_1$–$C_5$). Based on G*Power 3~\cite{faul2007g} calculations with a conservative effect size ($f = 0.3$), $\alpha = .05$, and 80\% power, we determined a minimum sample size of 10 participants. We conducted an IRB-approved experiment with 16 participants (8 female; mean age $24.25$ years, $\sigma = 3.68$) who had minimal dance experience ($\mu = 0.13$ years, $\sigma = 0.50$). None of the participants in this experiment had taken part in the preceding survey~(\S\ref{subsec:pre-survey}). Detailed demographics appear in Appendix~\ref{appendix:demographics}. The experimental setup consisted of a 3~m~$\times$~3~m learning space equipped with a smartphone (Samsung Galaxy A90 5G\footnote{\url{https://en.wikipedia.org/wiki/Samsung_Galaxy_A90_5G}}) for video playback and a mirror (1.0~m~$\times$~1.7~m) for visual feedback. Each instructional video showed one 8-count sequence with a mean duration of 4.8 seconds ($\sigma = 0.84$). Participants practiced each video for five minutes. The presentation order of learning videos was counterbalanced across participants to prevent order effects. Following each learning session, participants completed three assessments: 1) NASA--TLX~\cite{hart1988development} for workload measurement~\cite{radhakrishnan2023investigating}, 2) self-efficacy~\cite{xu2025effects, anggraeni2024exploring}, and 3) perceived dance difficulty. We recorded participants' learned sequences using a Microsoft Azure Kinect\footnote{\url{https://azure.microsoft.com/en-us/products/kinect-dk}} for quantitative performance evaluation. The complete study session lasted approximately 90 minutes, with each participant receiving \$40 compensation.

\begin{figure*}[t!]
    \centering
    \includegraphics[width=\linewidth]{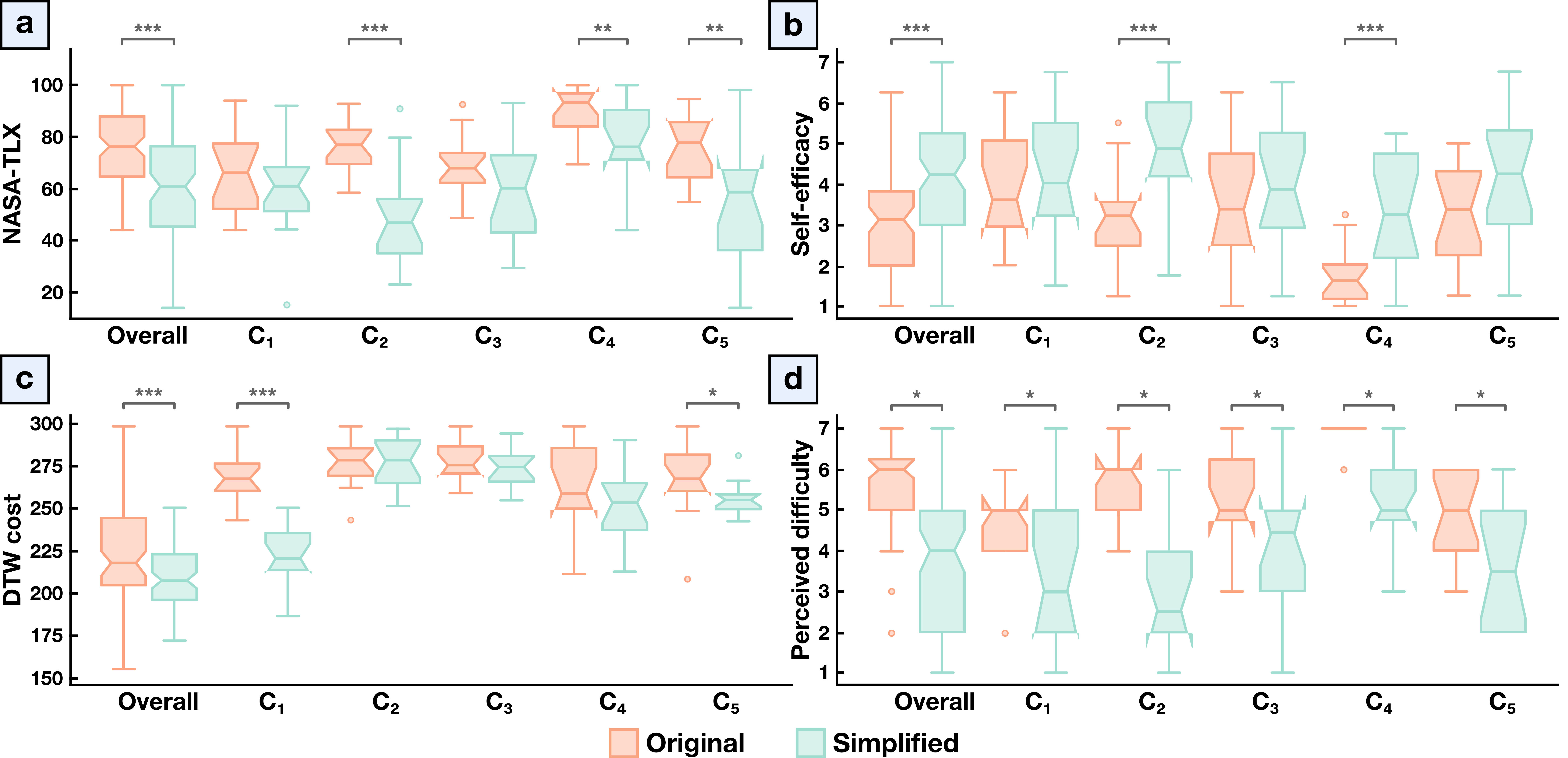}
    \caption{Overall and criterion-wise $(C_1$--$C_5)$ comparisons between \originalbox{original} and \simplifiedbox{simplified} conditions using notched boxplots with significance annotations ($^{*}p<.05$, $^{**}p<.01$, $^{***}p<.001$). \subfigbox{a} workload~(NASA--TLX)~($\downarrow$); \subfigbox{b} self-efficacy~($\uparrow$); \subfigbox{c} objective performance~(motion similarity; DTW cost)~($\downarrow$); \subfigbox{d} perceived difficulty~($\downarrow$).}
    \label{fig:learner-study-result}
    \vspace{-0.5cm}
\end{figure*}

\subsubsection{Results}\label{subsubsec:eval-of-simplified-dance-with-novices}
To establish the overall statistical framework, we first conducted 2-way repeated-measures ANOVA for each metric with condition (\originalbox{original} vs.\ \simplifiedbox{simplified}) and criteria ($C_1$--$C_5$) as within-subject factors. All metrics showed significant main effects for both condition (all $p < .001$) and criteria (all $p < .001$), and critically, significant condition $\times$ criteria interactions: workload ($F(4,60) = 4.20$, $p = .005$, $\eta_p^2 = .22$), self-efficacy ($F(4,60) = 4.48$, $p < .005$, $\eta_p^2 = .23$), objective performance ($F(4,60) = 11.28$, $p < .001$, $\eta_p^2 = .43$), and perceived difficulty ($F(4,60) = 3.36$, $p = .015$, $\eta_p^2 = .18$). These significant interactions indicate that simplification effects vary across criteria, justifying criterion-specific follow-up analyses.

For follow-up analyses, we examined the normality of paired differences using the Shapiro--Wilk test~\cite{shaphiro1965analysis}. When normality held, we applied paired-samples \textit{t}-tests~\cite{hsu2014paired}. Otherwise, we used Wilcoxon signed-rank tests~\cite{woolson2007wilcoxon}. Effect sizes are reported as Cohen's $d_z$ for paired \textit{t}-tests and rank--biserial correlation $r$ for Wilcoxon tests. Figure~\ref{fig:learner-study-result} presents notched boxplots for overall and criterion-specific ($C_1$--$C_5$) comparisons.

\textbf{Workload.} On the weighted NASA--TLX scores, the overall participant-level mean decreased from $75.93{\pm}14.38$ (\originalbox{original}) to $60.76{\pm}20.89$ (\simplifiedbox{simplified}). The \simplifiedbox{simplified} condition significantly reduced scores compared to the \originalbox{original} condition across all sessions ($\Delta = -15.17$, 95\% confidence internal~(CI) $[-19.61, -10.73]$\footnote{Note: $\Delta$ means \originalbox{original}-\simplifiedbox{simplified}.}; Wilcoxon $W = 411$, $p < .001$, $r = -.65$). Session-wise analyses revealed significant reductions for $C_2$ ($78.06{\pm}9.86$ to $51.63{\pm}18.60$; \textit{t}(15) $= -5.55$, $p < .001$, $d_z = -1.39$), $C_4$ ($89.72{\pm}9.47$ to $77.50{\pm}15.47$; Wilcoxon $W = 7$, $p = .002$, $r = -.78$), and $C_5$ ($75.58{\pm}13.10$ to $54.18{\pm}23.97$; \textit{t}(15) $= -3.76$, $p = .002$, $d_z = -.94$). $C_3$ showed a marginal trend ($69.64{\pm}12.56$ to $60.17{\pm}19.37$; \textit{t}(15) $= -2.01$, $p = .063$, $d_z = -.50$), while $C_1$ did not reach statistical significance ($66.65{\pm}15.19$ to $60.32{\pm}18.11$; \textit{t}(15) $= -1.38$, $p = .188$, $d_z = -.34$).

\textbf{Self-efficacy.} Participants reported significantly higher confidence with \simplifiedbox{simplified} dances ($\Delta = +0.93$, 95\% CI $[0.62, 1.24]$; Wilcoxon $W = 479$, $p < .001$, $r = .60$). Significant gains emerged for $C_2$ (\textit{t}(15) $= 5.55$, $p < .001$, $d_z = 1.39$) and $C_4$ (\textit{t}(15) $= 5.24$, $p < .001$, $d_z = 1.31$), with marginal trends in $C_1$ (Wilcoxon $W = 37$, $p = .113$, $r = .40$) and $C_5$ (Wilcoxon $W = 30$, $p = .069$, $r = .46$). $C_3$ did not reach statistical significance (\textit{t}(15) $= 0.85$, $p = .410$, $d_z = .21$).

\textbf{Objective performance.} We computed the dynamic time warping~(DTW)-based similarity score as an objective performance following prior works~\cite{laattala2024wave, jo2023flowar}. DTW temporally compares the learner's sequence to the ground truth and accumulates aligned pose-distance errors; a lower cost indicates higher spatiotemporal similarity. We derived similarity scores using DanceDTW~\cite{han2025choreocraft}, which is specialized for compensating for inter-participant physical differences. Then, we conducted one-sided paired tests for the directional hypothesis of $\text{cost}_\text{\simplifiedbox{simp.}} $<$ \text{cost}_\text{\originalbox{orig.}}$. In the overall analysis, the simplified condition yielded a significantly lower cost (Wilcoxon $W = 704$, $p_\text{one} < 10^{-5}$, $r = -.49$, $\Delta = -15.88$; $\overline{\text{cost}}_\text{\originalbox{orig.}}=222.73$, $\overline{\text{cost}}_\text{\simplifiedbox{simp.}}=206.84$). Session-wise tests showed a significance in $C_1$ (\textit{t}(15) $= -9.06$, $p_\text{one} < 10^{-7}$) and $C_5$ (\textit{t}(15) $= -2.12$, $p_\text{one} < .05$), while $C_2$--$C_4$ were not significant.

\textbf{Perceived difficulty.} The \simplifiedbox{simplified} condition yielded markedly lower perceived difficulty ($\Delta = -1.59$, 95\% CI $[-1.96, -1.21]$; Wilcoxon $W = 223$, $p < .001$, $r = -.72$). All five criteria showed significant reductions: $C_2$ (\textit{t}(15) $= -7.03$, $p < .001$, $d_z = -1.76$), $C_4$ (\textit{t}(15) $= -5.93$, $p < .001$, $d_z = -1.48$), $C_3$ (Wilcoxon $W = 9$, $p < .01$, $r = -.65$), $C_1$ (\textit{t}(15) $= -2.51$, $p < .05$, $d_z = -.63$), and $C_5$ (\textit{t}(15) $= -2.56$, $p = .022$, $d_z = -.64$).

\textbf{Order effects.} We compared within-session deltas (\simplifiedbox{simplified} $-$ \originalbox{original}) between participants whose first exposure was \originalbox{original}$\rightarrow$\simplifiedbox{simplified} vs.\ \simplifiedbox{simplified}$\rightarrow$\originalbox{original}. No significant differences emerged for workload ($p = .199$) or confidence ($p = .559$). Objective performance likewise showed no order effect (Mann--Whitney $U = 908$, $p = .301$), indicating that improvements stem from the dance motion manipulation rather than presentation order.

\subsection{Summary and Design Implications}\label{subsec:discussion-implications}
The results of our formative study provide two critical implications for the computational design. First, the user study confirms that the simplification criteria identified by experts~($C_1$--$C_5$) are valid pedagogical constructs; reducing workload and complexity along these dimensions significantly improves novice performance and confidence. Second, the paired dataset we constructed serves as a reliable ground truth for modeling these criteria. However, applying these criteria manually is labor-intensive and requires expert intervention. To scale these pedagogical benefits, we need to translate these qualitative insights into computational models.

\section{Defining Computational Metrics for Dance Motion Complexity}\label{sec:computational-metrics}
Drawing from the qualitative insights on challenging movements identified in our workshop, we developed computational measures to quantify each complexity criterion. The core challenge in this translation is bridging the gap between the subjective, pedagogical descriptions of choreographers and objective kinematic features extracted from motion data. To address this, we mapped the five identified criteria~($C_1$--$C_5$) to specific physical properties. These measures are designed not just to describe motion, but to specifically detect the ``breakdown points'' for novices validated in \S\ref{sec:workshop}.

\subsection{Complexity Measure Formulation}\label{subsec:complexity-metrics}
Each complexity measure combines multiple kinematic components with different physical units. To integrate these heterogeneous terms into a single score, we apply weighting coefficients that control each component's contribution. These coefficients were optimized via Bayesian optimization~\cite{garnett2023bayesian} to maximize classification accuracy on expert-annotated complexity labels.

\subsubsection{$C_1$: Steps and footwork}
Novices struggled with footwork due to its unfamiliarity and high variability. To capture this, we quantify four aspects: how fast feet move (velocity), how unpredictable the movements are (entropy), how much ground is covered (horizontal range), and how frequently foot contacts change  (transitions). To quantify this complexity, we combine four complementary measures:

\begin{equation}\label{eq:c1}
C_1 = \bar{V}_f + \alpha_1 \, H(\|\mathbf{j}^v_{1:T}[f]\|) + \alpha_2 \, R_{XZ}(\mathbf{j}^p_{1:T}[f]) + \alpha_3 \, \tau_f,
\end{equation}

where $\bar{V}_f = \frac{1}{T}\sum_{t=1}^T(\|\mathbf{j}^v_t[f_r]\| + \|\mathbf{j}^v_t[f_l]\|)$ is the mean foot velocity (m/s) capturing overall movement speed, with $f_r$ and $f_l$ denoting right and left foot joint indices; $H(\|\mathbf{j}^v_{1:T}[f]\|)$ is the entropy (nats) of discretized foot velocity magnitudes, measuring movement unpredictability (high entropy indicates erratic patterns, low entropy indicates repetitive motion)~\cite{huang2021gait, coates2020entropy}; $R_{XZ}(\mathbf{j}^p_{1:T}[f])$ is the horizontal range (m) quantifying spatial coverage; and $\tau_f = \frac{1}{T}\sum_{t=2}^T \mathbb{I}[\mathbf{c}^f_t \neq \mathbf{c}^f_{t-1}]$ is the contact transition rate (dimensionless) capturing step frequency by counting contact changes. Here $\mathbf{j}^v_t[j]$ denotes the 3D velocity vector (m/s) of joint $j$ at frame $t$, $\mathbf{j}^p_t[j]$ denotes the 3D position vector (m), $T$ is the total number of frames, and $\mathbf{c}^f_t \in \{0,1\}^4$ is the binary foot contact vector. The weights $\alpha_1=1.5$, $\alpha_2=0.05$, $\alpha_3=15.0$ balance each term's contribution, with the high $\alpha_3$ weight emphasizing that step frequency is the most discriminative indicator of footwork complexity. The entropy and range functions are:

\begin{align}
    H(X) &= -\sum_{k} p_k \ln p_k, \\
    R_{XZ}(\mathbf{j}^p) &= \sqrt{(\max(x) - \min(x))^2 + (\max(z) - \min(z))^2},
\end{align}

where $p_k$ is the probability of velocity magnitude falling in bin $k$ (10 uniform bins), and $x$, $z$ denote horizontal coordinates.

\subsubsection{$C_2$: Dense movement}
Dense movements pack many actions into short time windows, overwhelming novices with information overload. We measure this by tracking both the speed of limb movements and how abruptly they change direction (acceleration), normalizing across joints to ensure fair comparison. We define dense movement through normalized velocities and smoothed accelerations:

\begin{equation}
    C_2 = \frac{1}{T}\sum_{t=1}^T\sum_{j\in\mathcal{J}_{\text{limbs}}}\frac{\|\mathbf{j}^v_t[j]\|}{\sigma_j} + \beta \cdot \text{median}_{t,j}\left(\|\hat{\mathbf{j}}^a_t[j]\|\right),
\end{equation}

where the first term is the mean normalized limb velocity (dimensionless) quantifying sustained movement intensity across all limbs, with $\mathcal{J}_{\text{limbs}}$ covering all arm and leg joints (knee, ankle, foot, shoulder, elbow, wrist) and $\sigma_j$ being the sequence-wide velocity standard deviation (m/s) of joint $j$ (normalizing each joint's contribution); and the second term is the median acceleration magnitude (m/s$^2$) capturing abrupt direction changes (high acceleration = jerky motion), weighted by $\beta=0.005$ to match the scale of the velocity term. The smoothed acceleration $\hat{\mathbf{j}}^a_t[j]$ is:

\begin{equation}
    \hat{\mathbf{j}}^a_t[j] = \hat{\mathbf{j}}^v_{t+1}[j] - 2\hat{\mathbf{j}}^v_t[j] + \hat{\mathbf{j}}^v_{t-1}[j],
\end{equation}

where $\hat{\mathbf{j}}^v_t[j]$ is the Savitzky-Golay (SG) filtered velocity to reduce noise~\cite{schafer2011savitzky, mccay2019establishing}. The median operator over all joint-frame pairs provides robustness to outliers~\cite{massart1986least, rousseeuw1991tutorial}.

\subsubsection{$C_3$: Rotations and direction changes}
Body rotations challenge balance and spatial orientation, especially when learners rely on mirrors that flip their perspective. We quantify rotation complexity through three components: total rotation angle (how much the body turns), angular velocity (how fast it rotates), and angular acceleration (how suddenly rotation speed changes). Our measure combines total rotation, angular velocity, and angular acceleration:

\begin{equation}\label{eq:c3}
    C_3 = \gamma_1 \cdot \frac{|\theta_T - \theta_1| \bmod 2\pi}{\pi} + \gamma_2 \cdot \mathbb{E}[|\Delta \hat{\theta}_t|] + \gamma_3 \cdot \mathbb{E}[|\Delta^2 \hat{\theta}_t|],
\end{equation}

where $\theta_t$ (radians) is the pelvis yaw angle at frame $t$, $\hat{\theta}_t$ is the SG-filtered angle, and the terms represent: 1) normalized total rotation (dimensionless, range $[0,2]$) measuring cumulative turning from start to end, weighted by $\gamma_1=0.3$; 2) mean angular velocity magnitude $|\Delta \hat{\theta}_t| = |\hat{\theta}_{t+1} - \hat{\theta}_t|$ (rad/frame) capturing rotation speed, weighted by $\gamma_2=1.0$; and 3) mean angular acceleration magnitude $|\Delta^2 \hat{\theta}_t| = |\Delta \hat{\theta}_{t+1} - \Delta \hat{\theta}_t|$ (rad/frame$^2$) penalizing jerky rotations that challenge balance control, weighted by $\gamma_3=0.5$. $\mathbb{E}[\cdot]$ denotes the temporal mean.

\subsubsection{$C_4$: Multi-limb coordination}
Moving different body parts independently is cognitively demanding---choreographers noted learners feel like ``robots'' when attempting this. We capture coordination difficulty by measuring how differently upper and lower body move: high variance in their movement intensities indicates independent, desynchronized patterns that challenge novices. We quantify multi-limb coordination through movement intensity differences:

\begin{align}
    C_4 &= \text{Var}(I_t^{\text{upper}} - I_t^{\text{lower}}) \cdot \mathbb{I}[\min(\mu_{\text{upper}}, \mu_{\text{lower}}) > \delta], \\
    \text{where} \quad I_t^{\text{upper}} &= \frac{1}{|\mathcal{J}_{\text{upper}}|} \sum_{j \in \mathcal{J}_{\text{upper}}} \|\mathbf{j}^v_t[j]\|, \quad I_t^{\text{lower}} = \frac{1}{|\mathcal{J}_{\text{lower}}|} \sum_{j \in \mathcal{J}_{\text{lower}}} \|\mathbf{j}^v_t[j]\|.
\end{align}

Here $\mathcal{J}_{\text{upper}}$ includes upper body joints (collars, shoulders, elbows, wrists), $\mathcal{J}_{\text{lower}}$ includes lower body joints (hips, knees, ankles, feet), and $I_t$ represents movement intensity (m/s) at frame $t$. The variance $\text{Var}(\cdot)$ ((m/s)$^2$) of intensity differences captures desynchronization: high variance indicates independent movement patterns (upper body moving fast while lower body is slow, or vice versa), while low variance suggests synchronized motion (both moving similarly). The indicator function $\mathbb{I}[\cdot]$ activates only when both segments show sufficient movement (means $\mu_{\text{upper}}, \mu_{\text{lower}} > \delta = 0.01$ m/s), ensuring the measure focuses on actual coordination challenges rather than static poses~\cite{dubois2023guide}.

\subsubsection{$C_5$: Asymmetry of bilateral limbs}
Moving left and right sides differently is cognitively challenging---choreographers compared it to ``drawing a square with one hand while drawing a triangle with the other''. We measure asymmetry by comparing velocity and spatial position differences between paired joints, penalizing both instantaneous asymmetry and cases where one side dominates movement. We combine instantaneous asymmetry and movement imbalance to measure left-right asymmetry:

\begin{equation}
C_5 = \left(\frac{1}{T} \sum_{t=1}^T A_t\right) \cdot \left(1 + \lambda \cdot \mathbb{I}\left[ \frac{\min(V_L, V_R)}{\max(V_L, V_R) + \epsilon} < \delta \right] \right),
\end{equation}

where the asymmetry score $A_t$ (m/s and m) at each frame is computed from velocity and position differences:

\begin{equation}
    A_t = \sum_{j \in \mathcal{P}} w_j \left(\Delta v_t[j] + 0.5 \cdot \Delta p_t[j]\right),
\end{equation}

with velocity asymmetry $\Delta v_t[j] = |\|\mathbf{j}^v_t[j_L]\| - \|\mathbf{j}^v_t[j_R]\||$ (m/s) measuring dynamic differences and position asymmetry $\Delta p_t[j] = \|\mathbf{j}^p_t[j_L] - \text{mirror}(\mathbf{j}^p_t[j_R])\|$ (m) measuring spatial differences. Total velocities (m/s) are:

\begin{align}
    V_L = \sum_{t=1}^{T} \sum_{j \in \mathcal{P}} \|\mathbf{j}^v_t[j_L]\|, \quad V_R = \sum_{t=1}^{T} \sum_{j \in \mathcal{P}} \|\mathbf{j}^v_t[j_R]\|.
\end{align}

$\mathcal{P}$ denotes paired joints (ankle, knee, hip, shoulder, elbow, wrist), $j_L$ and $j_R$ represent left/right instances of joint $j$, and $w_j$ are joint-specific weights inversely proportional to average joint movement magnitude (ensuring equal contribution). The first factor captures instantaneous asymmetry: the velocity term measures dynamic asymmetry (different speeds), while the position term measures spatial asymmetry (different locations). The second factor is a penalty multiplier with weight $\lambda=0.5$ that increases complexity when one side dominates movement (ratio $< \delta=0.01$; $\epsilon=10^{-6}$ prevents division by zero), ensuring one-sided movements are properly penalized~\cite{siebers2021comparison}.

\subsection{Complexity Measures Evaluations}\label{subsec:complexity-eval}
To validate our complexity measures, we trained five binary classifiers---one for each criterion ($C_1$--$C_5$)---to distinguish sequences containing the target complexity factor from those that do not. For example, given an arbitrary motion sequence, the $C_1$ classifier predicts whether that sequence involves notable steps and footwork complexity. We employed gradient boosting~\cite{friedman2001greedy} and XGBoost~\cite{chen2016xgboost}, which perform well in movement analysis~\cite{d2021keep, noh2021xgboost, salam2018evaluating}. 
% We augmented data three ways to ensure robustness based on prior work~\cite{jang2022motion}: 1) horizontal flipping to introduce lateral variability, 2) temporal scaling ($\{0.8, 0.9, 1.1, 1.2\}$) to simulate differences in execution speed, and 3) cropping 60–90\% of the sequence to approximate partial observations. Augmentation increased the amount of training data by eight-fold. Data were split 8:1:1 for train, validation, and test sets.

Results (Table~\ref{tab:complexity-evaluation}) show that all measures achieved classification accuracies above 91\% and AUC scores exceeding 0.98. These results indicate that each measure captures distinct aspects of movement complexity with high specificity—that is, classifiers trained on one complexity dimension reliably distinguish sequences that differ along that dimension. Detailed model configurations are provided in Appendix~\ref{appendix:configs-for-classifiers}.
% To assess whether these measures also align with human perception, we examined data from our learner study (\S\ref{subsubsec:eval-of-simplified-dance-with-novices}): reductions targeted by each measure corresponded to significant decreases in participants' perceived difficulty, providing human-centered validation of the proposed measures. Together, these findings confirm that our measures are both specific to their intended complexity dimensions and aligned with learners' subjective experience. Detailed model configurations are provided in Appendix~\ref{appendix:configs-for-classifiers}.

\begin{table}[h]
\centering
\renewcommand{\arraystretch}{1.15}
\small\sffamily
\caption{Challenging movement classification performance with proposed complexity measures.}
\label{tab:complexity-evaluation}
    \begin{tabular}{lccr}
    \hline
    \textbf{Complexity measure} & \textbf{Algorithm} & \textbf{Accuracy} & \textbf{AUC} \\
    \hline\arrayrulecolor{black!10}
    $C_1$ (Steps and footwork) & Gradient Boost & 93.34\% & 0.9891 \\
    \hline
    $C_2$ (Dense movement) & XGBoost & 94.30\% & 0.9930 \\
    \hline
    $C_3$ (Rotations and direction changes) & Gradient Boost & 91.90\% & 0.9814 \\
    \hline
    $C_4$ (Multi-limb coordination) & Gradient Boost & 92.58\% & 0.9872 \\
    \hline
    $C_5$ (Asymmetry of bilateral limbs) & Gradient Boost & 97.65\% & 0.9969 \\
    \arrayrulecolor{black!30}
    \hline
    \textbf{Average} & & \textbf{93.95\%} & \textbf{0.9895} \\
    \arrayrulecolor{black}
    \hline
    \end{tabular}
% \vspace{-0.5cm}
\end{table}

\section{Dance Motion Simplification Methodology}\label{sec:methodology}
To reduce the complexity of dance motion for novice learners, we employ rule-based and learning-based approaches. These two approaches are the first computational frameworks designed specifically for dance motion simplification, a task previously unaddressed in motion manipulation literature \cite{bruderlin1995motion, gleicher1997motion, zhang2024motiondiffuse}. Both approaches are designed based on empirically derived complexity measures and simplification strategies introduced in \S\ref{subsec:complexity-metrics}, which were established through workshops with choreographers and learners. To aid understanding of how each equation operates in practice, we provided pseudocode in Appendix~\ref{appendix:pseudocode} for reference.
% Together, these methods offer balanced benefits: rule-based provides clarity and precision, while learning-based ensures adaptability and scalability.

\subsection{Rule-Based Approach}\label{subsec:rule-based-model}
The rule-based method provides direct, interpretable control over motion parameters using established editing techniques~\cite{bruderlin1995motion, witkin1995motion}. This lightweight and training-free approach decomposes motion into elementary components and applies operations to adjust features like velocity, joint range, and spatial extent. It enforces anatomical constraints while offering transparent simplification. However, by focusing on low-level kinematics, this approach is best suited for structural adjustments, such as reducing spatial reach and speed, rather than capturing high-level stylistic qualities.

\begin{figure}[t]
    \centering
    \includegraphics[width=\linewidth]{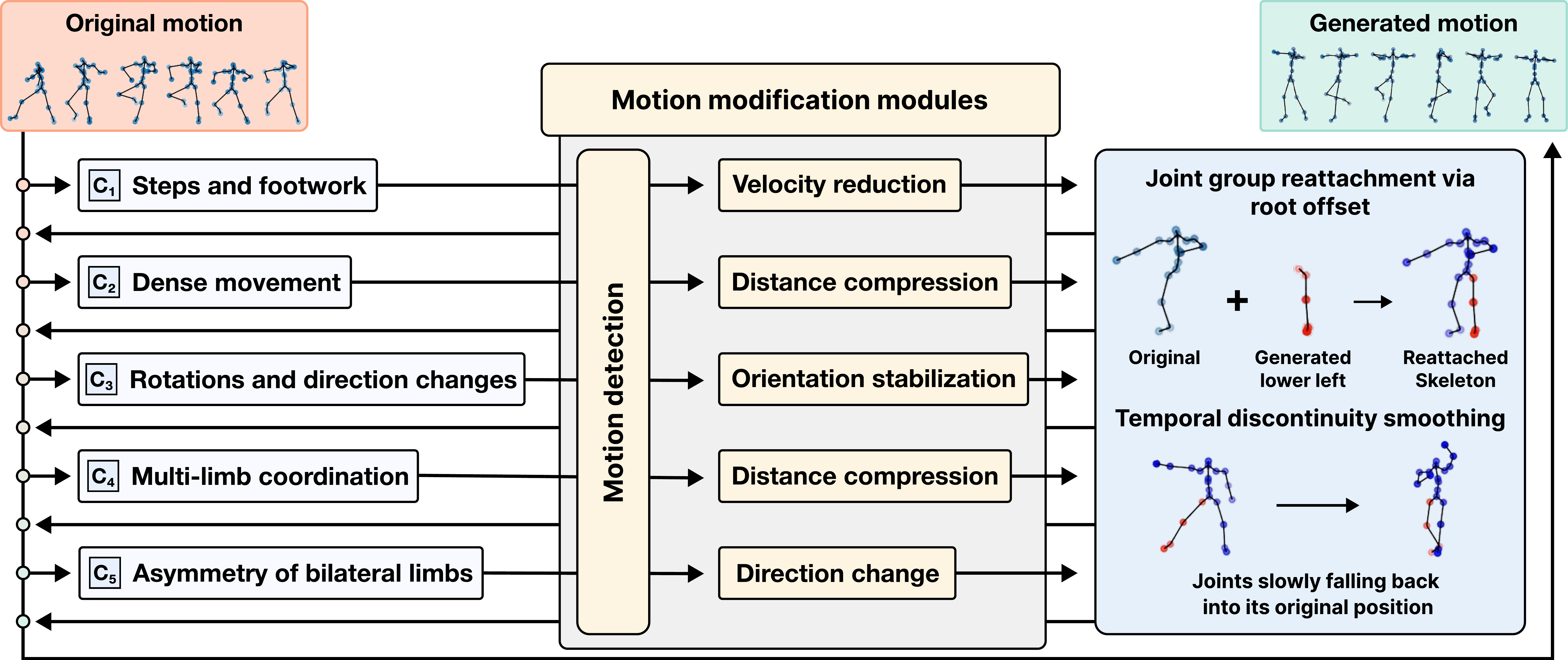}
    \caption{Overview of the rule-based approach. Beginning with an original dance sequence, the model detects motion features aligned with five complexity measures. Each identified complex movement activates a corresponding simplification rule: velocity reduction in lower limbs for $C_1$, distance compression for $C_2$ and $C_4$, orientation stabilization for $C_3$, and direction change for $C_5$. Two post-processing modules refine the output: 1) joint group reattachment via root offset to reconnect simplified limb groups to the skeleton, and 2) temporal discontinuity smoothing to interpolate joint positions.}
    \label{fig:rule-model-structure}
    \vspace{-0.5cm}
\end{figure}

\subsubsection{Implementation details} As shown in Figure~\ref{fig:rule-model-structure}, the rule-based model first computes complexity scores $C_1$--$C_5$, then detects intervals matching each measure’s characteristics. The pipeline sequentially applies targeted modifications: velocity reduction for steps and footwork ($C_1$), distance compression for dense movements and multi-limb coordination ($C_2$, $C_4$), orientation stabilization for rapid rotations and direction changes ($C_3$), and directional change for bilateral asymmetries ($C_5$). Two post-processing modules further refine results: joint group reattachment via root offset maintains limb alignment, and temporal discontinuity smoothing restores natural joint trajectories. This pipeline yields simplified dance motion while maintaining structural integrity.

\textbf{Motion detection.} We treat each joint trajectory as a sequence of short, \emph{monotonic} motion pieces: we first cut the per-axis position signal into intervals where the joint keeps moving in a consistent direction (ignoring sub-$\varepsilon$ jitter), and then merge axis-wise pieces that co-occur in time to form a single 3D motion trend. This yields compact motion events that explicitly capture \emph{when} a joint moves (duration), \emph{how far} it moves (travel distance), and \emph{in which direction} it moves (3D sign pattern). The motion detection module analyzes the world coordinates of SMPL joints, obtained from TRAM model~\cite{wang2024tram}, to determine the direction, distance, and duration of their motion. For each joint and axis, the module first computes frame-wise displacements and automatically segments the full sequence into short intervals where the displacement keeps a consistent sign. Within each such locally identified interval $[s, e]$, where $s$ and $e$ denote the start and end frame indices, we define a joint's motion along axis $x$ to be \emph{increasing} if

\begin{equation}
    \mathbf{j}^p_{t}[j]_x \leq \mathbf{j}^p_{t+1}[j]_x - \varepsilon,
    % \; \forall \, t \in [s, e-1], 
    \; \text{and \emph{decreasing} if} \;
    \; \mathbf{j}^p_{t}[j]_x \geq \mathbf{j}^p_{t+1}[j]_x + \varepsilon,
    \; \forall \, t \in [s, e-1].
\end{equation}

The same conditions apply analogously for the \(y\) and \(z\) axes.

To group motions spanning multiple axes, we use a sweep-line–inspired algorithm~\cite{cormen2009introduction} that merges temporally overlapping axis trends, each represented as $(s, e, \vec{v})$, where $\vec{v} \in \{-1, 0, 1\}^3$ encodes direction along $(x, y, z)$. We sort trends by start index and use a virtual sweep line to merge intervals whose overlap ratio exceeds threshold $\alpha$:

\begin{equation}
    \frac{|[s_1, e_1] \cap [s_2, e_2]|}{\min(e_1 - s_1, \, e_2 - s_2)} \geq \alpha.
\end{equation}

When two trends are merged, we unify the time span and sum direction vectors to create a combined trend. Empirically, $\varepsilon = 0.2$ and $\alpha=0.5$ gave optimal results for our dataset (\S\ref{subsec:dataset-configuration}).

\textbf{Distance compression.} Distance compression shrinks a joint's motion \emph{amplitude} by multiplying every frame-to-frame step by $k$, so the joint follows the same trajectory at the same timing but with a smaller spatial excursion.
We apply this rule to complexity measures that are dominated by excessive travel distance and large spatial reach---e.g., dense movements and multi-limb coordination ($C_2$, $C_4$)---because reducing amplitude directly lowers spatial complexity while preserving rhythm. For a joint $j$ and interval $[s,e]$, we scale the frame-wise displacement by a factor $k\in[0,1]$:

\begin{equation}
    \mathbf{j}^{p,\textit{gen}}_{t}[j] =
    \mathbf{j}^{p,\textit{gen}}_{t-1}[j] + k\big(\mathbf{j}^{p}_{t}[j]-\mathbf{j}^{p}_{t-1}[j]\big),\quad t=s+1,\dots,e,
\end{equation}
with $\mathbf{j}^{p,\textit{gen}}_{s}[j]=\mathbf{j}^{p}_{s}[j]$. This preserves the temporal rhythm while reducing travel distance.

\textbf{Velocity reduction.} Velocity reduction makes a fast motion look slower by redistributing each original step over $\lambda$ smaller sub-steps \emph{within the same fixed timeline}, i.e., we insert intermediate poses so the joint advances more gradually without changing the overall motion path. We use this rule for step- and footwork-dominated complexity (e.g., $C_1$), where rapid joint displacements create dense, hard-to-follow movements for novices.
We slow down a motion segment using a fixed-length time-warp within the original timeline. Given an integer slow-down factor $\lambda>1$ and an interval $[s,e]$, we map each original step $(s+k)\to(s+k+1)$ to $\lambda$ interpolated frames:
\begin{equation}
    \mathbf{j}^{p,\textit{gen}}_{f}[j] =
    \mathbf{j}^{p}_{s+k}[j] + \frac{i}{\lambda}\Big(\mathbf{j}^{p}_{s+k+1}[j]-\mathbf{j}^{p}_{s+k}[j]\Big), \quad f = s + k\lambda + i,\quad i=0,\dots,\lambda-1.
\end{equation}

We set the final keyframe to match the original endpoint and apply the edit only if the stretched endpoint does not exceed the sequence length; otherwise, we skip this rule for the segment.

\textbf{Directional change.} This rule enforces left--right consistency: when the two wrists perform temporally aligned motions but in opposite directions, we mirror the right-arm motion so that both sides move in a coherent direction.
Instead of mirroring absolute positions, we flip the sign of the \emph{frame-to-frame displacement} only along the disagreeing axes, which preserves the motion's timing and local magnitude while correcting directional inconsistency (targeting bilateral-asymmetry complexity such as $C_5$). To reduce bilateral asymmetry, we compare overlapping motion trends from the left and right wrist. When two trends overlap sufficiently and their direction indicators disagree, we construct a flip vector $\vec{v}\in\{-1,+1\}^3$ that flips only the axes whose trend directions differ, and apply it to the right-arm joint chain over the overlapped window:
\begin{equation}
    \mathbf{j}^{p,\textit{gen}}_{f}[j] = \mathbf{j}^{p}_{s}[j] + \sum_{t=s+1}^{f}\big(\vec{v}\odot(\mathbf{j}^{p}_{t}[j]-\mathbf{j}^{p}_{t-1}[j])\big).
\end{equation}
where $\odot$ denotes element-wise multiplication. This preserves local magnitudes and timing while mirroring selected axes to align bilateral motion directions.

\textbf{Orientation stabilization.} This rule stabilizes the dancer's global facing direction while preserving the relative limb configuration. We rotate the whole skeleton around the pelvis so that the body faces a consistent heading across frames, reducing rapid yaw changes (e.g., $C_3$). At each frame $f$, we estimate the current yaw $\psi_f$ from a pelvis-centered reference vector $\mathbf{d}_f = \mathbf{j}^p_f[\text{ref}] - \mathbf{j}^p_f[\text{pelvis}]$ in the XZ plane, and compute a frame-wise correction $\theta_f = \psi_{\text{target}} - \psi_f$. We then rotate all joints around the pelvis by $\theta_f$:
\begin{equation}
    \mathbf{j}^{p,\textit{gen}}_{f}[j]
    = \mathbf{R}_y(\theta_f)\!\left( \mathbf{j}^p_{f}[j] - \mathbf{j}^p_{f}[\text{pelvis}] \right)
    + \mathbf{j}^p_{f}[\text{pelvis}],
\end{equation}
where $\mathbf{R}_y(\cdot)$ denotes the Y-axis rotation matrix. By construction, this cancels instantaneous yaw, so the resulting facing direction becomes approximately $\psi_{\text{target}}$ at every frame.

\textbf{Post-simplification correction of discontinuities.}
Editing only a subset of joints often breaks kinematic coherence: the edited limb chain may drift away from the torso, and the boundary frame can exhibit a visible “snap” back to the unedited motion. We therefore apply two corrections in a single post-processing routing: 1) root locking for joint-group reattachment and 2) fade-out offset smoothing to remove boundary pops.

\textbf{Joint-group reattachment via root locking.}
We apply each rule to an ordered joint chain $\mathcal{J}$ (e.g., a leg or arm), where the first joint is treated as the chain root~(proximal anchor joint)~$r$. For every frame $f\in[s,e]$, we compute a translation that aligns the edited root back to the original root position,
\begin{equation}
\Delta_f = \mathbf{j}^p_f[r]-\mathbf{j}^{p,\textit{gen}}_f[r],
\end{equation}
and add it to all joints in the chain:
\begin{equation}
\mathbf{j}^{p,\textit{gen}}_f[j] \leftarrow \mathbf{j}^{p,\textit{gen}}_f[j]+\Delta_f,\quad \forall j\in\mathcal{J}.
\end{equation}
Intuitively, this “pins” the limb chain to the body at its root while preserving the relative edits within the chain.

\textbf{Temporal discontinuity smoothing.}
Even after root locking, the last edited frame $e$ can be spatially offset from the next unedited frame ($e{+}1$), producing a jump. To remove it, we translate the future segment by the endpoint offset and gradually fade it out.
For each edited joint $j\in\mathcal{J}$, define the endpoint offset
\begin{equation}
    o_e[j] = \mathbf{j}^p_e[j]-\mathbf{j}^{p,\textit{gen}}_e[j].
\end{equation}
For each future frame $f>e$, we 1) shift the generated joint by the remaining offset and 2) reduce the offset only when the original motion moves against it:
\begin{align}
    \mathbf{j}^{p,\textit{gen}}_f[j] &\leftarrow \mathbf{j}^{p,\textit{gen}}_f[j]-o_{f-1}[j],\\
    d_f[j] &= \mathbf{j}^p_f[j]-\mathbf{j}^p_{f-1}[j],\\
    o_f[j] &=
\begin{cases}
\texttt{clip0}(o_{f-1}[j]+d_f[j]), & \text{if } \texttt{sign}(d_f[j])=-\texttt{sign}(o_{f-1}[j])\\
o_{f-1}[j], & \text{otherwise,}
\end{cases}
\end{align}
where $\texttt{clip0}(\cdot)$ sets each axis to $0$ if the update would flip its sign (preventing overshoot). Intuitively, we first “attach” the unedited tail to the edited endpoint, then release this attachment as the original motion naturally closes the gap.

\subsubsection{Ablation study}\label{subsec:rule-ablation-study}
In our ablation study, we evaluate the generated dances using several metrics. Physical foot contact~$(PFC)$, assesses the physical plausibility of movements by measuring the proportion of frames where foot velocity remains below a threshold during ground contact; lower values indicate less unrealistic foot-sliding~\cite{tseng2023edge}. Since dance involves full-body movement and $PFC$ only considers the lower body, we also adopt physical body contact~$(PBC)$, which extends this assessment to the entire body~\cite{luo2024popdg}. To measure motion quality, we use $FID_k$ and $FID_g$ based on prior work~\cite{heusel2017gans}, which evaluate the difference between distributions of generated and ground truth dances in kinetic and geometric feature spaces, respectively; lower values indicate that generated dances more closely resemble the ground truth distribution. For motion diversity, $Dist_k$ and $Dist_g$ quantify the distributional spread of generated dances within kinetic and geometric feature spaces; higher values indicate greater variety in the generated outputs~\cite{luo2024popdg}. We used our collected dataset~(from \S\ref{subsec:dataset-configuration}) for two ablation studies~(\S\ref{subsec:rule-ablation-study} and \S\ref{subsec:learning-ablation-study}) and comparison study~(\S\ref{subsec:comparison}).

\begin{table}[h!]
\centering
\renewcommand{\arraystretch}{1.15}
\caption{Comparison of motion evaluation metrics when individually removing each complexity step ($C_1$–$C_5$) from the stair-like model structure. Arrows (↓, ↑, →) indicate whether lower, higher, or closer to ground truth (GT) is better.}
\vspace{-0.3cm}
\setlength{\tabcolsep}{4pt}
\label{tab:ablation-rule}
\small\sffamily
\resizebox{\linewidth}{!}{%
    \begin{tabular}{lcccccccccccc}
    \hline
    \textbf{Method} & $\boldsymbol{PFC~(\downarrow)}$ & $\boldsymbol{PBC~(\rightarrow)}$ & $\boldsymbol{FID_k~(\downarrow)}$ & $\boldsymbol{FID_g~(\downarrow)}$ & $\boldsymbol{Dist_k~(\uparrow)}$ & $\boldsymbol{Dist_g~(\uparrow)}$ & $\boldsymbol{C_1~(\downarrow)}$ & $\boldsymbol{C_2~(\downarrow)}$ & $\boldsymbol{C_3~(\downarrow)}$ & $\boldsymbol{C_4~(\downarrow)}$ & $\boldsymbol{C_5~(\downarrow)}$ \\
    \hline\arrayrulecolor{black!10}
    Original motion & 30.63 & -4.91 & 30.87 & 21.74 & 11.93 & 7.79  & 7.410 & 18.729 & 0.0843 & 0.2757 & 7.426 \\
    \hline
    Ground-truth & 22.56 & -2.72 & -- & -- & 9.34 & 7.47 & 6.168 & 17.099 & 0.0553 & 0.1644 & 5.490 \\
    \arrayrulecolor{black}
    \hline
    \multicolumn{13}{l}{\textbf{Step ablations}} \\
    \hline\arrayrulecolor{black!10}
    Full Model & 7.58 & 0.74 & \textbf{232.54} & 236.28 & 10.85 & 13.82 & 5.503 & \textbf{18.565} & \textbf{0.026} & 0.181 & 5.354 \\ 
    \hline
    \quad w/o C1 step & 7.56 & 0.75 & 232.63 & 236.27 & 10.87 & 13.80 & 5.535 & 18.587 & \textbf{0.026} & 0.178 & 5.407 \\
    \hline
    \quad w/o C2 step & 7.60 & 0.91 & 233.60 & 236.14 & 10.90 & 13.55 & 5.509 & 18.814 & \textbf{0.026} & 0.196 & 5.61 \\
    \hline
    \quad w/o C3 step & \textbf{7.29} & 1.04 & 335.04 & \textbf{216.09} & \textbf{13.60} & \textbf{13.97} &  \textbf{5.456} & 18.617 & 0.088 & \textbf{0.158} & 5.609 \\
    \hline
    \quad w/o C4 step & 7.64 & \textbf{0.68} & 233.10 & 236.19 & 10.89 & 13.75 & 5.596 & 18.642 & \textbf{0.026} & 0.176 & 5.489 \\
    \hline
    \quad w/o C5 step & 7.58 & 0.74 & 232.59 & 236.24 & 10.86 & 13.82 & 5.503 & 18.572 & \textbf{0.026} & 0.181 & \textbf{5.352} \\
    \arrayrulecolor{black}
    \hline
    \end{tabular}
    }
\vspace{-0.3cm}
\end{table}

\textbf{Overview.} To evaluate the contribution of each simplification step, we conducted an ablation study by individually removing the steps designed to reduce the $C_1$--$C_5$ measures from the rule-based model. As shown in Table~\ref{tab:ablation-rule}, the full model generally achieves low complexity scores across most measures, though individual removals occasionally yield marginally lower values on specific metrics. We discuss each step below.

\textbf{Steps and footwork step.} Removing the $C_1$ step leads to a slight increase in the $C_1$ score (5.535 vs.\ 5.503) as well as modest degradation in $C_2$ and $C_5$, suggesting that the footwork simplification provides a small but consistent benefit across related measures. The differences are minor, indicating that $C_1$ acts as a supporting rather than dominant component in the pipeline.

\textbf{Movement density step.} Among all steps, the removal of the $C_2$ step produces the most consistent degradation across complexity measures: $C_2$ rises from 18.565 to 18.814, $C_4$ from 0.181 to 0.196, and $C_5$ from 5.354 to 5.610. The $PBC$ score also worsens (0.91 vs.\ 0.74). These results suggest that $C_2$ plays a central role in reducing spatial complexity and broadly benefits other measures.

\textbf{Orientation stabilization step.} The removal of step $C_3$ has the most pronounced and mixed impact. On one hand, several metrics improve, including $PFC$ (7.29 vs.\ 7.58), $FID_g$ (216.09 vs.\ 236.28), and diversity scores ($Dist_k$, $Dist_g$). On the other hand, $FID_k$ degrades substantially (335.04 vs.\ 232.54), $PBC$ moves further from the ground truth (1.04 vs.\ 0.74), and $C_3$ itself reverts to near-original levels (0.088 vs.\ 0.026). This trade-off suggests that the current orientation stabilization module effectively reduces rotational complexity but may introduce kinetic artifacts that deviate from realistic motion distributions.

\textbf{Multi-limb coordination step.} Removing the $C_4$ step yields the best $PBC$ score (0.68), suggesting that compressing the spatial extent of limb groups can introduce slight body-contact artifacts. Meanwhile, the $C_4$ complexity score remains comparable (0.176 vs.\ 0.181), and other measures show only marginal changes. This indicates that the $C_4$ module has a targeted but limited effect within the full pipeline.

\textbf{Asymmetry reduction step.} The effect of removing the $C_5$ step is negligible: the $C_5$ score remains virtually unchanged (5.352 vs.\ 5.354), and all other metrics stay within a narrow margin of the full model. This suggests that the asymmetry module has a limited influence on overall simplification in the current pipeline, possibly because bilateral asymmetry occurs infrequently in our dataset. Nevertheless, we retain this step as it addresses a distinct perceptual quality---bilateral coordination---that may become more relevant for choreography with pronounced asymmetry.

\textbf{Takeaway.} The $C_2$ (movement density) step provides the most consistent benefit across measures. The $C_3$ (orientation stabilization) step is effective at reducing rotational complexity but introduces trade-offs in kinetic realism, indicating room for refinement. The remaining steps ($C_1$, $C_4$, $C_5$) offer targeted contributions, with their effects being relatively modest in the current dataset. Overall, the full pipeline represents a reasonable trade-off between complexity reduction and motion quality.

% As shown in Table~\ref{tab:ablation-rule} (see Appendix~\ref{appendix:rule-implementation} for complete results and analysis), all operations contribute meaningfully to complexity reduction. Distance compression ($C_2$, $C_4$) most strongly affects spatial compactness, while velocity reduction ($C_1$) successfully slows rapid movements. Orientation stabilization ($C_3$) presents a trade-off: while reducing rotational complexity, it impacts other quality metrics, suggesting the need for further refinement. The asymmetry module ($C_5$) shows more subtle effects. Overall, the full model achieves the most balanced performance across all metrics, confirming that each operation plays a distinct role in the simplification pipeline.

\subsection{Learning-Based Approach}\label{subsec:learning-based-model}
The learning-based method captures the high-level semantics of dance, such as stylistic coherence and musical synchronization, which are hard to model explicitly. We extend POPDG~\cite{luo2024popdg}, a state-of-the-art diffusion-based approach for music-conditioned dance motion generation~\cite{zhang2025danceeditor, zhao2025freedance}, by integrating a ControlNet module~\cite{zhang2023adding}. ControlNet provides a mechanism for learning additional conditioning pathways on top of the pre-trained diffusion model. 

In our model, the original (complex) motion sequence provides structural cues (e.g. pose trajectory, movements and timing), and ControlNet learns the mapping from these cues to their corresponding simplified versions, enabling the model to produce simplified dances that remain faithful to the original's movements and style. We also add complexity measures to the training loss, allowing adaptive simplification based on identified challenges (\S\ref{subsec:results-and-findings-from-FG}). This allows the model to learn non-linear relationships between difficulty, aesthetics, and rhythm, generating novice-friendly yet stylistically faithful dance variations.

\begin{figure}[t]
    \centering
    \includegraphics[width=\linewidth]{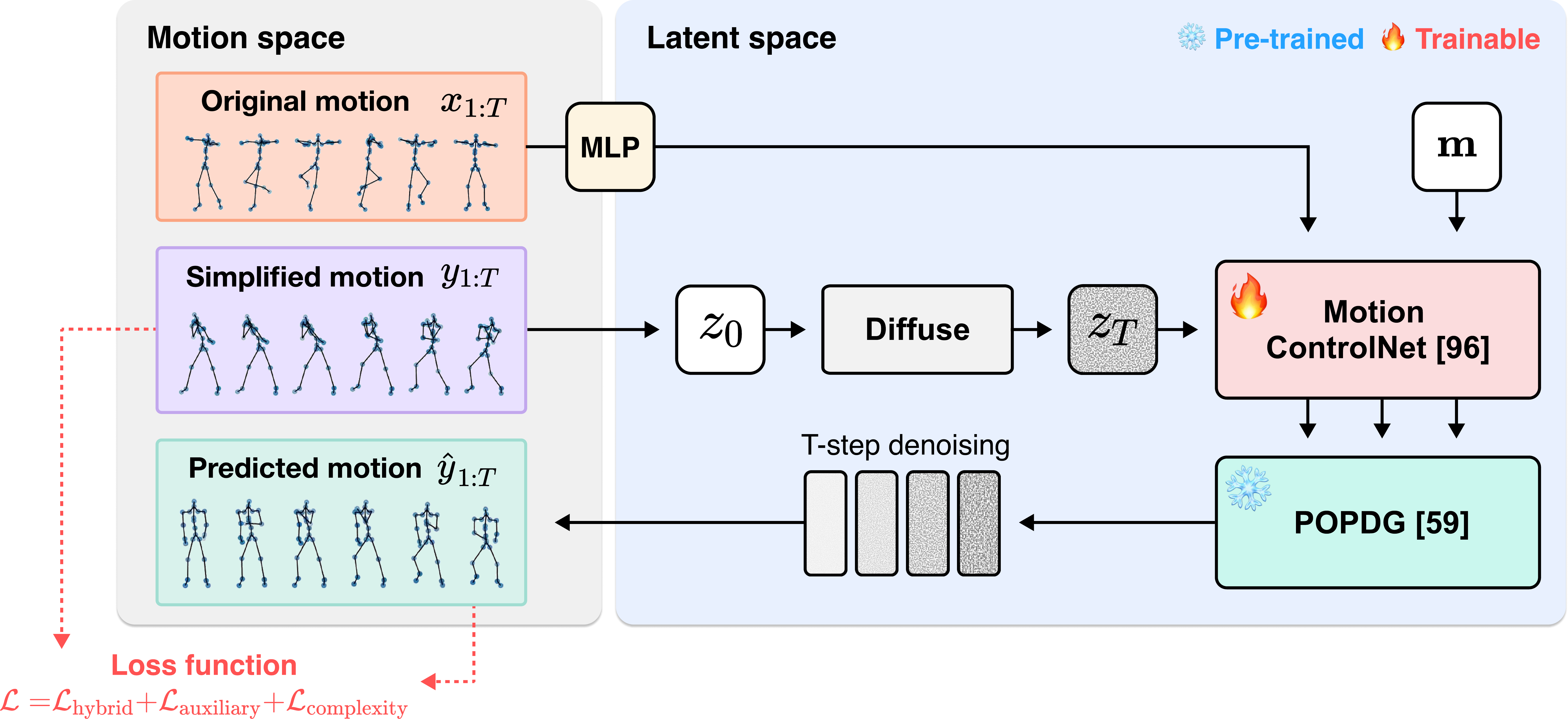}
    \caption{Overview of the learning-based approach. Given a pair of \texttt{\{original, simplified\}} dance sequences $\{\mathbf{x}, \mathbf{y}\} := \{x_{1:T}, y_{1:T}\}$ and music features $\mathbf{m}$ extracted with Jukebox~\cite{dhariwal2020jukebox}, we begin by adding $T$ steps of noise to the simplified dance to obtain the noisy sequence $z_T$ in the forward diffusion operation. The model learns to reverse this by denoising $z_T$, conditioned on the original motion and music. The original dance $x_{1:T}$ is encoded via a trainable multi-layer perceptron (MLP) and injected as a condition into the denoising process through ControlNet module~\cite{zhang2023adding}. At each time step $t$, the model predicts a simplified estimate $y_t$ and noises it back to $\hat{z}_{t-1}$, repeating until $t=1$. The generated output $\hat{y}_{1:T}$ is optimized to 1) match the simplified distribution ($\mathcal{L}_\text{hybrid}$), 2) generate plausible outputs ($\mathcal{L}_\text{auxiliary}$), and 3) effectively simplify ($\mathcal{L}_\text{complexity}$).}
    \label{fig:learning-model-structure}
    \vspace{-0.5cm}
\end{figure}

\subsubsection{Implementation details} We frame dance motion simplification as a conditional generative modeling task. We represent dance poses in the 24-joint SMPL format~\cite{loper2023smpl}, starting with a root translation followed by the 6-DOF rotation representation~\cite{zhou2019continuity} for each joint and binary contact labels for feet, joint and neck similar to POPDG: $\mathbf{p} \in \mathbb{R}^{156 =3 + 24 \times 6 + 9}$.
Let $\mathbf{x} \sim \mathcal{X} \subset \mathbb{R}^{T \times 156}$ denote original motion and $\mathbf{y} \sim \mathcal{Y} \subset \mathbb{R}^{T \times 156}$ the simplified motion. $\mathbf{m} \sim \mathcal{M} \subset \mathbb{R}^{4800}$ are optional musical features from frozen Jukebox~\cite{dhariwal2020jukebox}. We aim to model the conditional distribution $p(\mathbf{y} \mid \mathbf{x}, \mathbf{m})$, transferring dances from the complex domain $\mathcal{X}$ to the simplified domain $\mathcal{Y}$.

\textbf{Model structure and simplification process.} We adopt the improved denoising diffusion probabilistic model~(iDDPM) \cite{nichol2021improved} to model the distribution of dance motion data. In the forward noising process,  a clean dance sequence $\mathbf{x_0}$ is progressively perturbed into a series of latent variables $\{z_t\}$ according to the marginal Markov chain:

\begin{equation}
    q(z_t \mid \mathbf{x}) \sim \mathcal{N}(\sqrt{\bar{a_t}\mathbf{x}},  \mathbf{\Sigma}_t),
\end{equation}

where the variance $\Sigma(\mathbf{x}, t)$ is parameterized as:

\begin{equation}
     \Sigma(\mathbf{x}, t) = \exp(v\log\beta_t + (1-v)\log\bar{\beta}_t),
\end{equation}

with $\bar{\alpha}_t \in [0, 1]$ and $\beta_t$ as hyperparameters, and $v$ learned by the model.

POPDG learns the reverse denoising process, estimating $\hat{\mathbf{x}}_\theta(z_t, t, \mathbf{m}) \approx \mathbf{x}$ with model parameters $\theta$ for every time step $t$.
For our task, the goal is instead to reconstruct the simplified motion $\mathbf{y}_0$ given both the input dance $\mathbf{x}$ and music features $\mathbf{m}$:

\begin{equation}
    \hat{\mathbf{y}}_\theta(z_t, t, \mathbf{m}, \mathbf{x}) \approx \mathbf{y}_0.
\end{equation}

To achieve this, we adopt ControlNet~\cite{zhang2023adding} to condition the denoising process on the complex input motion $\mathbf{x}$. ControlNet has demonstrated success in preserving content while enabling flexible manipulation across various domains, from image stylization~\cite{wang2024instantstyle, ye2025stylemaster} to video editing~\cite{chen2023controlstyle}. Following this approach, we encode the original motion $\mathbf{x}$ through an MLP and inject it into the denoiser via residual connections~\cite{he2016deep}, allowing our model to maintain the dance's stylistic essence while modulating its complexity. Figure~\ref{fig:learning-model-structure} illustrates this architecture.

\textbf{Loss functions.} Following Ho et al. \cite{ho2020denoising}, we optimize our model with the ``simple'' objective:

\begin{equation}\label{eq:L_simple}
     \mathcal{L}_{\text{simple}} = \mathbb{E}_{\mathbf{y},t}[ \lvert\lvert \mathbf{y} - \hat{\mathbf{y}}_\theta(z_t,t,\mathbf{m},\mathbf{x}) \rvert\rvert_2 ^ 2 ].
\end{equation}

However, the role of variance is neglected in this equation. Therefore, we follow iDDPM to add an additional term for variational lower bound loss:

\begin{equation}\label{eq:L_vlb}
     \mathcal{L}_{\text{vlb}} = \mathbb{E}_{t\sim p_t} \left\lceil \frac{L_t}{p_t} \right\rceil \quad \text{where } p_t \propto \sqrt{E[L_t^2]} \text{ and } \sum p_t=1, \ L_t := D_{KL}(q(y_T|y_0) ||p(y_T))
\end{equation}

Here, we can define the new hybrid objective:

\begin{equation}\label{eq:L_hybrid}
    \mathcal{L}_{\text{hybrid}} = \mathcal{L}_{\text{simple}} + \lambda\mathcal{L}_{\text{vlb}}.
\end{equation}

To increase the realism of the generated movements, we adopt auxiliary losses similar to those in POPDG and EDGE~\cite{tseng2023edge} to augment physical realism by incorporating terms on joint positions~(Eq.~\ref{eq:L_joint}), velocities, accelerations~(Eq.~\ref{eq:L_va}), and foot and body contact~(Eq.~\ref{eq:L_body}):

\begin{equation}\label{eq:L_joint}
    \mathcal{L}_{\text{joint}} = \frac{1}{N} \sum_{i=1}^{N}\| (FK(\mathbf{y}^{(i)}) - FK(\hat{\mathbf{y}}^{(i)}) \|_2^2,
\end{equation}

\begin{equation}\label{eq:L_va}
    \mathcal{L}_{{\text{vel.,acc.}}} = \frac{1}{N} \sum_{i=1}^{N} \|(\mathbf{y}'^{(i)} - \hat{\mathbf{y}}'^{(i)})\|^2_2 + \|(\mathbf{y}''^{(i)} - \hat{\mathbf{y}}''^{(i)})\|_2^2,
\end{equation}

\begin{equation}\label{eq:L_body}
    \mathcal{L}_{\text{body}} = \frac{1}{N-1} \sum_{i=1}^{N-1}\| (FK(\hat{\mathbf{y}}^{(i+1)}) - FK(\hat{\mathbf{y}}^{(i)})) \cdot \hat{\mathbf{c}}^{(i)} \|_2^2,
\end{equation}

\begin{equation}\label{eq:L_aux}
    \mathcal{L}_{\text{auxiliary}} = \lambda_{\text{vlb}}\mathcal{L}_{\text{vlb}} + \lambda_{\text{vel., acc.}}\mathcal{L}_{\text{vel., acc.}} + \lambda_{\text{joint}}\mathcal{L}_{\text{joint}}.
\end{equation}

Where $N$ denotes the total number of frames, $FK(\cdot)$ denotes the forward kinematics function that converts SMPL joint parameters to 3D joint positions, $\mathbf{c}^{(i)}$ is the predicted contact label at frame $i$, $y'$ and $y''$ denote the first and second derivatives (vel. and acc.) respectively. 

In addition to the aforementioned losses, to encourage our model to simplify motions according to the five challenge categories defined in \S\ref{subsec:results-and-findings-from-FG}, we add the $C_1$ to $C_5$ measures as losses as defined in \S\ref{subsec:complexity-metrics}:

\begin{equation}\label{eq:L_complexity}
    \mathcal{L}_{\text{complexity}} = \lambda_{C_1}\mathcal{L}_{C_1} + \lambda_{C_2}\mathcal{L}_{C_2} + \lambda_{C_3}\mathcal{L}_{C_3} + \lambda_{C_4}\mathcal{L}_{C_4} + \lambda_{C_5}\mathcal{L}_{C_5}.
\end{equation}

Here, we formulate the overall training loss as the sum of three components:

\begin{equation}\label{eq:L}
    \mathcal{L} = \mathcal{L}_{\text{hybrid}} + \mathcal{L}_{\text{auxiliary}} + \mathcal{L}_{\text{complexity}}.
\end{equation}

\textbf{Sampling and guidance.}
Given music and a complex dance sequence, our model learns to denoise dance sequences to the simplified domain from time $t = T$ to $t = 1$. At each denoising time step $t$, the model predicts the denoised sample in the simplified domain and noises it back to time step $t-1$: $\hat{z}_{t-1} \sim q(\hat{\mathbf{y}}_\theta(\hat{z}_t, t, \mathbf{m}, \mathbf{x}), t-1)$. For controllable simplification, we use classifier-free guidance~\cite{ho2021classifierfree}, randomly dropping conditioning music or dance inputs and replacing them with noise. Usually, when the constraints are independent, the denoising process is modeled as in Liu et al. \cite{liu2022compositional}. However, since there is a clear correlation between the input dance motion and the given accompanying music, we follow Cho et al. \cite{cho2025enhanced} to model the output as:

\begin{equation}
    \tilde{\mathbf{y}}(\hat{z}_t, t,\mathbf{m}, \mathbf{x})
    = \hat{\mathbf{y}}(\hat{z}_t, t) 
    + w \cdot [
    \chi \cdot ( \hat{\mathbf{y}}(\hat{z}_t, t, \mathbf{m}, \mathbf{x}) - \hat{\mathbf{y}}(\hat{z}_t, t) )
    + (1 - \chi) \cdot \sum_{\mathbf{k} =\{\mathbf{m}, \mathbf{x}\}}\beta_i(\hat{\mathbf{y}}(\hat{z}_t, t, \mathbf{k}) - \hat{\mathbf{y}}(\hat{z}_t, t)
    ],
\end{equation}

where $w$ controls the strength of conditioning, $\chi \in [0, 1]$ controls the trade-off between joint and independent conditioning, $\beta_\mathbf{m} \in [0, 1]$ and $\beta_\mathbf{x} = 1 - \beta_\mathbf{m}$ control the conditional weight for the music and input dance motion features, respectively.

At inference time, we can control the degree of simplification and the realism of the model's output by varying the guidance strength $w$. Similarly, we can also modulate the relative importance of music through $\beta_\mathbf{m}$, where setting $\beta_\mathbf{m} = 0$ allows the model to simplify the input motion without any musical information.

\textbf{Training setup.} The entire training process took around 19 hours on a single NVIDIA RTX 6000 Ada Generation GPU\footnote{\url{https://www.nvidia.com/en-us/products/workstations/rtx-6000/}} with a batch size of 50. Since we train our model on annotated sequences with a shorter duration than POPDG's context length, we slow down the motions by a factor of 2 during training and rescale the generated outputs back to their original speed.

We set the default hyperparameters to $w = 3.5,\chi = 0.9, \beta_x = 0.5$, and $\beta_m = 0.5$. For evaluation, we rely on the same segmented sequences, which requires using a smaller stride of 1.0 seconds (vs. POPDG's 2.5 sec. stride). However, this modification introduces minor discontinuities at slice boundaries. To mitigate this, in addition to the standard interpolation across overlapping slides in POPDG, we also perform translation and rotation smoothing, ensuring smooth, consistent results while maintaining long-context generation.

\subsubsection{Ablation Study}\label{subsec:learning-ablation-study}
For the learning-based model, we conducted ablation studies on various metrics as introduced in~\S\ref{subsec:rule-ablation-study}. Table~\ref{tab:learning-ablations} shows the impact of different modules and hyperparameters on the model performance.

\textbf{MLP motion encoder.} A simple MLP encoder is able to identify the important structural cues from the input dance motion and relay it to ControlNet, as reflected in lower $PFC$ and $PBC$ scores.

\textbf{Complexity loss.} For $C_2$--$C_4$, the losses helped the model learn simplification patterns corresponding to the challenge categories identified in \S\ref{subsec:results-and-findings-from-FG}. However, foot movement complexity $C_1$ remains difficult for the model to capture, indicating that a more suitable measure may be needed.

\textbf{Condition on music.} Removing music as a conditioning signal resulted in outputs that were less style-preserving and exhibited weaker, less pronounced movements. This suggests that the model struggles to maintain coherent dynamics when lacking music information. Nevertheless, our model retains a baseline capacity for motion simplification even when explicit musical information is unavailable.

\begin{table}[htb!]
\centering
\renewcommand{\arraystretch}{1.1}
\caption{Ablation studies on motion evaluation metrics. Comparison across different components and configurations. Arrows (↓, ↑, →) indicate whether lower, higher, or closer to ground truth (GT) is better.}
\label{tab:learning-ablations}
\small\sffamily
\resizebox{\linewidth}{!}{%
    \begin{tabular}{lcccccccccccc}
    \hline
    \textbf{Method} & $\boldsymbol{PFC~(\downarrow)}$ & $\boldsymbol{PBC~(\rightarrow)}$ & $\boldsymbol{FID_k~(\downarrow)}$ & $\boldsymbol{FID_g~(\downarrow)}$ & $\boldsymbol{Dist_k~(\uparrow)}$ & $\boldsymbol{Dist_g~(\uparrow)}$ & $\boldsymbol{C_1~(\downarrow)}$ & $\boldsymbol{C_2~(\downarrow)}$ & $\boldsymbol{C_3~(\downarrow)}$ & $\boldsymbol{C_4~(\downarrow)}$ & $\boldsymbol{C_5~(\downarrow)}$ \\
    \hline\arrayrulecolor{black!10}
    Original motion & 30.63 & -4.91 & 30.87 & 21.74 & 11.93 & 7.79  & 7.410 & 18.729 & 0.0843 & 0.2757 & 7.426 \\
    \hline
    Ground-truth & 22.56 & \textbf{-2.72} & -- & -- & 9.34 & 7.47 & 6.168 & 17.099 & 0.0553 & 0.1644 & 5.490 \\
    \arrayrulecolor{black}
    \hline
    \multicolumn{13}{l}{\textbf{Architecture ablations}} \\
    \hline\arrayrulecolor{black!10}
    Learning-based (Full) & 2.31 & 1.98 & 43.93 & \textbf{30.95} & 7.02 & 9.31 & 6.914 & \textbf{14.372} & 0.0216 & 0.0323 & 3.152 \\
    \hline
    \quad w/o MLP encoder & 2.74 & 3.36 & \textbf{43.89} & 31.28 & 7.14 & \textbf{9.77} & 6.269 & 15.676 & 0.0127 & 0.0306 & 3.081 \\
    \hline
    \quad w/o complexity loss & \textbf{0.90} & 5.76 & 48.57 & 45.85 & \textbf{7.56} & 9.56 & \textbf{4.451} & 17.385 & 0.0751 & 0.1305 & 3.935 \\
    \hline
    \quad w/o music condition & 1.37 & \textbf{1.86} & 71.37 & 34.12 & 4.42 & 8.24 & 6.067 & 15.134 & \textbf{0.00936} & \textbf{0.02137} & \textbf{2.832} \\
    \arrayrulecolor{black}
    \hline
    \multicolumn{13}{l}{\textbf{Guidance weight analysis} ($w$)} \\
    \hline\arrayrulecolor{black!10}
    \quad $w = 2.0$ & \textbf{1.06} & 2.03 & 60.38 & 152.71 & 6.20 & \textbf{12.05} & \textbf{4.352} & 15.062 & \textbf{0.00921} & \textbf{0.01341} & \textbf{2.201} \\
    \hline
    \quad $w = 3.0$ & 1.88 & 1.97 & 48.30 & 47.23 & 6.80 & 9.84 & 6.197 & 14.506 & 0.0181 & 0.0257 & 2.891 \\
    \hline
    \quad $w = 3.5$ & 2.46 & 1.99 & 42.00 & 26.67 & 7.13 & 9.22 & 7.237 & 14.415 & 0.0226 & 0.0352 & 3.282 \\
    \hline
    \quad $w = 4.0$ & 3.19 & 1.89 & 35.46 & 14.22 & 7.48 & 8.60 & 8.211 & \textbf{14.336} & 0.0274 & 0.0498 & 3.671 \\
    \hline
    \quad $w = 5.0$ & 5.03 & 1.54 & 23.53 & \textbf{8.09} & 8.43 & 8.16 & 10.294 & 14.379 & 0.0406 & 0.0825 & 4.573 \\
    \hline
    \quad $w = 6.0$ & 7.43 & \textbf{0.710} & \textbf{16.53} & 9.55 & \textbf{9.52} & 7.96 & 12.475 & 14.671 & 0.0530 & 0.1280 & 5.501 \\
    \arrayrulecolor{black}
    \hline
    \end{tabular}
    }
\vspace{-0.3cm}
\end{table}

\textbf{Guidance strength.} Except for $C_2$, increasing guidance strength monotonically increased output complexity and feature preservation relative to the original motion. This highlights a trade-off between the simplification and the style and content preservation from the original dance, and also shows the potential for controllable simplification.
\section{Evaluation of Dance Motion Simplification Methods}\label{sec:evalution}
In this section, we evaluate our simplification methods through three complementary perspectives, each addressing different aspects of the simplification quality. First, we conduct a comparison study~(\S\ref{subsec:comparison}) using computational metrics to objectively assess motion quality and simplification effectiveness. Second, we perform an expert assessment~(\S\ref{subsec:expert-assessment}) with professional choreographers to evaluate qualitative factors that computational metrics cannot capture, such as artistic naturalness and style preservation. Finally, we conduct a learner study~(\S\ref{subsec:learner-study}) with novice dancers to measure the ultimate goal of our system: learning effectiveness in real-world practice scenarios.

\subsection{Comparison Study}\label{subsec:comparison}

We compare our two methods: The rule-based~(\S\ref{subsec:rule-based-model}) and learning-based approaches~(\S\ref{subsec:learning-based-model}) across multiple metrics, as detailed in \S\ref{subsec:rule-ablation-study}. Table~\ref{tab:comparison} shows the performance of our two approaches.

\subsubsection{Realism, physical plausibility, and motion diversity.} 
The rule-based model often produces artifacts when simplifying complex motions, particularly those involving whole-body rotations. These artifacts include unnatural joint-to-joint lengthening or crisscrossing limbs, characterized by higher $FID_k$ and $FID_g$ scores. Despite these issues, rule-based model excels in selectively simplifying key aspects of motion--such as specific joints or simplification criteria--while remaining faithful to the original sequence. It also yields more realistic body contact and more diverse outputs, whereas the learning-based model produces motions that are closer to the ground-truth simplification as it learns the underlying distribution of simplified dances. These differences suggest that the rule-based method is better suited for applications requiring precise control over specific simplification aspects, while the learning-based method excels when the goal is to approximate expert-created simplifications.

\subsubsection{Motion complexity.} 
The learning-based model outperforms the rule-based model in $C_2$--$C_5$, though it underperforms in $C_1$. This suggests it can capture the non-linear relationships underlying the simplification strategies in each criterion, resulting in a stronger overall simplification. However, while we can control the simplification rate via guidance, this control is global. In contrast, the rule-based approach allows fine-grained control, capable of simplifying a dance according to any combination of criteria $C_1$--$C_5$. This trade-off between simplification strength and controllability reflects a fundamental design choice: global optimization versus compositional flexibility.

\begin{table}[h!]
\centering
\renewcommand{\arraystretch}{1.15}
\caption{Comparison of motion evaluation metrics across different methods. Arrows (↓, ↑, →) indicate whether lower, higher, or closer to ground truth (GT) is better.}\label{tab:comparison}
\vspace{-0.3cm}
\small\sffamily
\resizebox{\linewidth}{!}{%
    \begin{tabular}{lccccccccccc}
    \hline
    \textbf{Method} & $\boldsymbol{PFC~(\downarrow)}$ & $\boldsymbol{PBC~(\rightarrow)}$ & $\boldsymbol{FID_k~(\downarrow)}$ & $\boldsymbol{FID_g~(\downarrow)}$ & $\boldsymbol{Dist_k~(\uparrow)}$ & $\boldsymbol{Dist_g~(\uparrow)}$ & $\boldsymbol{C_1~(\downarrow)}$ & $\boldsymbol{C_2~(\downarrow)}$ & $\boldsymbol{C_3~(\downarrow)}$ & $\boldsymbol{C_4~(\downarrow)}$ & $\boldsymbol{C_5~(\downarrow)}$ \\
    \hline
    Original motion & 30.63 & -4.91 & 30.87 & 21.74 & 11.93 & 7.79 & 7.410 & 18.729 & 0.0843 & 0.2757 & 7.426 \\
    \arrayrulecolor{black!30}
    \hline
    \arrayrulecolor{black!10}
    Ground-truth & 22.56 & -2.72 & -- & -- & 9.34 & 7.47 & 6.168 & 17.099 & 0.0553 & 0.1644 & 5.490 \\
    \hline
    Rule-based & 7.58 & \textbf{0.74} & 232.54 & 236.28 & \textbf{10.85} & \textbf{13.82} & \textbf{5.503} & 18.565 & 0.0263 & 0.1806 & 5.354 \\
    \hline
    Learning-based & \textbf{2.31} & 1.98 & \textbf{43.93} & \textbf{30.95} & 7.02 & 9.31 & 6.914 & \textbf{14.372} & \textbf{0.0216} & \textbf{0.0323} & \textbf{3.152} \\
    \arrayrulecolor{black}
    \hline
    \end{tabular}
    }
\vspace{-0.3cm}
\end{table}

\subsection{Expert Assessment}\label{subsec:expert-assessment}
\subsubsection{Experiment design}
We conducted an expert assessment of the simplified dance motions derived from two approaches, rule-based and learning-based methods. Using G*Power 3~\cite{faul2007g}, we determined a minimum sample size of 20 participants with a conservative effect size ($f=0.3$), $\alpha=.05$, and 80\% power, and accordingly recruited 20 participants (10 female, mean age 28.2 years, $\sigma=2.88$) through snowball sampling. Participants have 9.4 years of choreographic experience ($\sigma=3.82$), and 6.85 years of teaching experience ($\sigma=3.1$)~(detailed demographics are available in Appendix~\ref{appendix:demographics}), and they did not participate in the focus groups described in \S\ref{subsec:focus-group}. The purpose of the evaluation was to compare and assess the motions simplified by the rule-based method~(\rbbox{rule}) and the learning-based~(\lbbox{learning}) method against the ground-truth simplifications~(\gtbox{GT}) created by choreographers during focus groups when given original dance motions. Participants performed a task of viewing and comparing a total of 75 motions (5 criteria~$\times$~5 dance motion~$\times$~3 methods). The videos were presented in a counterbalanced order to prevent order effects. After viewing the generated simplified motions, participants evaluated them on a 7-point Likert scale for: 1) motion naturalness, 2) simplification degree, and 3) preservation of the original dance motion~(style preservation). We conducted the study through an online survey using Tally Forms\footnote{\url{https://tally.so/}}, with the study lasting less than an hour. This study was conducted under IRB approval, and participants were compensated \$40.

\subsubsection{Results}\label{subsubsec:result-expert-assessment}
\begin{figure*}[htp!]
    \centering
    \includegraphics[width=0.8\linewidth]{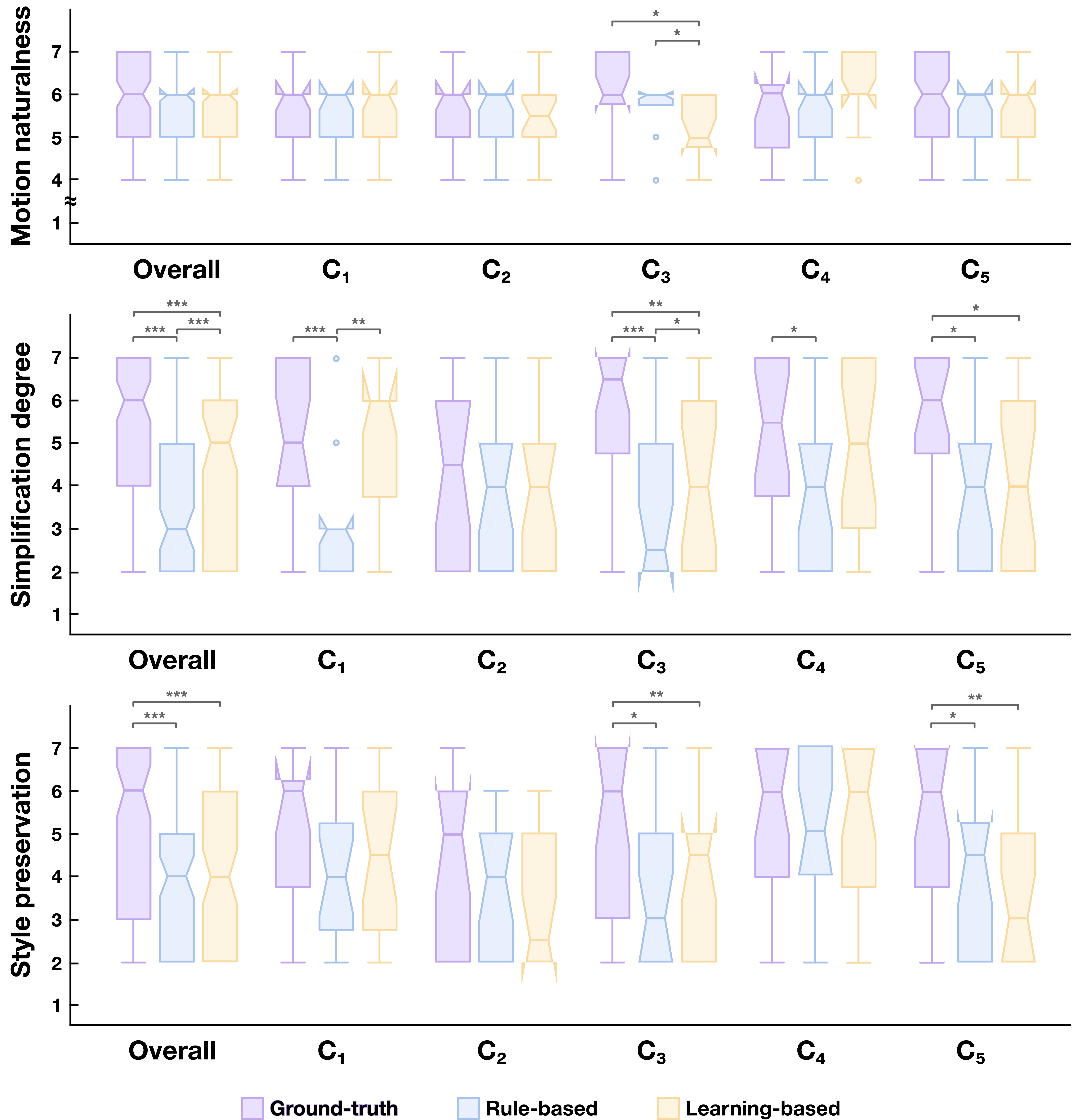}
    \caption{Overall and criterion-wise expert assessments. Notched boxplots (\gtbox{GT} / \rbbox{Rule} / \lbbox{Learning}) for \emph{Motion naturalness}, \emph{Simplification degree}, and \emph{Style preservation} with significance annotations ($^{*}p<.05$, $^{**}p<.01$, $^{***}p<.001$).}
    \label{fig:expert-assessment}
    \vspace{-0.5cm}
\end{figure*}

This study collected 1,500 responses~(25~items~$\times$~3~questions~$\times$~20~participants). For each participant–item pair, we compared \{\gtbox{GT}, \rbbox{rule}, \lbbox{learning}\} using a within-subjects design (Friedman omnibus followed by Wilcoxon signed-rank with Holm correction~\cite{zimmerman1993relative}). Figure~\ref{fig:expert-assessment} shows notched boxplots with significance.

\textbf{Motion naturalness.} We found a significant omnibus effect (Friedman \(\chi^2(2)=9.49\), \(p=.0087\)). Pairwise tests showed \gtbox{GT} and \rbbox{rule} both rated more natural than \lbbox{learning} (\gtbox{GT} \(>\) \lbbox{learning}: \(p_{\text{Holm}}=.0012\), \(r\approx.16\); \rbbox{rule} \(>\) \lbbox{learning}: \(p_{\text{Holm}}=.0109\), \(r\approx.12\)), while \gtbox{GT} vs.\ \rbbox{rule} did not reach statistical significance (\(p=.187\)). Descriptively, \gtbox{GT} \(\mu=5.74\), \rbbox{rule} \(\mu=5.67\), \lbbox{learning} \(\mu=5.54\). Effects were small. Criterion-wise, most criteria showed no reliable differences; an exception was $C_3$, where both \gtbox{GT} and \rbbox{rule} were rated more natural than \lbbox{learning} (Holm $p<.05$). The small differences in naturalness between \gtbox{GT} and automated methods can be explained by the physical constraints and motion smoothing applied in both automated approaches, which ensure baseline physical plausibility. The small effect sizes and high absolute ratings across all methods indicate that both automated approaches produce choreographically acceptable motions, though the learning-based method's artifacts in complex footwork are perceptible to expert eyes.

\textbf{Simplification degree.} The omnibus test was highly significant (\(\chi^2(2)=203.64\), \(p<.001\)). All pairwise differences were significant: \gtbox{GT} \(>\) \rbbox{rule} (\(p_{\text{Holm}}<.001\), \(r\approx.54\)), \gtbox{GT} \(>\) \lbbox{learning} (\(p_{\text{Holm}}<.001\), \(r\approx.51\)), and \lbbox{learning} \(>\) \rbbox{rule} (\(p_{\text{Holm}}=.00027\), \(r\approx.16\)). Descriptively, \gtbox{GT} \(\mu=5.21\), \lbbox{learning} \(\mu=4.20\), \rbbox{rule} \(\mu=3.88\). Effects were medium for contrasts with \gtbox{GT} and small for \lbbox{learning} vs.\ \rbbox{rule}. \gtbox{GT} consistently outperformed \rbbox{rule} across $C_1$, $C_3$, $C_4$, and $C_5$; \gtbox{GT} also exceeded \lbbox{learning} at $C_3$ and $C_5$, while \lbbox{learning} surpassed \rbbox{rule} at $C_1$ and $C_3$ (all Holm $p<.05$). These results reveal that neither automated method simplifies as aggressively as human choreographers, with the rule-based approach being particularly conservative—prioritizing motion preservation over complexity reduction.

\textbf{Style preservation.} We observed a strong omnibus effect (\(\chi^2(2)=163.80\), \(p<.001\)). \gtbox{GT} outperformed both alternatives (\gtbox{GT} \(>\) \rbbox{rule}: \(p_{\text{Holm}}<.001\), \(r\approx.38\); \gtbox{GT} \(>\) \lbbox{learning}: \(p_{\text{Holm}}<.001\), \(r\approx.59\)), and \rbbox{rule} \(>\) \lbbox{learning} (\(p_{\text{Holm}}=.0015\), \(r\approx.14\)). Descriptively, \gtbox{GT} \(\mu=5.14\), \rbbox{rule} \(\mu=4.20\), \lbbox{learning} \(\mu=3.91\). Effects were large for \gtbox{GT} vs.\ \lbbox{learning}, medium for \gtbox{GT} vs.\ \rbbox{rule}, and small for \rbbox{rule} vs.\ \lbbox{learning}. \gtbox{GT} was superior Overall and within $C_3$ and $C_5$ (vs.\ both \rbbox{rule} and \lbbox{learning}); other criteria showed no reliable differences after correction. The large effect size between \gtbox{GT} and \lbbox{learning} ($r\approx.59$) highlights a key limitation of data-driven generation: while it produces smooth motions, it sometimes loses the stylistic nuances that characterize the original choreography, whereas the rule-based method better maintains these artistic signatures despite more conservative simplification.

\subsection{Learner Study}\label{subsec:learner-study}
Expert assessments show that while both proposed methods received significantly lower ratings than ground-truth in certain metrics, their absolute scores remained near or above the midpoint of neutral. While these results indicate room for improvement compared to GT, a comprehensive evaluation should also examine how these simplifications perform in a learning context. Therefore, we conducted a complementary learner study examining learning effectiveness. 

\subsubsection{Experiment design and apparatus}
We validated the simplified dance motions derived from our approaches with novice learners. We employed a mixed-design, a between-subject design for evaluating three simplified motion conditions~(\gtbox{GT}, \rbbox{rule}, and \lbbox{learning}), and a within-subjects design for evaluating the criteria between original and simplified dance motion. The evaluation was conducted with 18 novice learners with an average dance learning experience of 0.5 months ($\sigma=1.2$) (9 females, mean age 21.67 years, $\sigma=1.75$), who had not participated in the experiment described in \S\ref{subsec:eval-of-simplified-choreography}~(detailed demographics are available in Appendix~\ref{appendix:demographics}). Each participant was divided into three groups (\gtbox{GT}, \rbbox{rule}, \lbbox{learning}) with 6 participants (3 females) per group. The experimental environment and equipment settings were identical to \S\ref{subsec:eval-of-simplified-choreography}, where participants learned each 8-count choreography sequence for 5 minutes in a 3~m~$\times$~3~m space using a smartphone and full-body mirror. The experimental validation measures also remained the same, using NASA--TLX to assess learning load, self-efficacy, and perceived difficulty, and recording participants' choreography performance after the learning session to evaluate objective performance. This experiment was also conducted over 90 minutes under IRB approval, and participants were compensated \$40.

\subsubsection{Results}
\begin{figure*}[htp!]
    \centering
    \includegraphics[width=0.92\linewidth]{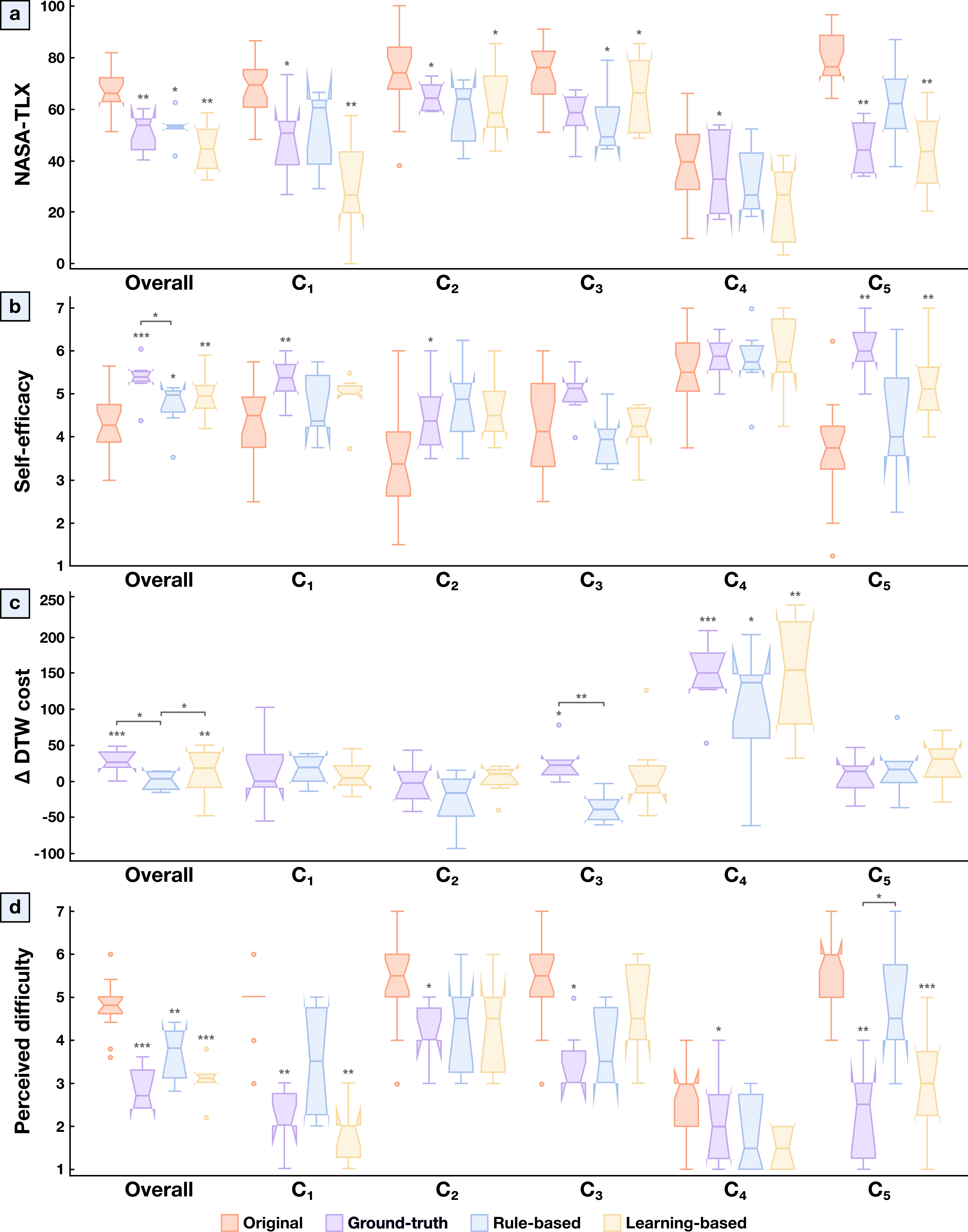}
    \caption{Overall and criterion-wise $(C_1$--$C_5)$ comparisons between \originalbox{original} and simplified--\{\gtbox{GT}, \rbbox{rule-based}, and \lbbox{learning-based}\} conditions using notched boxplots with significance annotations ($^{*}p<.05$, $^{**}p<.01$, $^{***}p<.001$). Significance annotations on each boxplot indicate within-method paired tests (vs. each participant's \originalbox{original} or, for objective performance, tests of $\Delta$ against zero. \subfigbox{a} workload~(NASA--TLX); \subfigbox{b} self-efficacy; \subfigbox{c} objective performance; \subfigbox{d} perceived difficulty.}
    \label{fig:learner-study-2-result}
    \vspace{-0.5cm}
\end{figure*}

For each participant, we calculated the difference between original and simplified conditions ($\Delta =$ \originalbox{original} $-$ \simplifiedbox{simplified}). Positive $\Delta$ values indicate improvement for workload, movement errors, and perceived difficulty (lower is better), while negative $\Delta$ values indicate improvement for self-efficacy (higher is better). Table~\ref{tab:learner-2-results} and Figure~\ref{fig:learner-study-2-result} present comprehensive statistical results across all methods and metrics. 
We first examined condition-by-method interactions, finding significant interactions for self-efficacy (Kruskal-Wallis $p=.004$) and objective performance ($p=.030$), with a marginal trend for perceived difficulty (ANOVA $F(2, 15)=2.87$, $p=.088$). These interactions suggest that simplification methods differ in their effectiveness across metrics.

\begin{table}[t!]
\centering
\renewcommand{\arraystretch}{1.1}
\caption{Simplification effects by method: Mean changes ($\Delta$), 95\% confidence intervals~(CIs), and effect sizes}
\label{tab:learner-2-results}
\vspace{-0.3cm}
\small\sffamily
\setlength{\tabcolsep}{4pt}
\begin{tabular}{c l c c r r r}
    \hline
    \multirow{2}{*}{\textbf{Metric}} & \multirow{2}{*}{\textbf{Method}} & \multicolumn{2}{c}{\textbf{Mean} $\Delta$ (\textbf{95\% CI})} & \multirow{2}{*}{\textbf{Test statistic}} & \multirow{2}{*}{$\boldsymbol{p}$\textbf{-value}} & \multirow{2}{*}{\textbf{Effect size}} \\
    \arrayrulecolor{black!40}
    \cline{3-4}
    & & \textbf{Value} & \textbf{CI} & & & \\
    \arrayrulecolor{black}
    \hline
    
    \multirow{4}{*}{\makecell[l]{Workload}} 
    & \gtbox{GT} & 20.86 & [9.16, 32.56] & $t(5)=4.58$ & .006 & $d_z=1.87$ \\
    \arrayrulecolor{black!10}
    \cline{2-7}
    & \rbbox{Rule} & 9.11 & [2.17, 16.04] & $t(5)=3.38$ & .020 & $d_z=1.38$ \\
    \cline{2-7}
    & \lbbox{Learning} & 21.31 & [13.09, 29.54] & $t(5)=6.66$ & .001 & $d_z=2.72$ \\
    \cline{2-7}
    & \textit{Between-method} & \multicolumn{2}{c}{--} & $F(2,15)=3.76$ & .048$^{*}$ & $\eta^2=0.33$ \\
    \arrayrulecolor{black!40}
    \hline
    
    \multirow{4}{*}{\makecell[l]{Self-efficacy}} 
    & \gtbox{GT} & $-$1.38 & [$-$2.09, $-$0.66] & $W=61$ & $<$.001 & $r=0.83$ \\
    \arrayrulecolor{black!10}
    \cline{2-7}
    & \rbbox{Rule} & $-$0.32 & [$-$0.60, $-$0.03] & $W=268$ & .049 & $r=0.38$ \\
    \cline{2-7}
    & \lbbox{Learning} & $-$0.60 & [$-$1.02, $-$0.18] & $W=210$ & .004 & $r=0.56$ \\
    \cline{2-7}
    & \textit{Between-method} & \multicolumn{2}{c}{--} & $H=7.48$ & .004$^{**}$ & -- \\
    \arrayrulecolor{black!40}
    \hline
    
    \multirow{4}{*}{\makecell[l]{Objective\\performance}} 
    & \gtbox{GT} & 38.10 & [22.64, 53.57] & $W=318$ & .003$^\dagger$ & $r=0.62$ \\
    \arrayrulecolor{black!10}
    \cline{2-7}
    & \rbbox{Rule} & 14.04 & [$-$4.30, 32.37] & $W=493$ & .094$^\dagger$ & $r=0.31$ \\
    \cline{2-7}
    & \lbbox{Learning} & 38.99 & [21.47, 56.51] & $W=318$ & .003$^\dagger$ & $r=0.62$ \\
    \cline{2-7}
    & \textit{Between-method} & \multicolumn{2}{c}{--} & $H=5.98$ & .030$^{***}$ & -- \\
    \arrayrulecolor{black!40}
    \hline
    
    \multirow{4}{*}{\makecell[l]{Perceived difficulty}} 
    & \gtbox{GT} & 2.10 & [1.34, 2.86] & $W=18$ & $<$.001 & $r=0.90$ \\
    \arrayrulecolor{black!10}
    \cline{2-7}
    & \rbbox{Rule} & 1.00 & [0.01, 1.99] & $W=180$ & .009 & $r=0.57$ \\
    \cline{2-7}
    & \lbbox{Learning} & 1.53 & [0.81, 2.26] & $W=142$ & $<$.001 & $r=0.69$ \\
    \cline{2-7}
    & \textit{Between-method} & \multicolumn{2}{c}{--} & $F(2,15)=2.87$ & .088 & $\eta^2=0.28$ \\
    \arrayrulecolor{black}
    \hline
    \end{tabular}
    \begin{tablenotes}
        \small
        \item Note: $\Delta =$ \originalbox{original} $-$ \simplifiedbox{simplified}. Positive values indicate a reduction in workload, DTW cost, and difficulty. Negative values indicate an increase in self-efficacy. $^\dagger$One-sided test. $^{*}$\lbbox{learning} > \rbbox{rule} ($p_{\text{Holm}}=.045$). $^{**}$\gtbox{GT} > \rbbox{rule} ($p_{\text{Holm}}=.006$). $^{***}$\gtbox{GT}, \lbbox{learning} > \rbbox{rule} (both $p_{\text{Holm}}=.048$).
    \end{tablenotes}
    \vspace{-0.5cm}
\end{table}

% \textbf{Workload.} On the weighted NASA--TLX scores, participant-level means (averaged across $C_1$--$C_5$) decreased in all methods: for \gtbox{GT}, from $73.63{\pm}10.79$ to $50.99{\pm}8.30$ ($\Delta=22.64$, 95\% CI $[8.64, 36.65]$; $t(5)=5.54$, $p<.001$, $d_z=1.03$); for \lbbox{learning}, from $66.17{\pm}4.51$ to $44.86{\pm}10.55$ ($\Delta=21.31$, 95\% CI $[13.09, 29.54]$; $t(5)=4.93$, $p<.001$, $d_z=0.92$); and for \rbbox{rule}, from $63.79{\pm}8.08$ to $52.73{\pm}6.60$ ($\Delta=11.06$, 95\% CI $[4.57, 17.55]$; $t(5)=2.39$, $p=.024$, $d_z=0.44$). Despite numerical differences among $\Delta$s, the between-method comparison on $\Delta$ did not reach statistical significance (ANOVA $F(2,15)=2.60$, $p=.107$). This suggests all three simplification approaches effectively reduce cognitive load for novices, indicating both automated methods achieved workload reductions comparable to \gtbox{GT}.
\textbf{Workload.} Using participant-level means (averaged across $C_1$--$C_5$), NASA--TLX decreased for all methods. For \gtbox{GT}, workload dropped from $71.85{\pm}7.39$ (\originalbox{original}) to $50.99{\pm}8.30$ (\simplifiedbox{simplified}) ($\Delta=20.86$, 95\% CI $[9.16, 32.56]$; $t(5)=4.58$, $p=.006$, $d_z=1.87$). For \rbbox{rule}, workload decreased from $61.84{\pm}8.11$ to $52.73{\pm}6.60$ ($\Delta=9.11$, 95\% CI $[2.17, 16.04]$; $t(5)=3.38$, $p=.020$, $d_z=1.38$). For \lbbox{learning}, workload decreased from $66.17{\pm}4.51$ to $44.86{\pm}10.55$ ($\Delta=21.31$, 95\% CI $[13.09, 29.54]$; $t(5)=6.66$, $p=.001$, $d_z=2.72$). Between-method comparisons on participant-level $\Delta$ indicated a method effect ($F(2,15)=3.76$, $p=.048$, $\eta^2=.334$), with Holm-corrected post-hoc tests showing larger reductions for \lbbox{learning} than \rbbox{rule} ($p_{\text{Holm}}=.045$).

\textbf{Self-efficacy.} The methods differed significantly in boosting learner confidence. \gtbox{GT} produced the largest gains ($\Delta=-1.38$, $r=.83$), significantly outperforming \rbbox{rule}-based improvements ($\Delta=-0.32$, $r=.38$; $p_{\text{Holm}}=.006$). \lbbox{Learning} showed intermediate gains ($\Delta=-0.60$, $r=.56$), not significantly different from either alternative. The significant advantage of \gtbox{GT} in boosting confidence suggests that expert-crafted simplifications better balance challenge and achievability, while the rule-based method's conservative simplification may leave learners feeling less accomplished.

\textbf{Objective performance.} Similar to \S\ref{subsubsec:eval-of-simplified-dance-with-novices}, DTW was used for measuring objective performance. DTW cost analysis revealed that \gtbox{GT} and \lbbox{learning} produced nearly identical improvements in movement accuracy ($\Delta=38.10$ and $38.99$ respectively, both $r=.62$), both significantly outperforming \rbbox{rule}-based simplification ($\Delta=14.04$, $r=.31$; both $p_{\text{Holm}}=.048$). Notably, \lbbox{learning} matched \gtbox{GT} in improving performance. This finding is particularly significant: despite receiving lower expert ratings for naturalness and style preservation (\S\ref{subsubsec:result-expert-assessment}), the learning-based method delivers equivalent learning outcomes for novices, suggesting that motion smoothness and simplification degree matter more for learning effectiveness than expert-perceived stylistic fidelity.

\textbf{Perceived difficulty.} All three methods reduced perceived difficulty relative to \originalbox{original}. \gtbox{GT} yielded the largest decrease ($\Delta=2.10$, $p<.001$, $r=.90$),  followed by \lbbox{learning} ($\Delta=1.53$, $p<.001$, $r=.69$) and \rbbox{rule} ($\Delta=1.00$, $p=.009$, $r=.57$). A between-method analysis on participant-level mean $\Delta$ showed a trend toward larger benefits for \gtbox{GT}/\lbbox{learning} than \rbbox{rule} (ANOVA $F(2,15)=2.87$, $p=.088$); however, pairwise contrasts did not reach corrected significance, so we refrain from strong claims about method-to-method differences for perceived difficulty. The consistent difficulty reduction across all methods, combined with the large effects for \gtbox{GT} and \lbbox{learning}, indicates that these approaches successfully make choreography feel more approachable to novices.

\textbf{Consistency and order effects.} The benefits were consistent across choreography criteria, with significant improvements in all five criteria for workload and perceived difficulty, and in four of five for self-efficacy and performance (all except $C_4$). We found no order effects (all $p\geq.430$), confirming that improvements stem from simplification rather than presentation sequence.

\subsection{Evaluation Summary and Takeaways}\label{subsec:eval-summary}
Across three complementary evaluations, we observed a consistent set of patterns about what each simplification approach does well and where it falls short.
The comparison study (\S\ref{subsec:comparison}) provided an objective view of motion quality and complexity reduction, the expert assessment (\S\ref{subsec:expert-assessment}) captured choreographic qualities beyond computational metrics, and the learner study (\S\ref{subsec:learner-study}) tested whether these simplifications translate into practice-time benefits for novices.

Overall, the \lbbox{learning}-based approach more closely tracked expert-created simplifications in terms of aggregate complexity reduction (especially $C_2$--$C_5$) and produced smoother motions with substantially lower $FID$ than the \rbbox{rule}-based approach, while the \rbbox{rule}-based approach offered stronger controllability and tended to better preserve structural properties such as contact-related realism and output diversity.
In expert ratings, both automated methods achieved high absolute scores near or above the neutral midpoint, but remained below the \gtbox{GT} upper bound in simplification degree and (more clearly) style preservation.
In the learner study, all methods reduced workload and perceived difficulty relative to the original motions; importantly, the \lbbox{learning}-based approach achieved objective performance gains comparable to \gtbox{GT}, whereas the \rbbox{rule}-based approach showed smaller improvements on performance and confidence-related measures.
These results suggest that the two automated approaches are complementary: one prioritizes distributional matching and learning effectiveness, and the other prioritizes compositional control and faithful structural preservation.

\begin{table}[t!]
\centering
\renewcommand{\arraystretch}{1.15}
\caption{Summary of evaluation outcomes across the three studies, focusing on the two automated methods.}
\label{tab:eval-summary}
\vspace{-0.25cm}
\scriptsize\sffamily
\setlength{\tabcolsep}{3pt}

\begin{tabular}{C{2.9cm} C{5.2cm} C{5.2cm}}
\hline
\textbf{Evaluation} & \textbf{\rbbox{Rule}-based} & \textbf{\lbbox{Learning}-based} \\
\hline

\textbf{Technical comparison~(\S\ref{subsec:comparison})} 
& Offers selective, criterion-level control and tends to preserve structural/contact-related properties and output diversity; however, it can introduce artifacts in complex rotations, reflected in higher $FID$.
Compared to \gtbox{GT}, it is more conservative in overall complexity reduction.
& Produces smoother motions with substantially lower $FID$ than \rbbox{rule} and achieves stronger reductions for $C_2$--$C_5$ (while underperforming on $C_1$).
It more closely tracks the distribution of \gtbox{GT} simplifications, though diversity can be lower. \\

\arrayrulecolor{black!20}\hline
\arrayrulecolor{black}

\textbf{Expert assessment~(\S\ref{subsec:expert-assessment})} 
& Rated comparably to \gtbox{GT} on motion naturalness overall, but lower on simplification degree and style preservation.
Relative to \lbbox{learning}, it tends to better preserve style while simplifying less aggressively.
& Received slightly lower naturalness in some criteria and lower style preservation than \rbbox{rule}, and remained below \gtbox{GT} on both simplification degree and style preservation.
It was perceived as simplifying more than \rbbox{rule} overall. \\

\arrayrulecolor{black!20}\hline
\arrayrulecolor{black}

\textbf{Learner study~(\S\ref{subsec:learner-study})} 
& Reduced workload and perceived difficulty relative to the original motions, but yielded smaller gains in self-efficacy and objective performance (and did not match \gtbox{GT}-level gains in our sample).
& Reduced workload and perceived difficulty, and achieved objective performance improvements comparable to \gtbox{GT} in our sample, despite lower expert ratings on style preservation.
Self-efficacy gains were intermediate between \rbbox{rule} and \gtbox{GT}. \\

\hline
\end{tabular}
\vspace{-0.35cm}
\end{table}

\section{Discussion, Limitations, and Future Work}\label{sec:discussion}
\subsection{Revisiting the Research Questions}

Our work set out to understand what makes choreography difficult for novice learners (RQ1), to derive generalizable criteria and strategies for simplifying challenging movements (RQ2), and to investigate whether dance simplification can be automated while preserving the essential characteristics of the original motion (RQ3).

For RQ1, the survey and focus group results revealed themes causing novices' difficulty in dance learning, including dense and rapid sequences, frequent directional changes and turns, complex multi-limb coordination, and asymmetric use of the body. These recurrent themes motivated our five complexity criteria, which capture footwork structure $(C_1)$, movement density $(C_2)$, rotational and directional changes $(C_3)$, multi-limb coordination $(C_4)$, and asymmetry $(C_5)$ as a compact representation of choreographic difficulty. Rather than being tied to a specific genre, these criteria emerged consistently across styles, suggesting that they capture cross-genre properties of choreographic complexity.

For RQ2, the workshop on dance motion simplification demonstrated that these criteria are not merely descriptive but can be used as actionable elements for simplification. Choreographers spontaneously referred to strategies that directly manipulate each criterion: reducing steps or travel $(C_1)$, inserting pauses $(C_2)$, smoothing turns $(C_3)$, decoupling arms from footwork $(C_4)$, or symmetrizing sequences $(C_5)$. When we re-encoded expert-created simplifications with our metrics, we observed systematic reductions in the targeted criteria, and these reductions aligned with novices’ perceived difficulty and performance outcomes. This suggests that $C_1$--$C_5$ and the associated strategies constitute a practically useful design space for simplifying dance motion.

For RQ3, our evaluation shows that both rule-based and learning-based simplification approaches can automate non-trivial parts of this process while preserving the style and musicality of the original choreography. The learning-based method more closely matches the overall distribution of expert simplifications and achieves larger reductions in most complexity dimensions, whereas the rule-based method offers finer control over which criteria are simplified and tends to preserve structural and contact-related properties better. Expert assessments and learner studies further indicate that automated simplifications can reduce workload, improve confidence and self-efficacy, and enhance objective performance. Taken together, these results suggest that automated simplification is feasible and beneficial, but that different modeling choices lead to distinct trade-offs.

\subsection{Alignment with Motor Learning Theories}

Our findings resonate with classic motor learning theories, such as the zone of proximal development~\cite{vygotsky1978mind} and the challenge point hypothesis~\cite{guadagnoli2004challenge}. Simplification allows learners to operate within a range of difficulty that is neither trivially easy nor overwhelmingly hard, reducing repeated failure while preserving enough challenge to foster learning. Across sessions, simplified conditions lowered learning load while improving self-efficacy and performance, indicating that learners were practicing near an ``optimal challenge'' region rather than at the extremes of frustration or boredom. This pattern aligns with prior work showing that task difficulty must be matched to the learner’s current capability to maximize learning and engagement.

Our framework can also be interpreted as an instance of progressive part practice~\cite{lee2025motor}, in which complex skills are decomposed into manageable components that are introduced and recombined over time. By selectively attenuating $C_1$–$C_5$, the system effectively structures practice around subsets of the full movement: for example, simplifying multi-limb coordination ($C_4$) while preserving basic footwork patterns ($C_1$), or reducing rotational complexity ($C_3$) before reintroducing rapid turns. The observed gains in performance and confidence, particularly when simplifying movement density and coordination, suggest that explicitly controlling these components offers a principled way to scaffold learning for novices.

\subsection{Strategies for Simplification Across Complexity Criteria}

Our analyses reveal that not all complexity dimensions are equally amenable to simplification. The largest performance gains were associated with reductions in step and footwork ($C_1$) and movement density ($C_2$), while rotation-related complexity ($C_3$) improved only marginally. This pattern suggests that some complexity dimensions are more tightly tied to genre-defining movements. In many styles, emblematic moves such as "Lock," "2-step," or "Waack" function as canonical building blocks. Choreographers reported that excessively simplifying these fundamentals risks eroding the recognizable identity of the genre, even if it would make the sequence technically easier.

When such genre-defining moves act as bottlenecks, simplification alone may be insufficient. Our results and expert feedback point to complementary strategies: dedicating prerequisite sessions to build foundational skills around core movements, or selectively simplifying only transitional segments while preserving key stylistic elements. For example, simplifying $C_2$ and $C_4$ around a difficult signature move can give learners more time and motor resources to focus on that move without diluting the choreography’s overall character. These observations highlight that effective simplification is not a uniform reduction of complexity, but a targeted rebalancing of where complexity resides within a sequence.

\subsection{Trade-offs Between Rule-based and Learning-based Models}

The two simplification approaches exhibit complementary strengths and limitations, each presenting distinct trade-offs between controllability, motion quality, and stylistic fidelity.

Rule-based method excels at fine-grained control over specific joints and kinematic parameters, enabling explicit constraints such as range of motion limits, floor contact conditions, and anatomical plausibility. Due to rule-based operations acting locally in time and space, errors can accumulate across multiple sequential edits, occasionally producing stiffness at segment boundaries. Moreover, higher-level stylistic qualities like phrase-level groove and micro-timing variations can be flattened. These limitations suggest opportunities for improvement through overlap-and-blend at edit boundaries, continuity constraints on motion, and phrase-aware rules operating at musical units to better preserve stylistic flow.

Learning-based method offers a strong macro-level understanding of movement semantics, achieving lower $FID$ scores and smoother phrasing than the rule-based approach. However, optimization toward distributional matching can dilute the fine-grained stylistic ``signatures'' that distinguish genres, such as the sharp accents of waack or the weighted groove of krump. Our expert assessment confirmed lower style preservation scores, particularly for rotations ($C_3$) and asymmetry ($C_5$). Incorporating style-aware objectives~\cite{jang2022motion} or content-style disentanglement architectures~\cite{aberman2020skeleton} could address these limitations.

Combining both approaches may leverage complementary strengths. A promising architecture uses the learning-based approach for an initial context-aware simplification, followed by rule-based post-processing to enforce constraints and correct localized artifacts. Beyond algorithmic solutions, interactive tool design can address residual artifacts and style concerns. For instance, instructors could identify segments where artifacts are perceptible and selectively apply rule-based smoothing, or adjust style-fidelity sliders to request outputs closer to the original choreography's character. 
% This human-in-the-loop approach positions automated simplification as a collaborative starting point rather than a final product as further discussed in \S\ref{subsec:design-implications}.

\subsection{Design Implications for Enhanced Dance Learning Experience}\label{subsec:design-implications}

The combination of empirical findings and expert feedback yields several design implications for choreography-authoring and learning-support systems. First, our results underscore the value of making complexity dimensions, such as $C_1$--$C_5$, explicit and controllable as first-class interaction primitives. Expert choreographers in our study expressed a desire for granular control over difficulty levels, including the ability to offer multiple simplification options for the same phrase and to isolate specific body parts (e.g., arms-only or footwork-only practice). Interfaces that expose simplification sliders or presets along interpretable dimensions could support progressive curricula in which difficulty is adapted session-by-session or even for individual movement sequences.

Second, our findings highlight the importance of further explanations alongside simplified motions, not just visually demonstrating them. Verbal guidance was repeatedly described as crucial for connecting visual demonstrations to learners’ internal understanding. Automated systems built on our framework could pair simplified motion with natural-language explanations---for example, using large language models~\cite{jiang2024motiongpt, wang2024motiongpt} to describe which components were reduced (e.g., ``we removed half turns and simplified arm patterns to focus on the basic groove'') and why. Such explanations can support strategy formation, help learners generalize beyond a single choreography, and make the simplification process more transparent.

Third, our work suggests opportunities for authoring tools that explicitly support collaboration between choreographers and models. For instance, a choreographer could specify which criteria are allowed to change, request alternative simplifications at different difficulty levels, or lock certain stylistic features while exploring variations in others. Batch simplification over a full class curriculum, combined with interactive refinement for selected key motion, could reduce preparation time while maintaining artistic control. Our findings suggest a design space where computational aids act as partners that propose simplification candidates, and human experts curate and adapt them to meet pedagogical goals.

\subsection{Limitations and Future Work}

Our study has several limitations that open up avenues for future research. First, our dataset exhibits imbalance across simplification criteria (e.g., many more pairs annotated for $C_2$ than for $C_5$), reflecting the real-world frequency of certain challenging patterns but also potentially biasing model performance. Future work should collect more examples of underrepresented simplifications, particularly those involving asymmetry and nuanced coordination, and examine how performance scales with more balanced data.

Also, the learner study captured short-term practice rather than long-term retention or transfer. We demonstrated that simplification can reduce workload and difficulty while improving performance within a limited number of sessions, but we did not measure how well learners retain simplified movements over time, how quickly they can transition from simplified to original versions, or how skills transfer to new choreographies. Longitudinal studies that track these outcomes would provide a more complete picture of how automated simplification affects skill acquisition.

Regarding motion representations and visualizations that relied on skeletal renderings, some learners found them abstract or hard to interpret. Future systems could enhance visual representations, such as stylized avatars or multi-view displays, and investigate how these choices influence learners’ understanding of simplified motion. Similarly, our approach was trained on a specific motion representation and dataset; adapting the framework to other motion formats or capture technologies (e.g., egocentric or sparse-sensor setups) remains an open challenge.
\section{Conclusion}\label{sec:conclusion}
This work presents the first comprehensive study of dance motion simplification for novice learners, addressing a gap in online dance education. Through research with 30 novices and 30 choreographers, we identified five key complexity factors and built a dataset of original and simplified sequences across 10 genres, enabling both rule- and learning-based simplification methods. The rule-based approach allows precise control over movement parameters, while the learning-based method, using a diffusion model, captures high-level dance motion semantics. Both methods reduce complexity while maintaining dance characteristics, as confirmed by technical evaluation and expert assessment with 20 choreographers, though human-generated simplifications still set the standard. Studies with 18 novices showed improved learning effectiveness, lower workload, higher self-efficacy, better performance, and reduced perceived difficulty. This work establishes a framework bridging professional choreography and novice learners, making dance more accessible while preserving its essence. Future research should address dataset imbalances and explore hybrid approaches, opening new possibilities for adaptive, scalable dance education technologies.

\bibliographystyle{ACM-Reference-Format}
\bibliography{main}

\newpage
\appendix
\section{Demographic Tables}~\label{appendix:demographics}

\begin{table*}[h]
\centering
\small\sffamily
\renewcommand{\arraystretch}{1.1}
\setlength{\tabcolsep}{4pt}
\caption{Participant demographics of novice learners for survey.}
\label{tab:demographics-survey}

\begin{subtable}[t]{0.49\linewidth}
\centering
% \caption{Participants P1--P15}
\begin{tabular}{c c c c}
\hline
\textbf{Alias} & \textbf{Gender} & \textbf{Age} & \textbf{Dance Learning (mo.)} \\
\hline
\arrayrulecolor{black!10}
P1  & Female & 38 & 1 \\ \hline
P2  & Male   & 22 & 1 \\ \hline
P3  & Female & 25 & 1 \\ \hline
P4  & Female & 26 & 3 \\ \hline
P5  & Female & 23 & 3 \\ \hline
P6  & Female & 22 & 2 \\ \hline
P7  & Female & 29 & 2 \\ \hline
P8  & Female & 22 & 3 \\ \hline
P9  & Female & 32 & 5 \\ \hline
P10 & Female & 27 & 6 \\ \hline
P11 & Male   & 21 & 1 \\ \hline
P12 & Female & 23 & 4 \\ \hline
P13 & Female & 48 & 3 \\ \hline
P14 & Female & 25 & None \\ \hline
P15 & Female & 17 & 1 \\
\arrayrulecolor{black}
\hline
\end{tabular}
\end{subtable}
\hfill
\begin{subtable}[t]{0.49\linewidth}
\centering
% \caption{Participants P16--P30}
\begin{tabular}{c c c c}
\hline
\textbf{Alias} & \textbf{Gender} & \textbf{Age} & \textbf{Dance Learning (mo.)} \\
\hline
\arrayrulecolor{black!10}
P16 & Female & 24 & 1 \\ \hline
P17 & Male   & 24 & 2 \\ \hline
P18 & Female & 40 & 5 \\ \hline
P19 & Female & 25 & 1 \\ \hline
P20 & Female & 40 & 1 \\ \hline
P21 & Male   & 38 & 1 \\ \hline
P22 & Female & 35 & 3 \\ \hline
P23 & Female & 26 & 1 \\ \hline
P24 & Female & 24 & 1 \\ \hline
P25 & Female & 36 & 1 \\ \hline
P26 & Female & 21 & 5 \\ \hline
P27 & Female & 26 & 5 \\ \hline
P28 & Female & 29 & None \\ \hline
P29 & Female & 22 & None \\ \hline
P30 & Female & 31 & 2 \\ \hline
\arrayrulecolor{black}
\hline
\end{tabular}
\end{subtable}
\end{table*}

\begin{table*}[h]
\centering
\small
\sffamily
\renewcommand{\arraystretch}{1.15}
\setlength{\tabcolsep}{4pt}
\caption{Participant demographics and learning order per criterion in learner study 1  (D/L=Dance Learning).}
\label{tab:demographics-learner_study_1}
\resizebox{\linewidth}{!}{%
\begin{tabular}{c c c c c c c c c}
\hline
\textbf{Alias} & \textbf{Age} & \textbf{Gender} & \textbf{D/L (mo.)} & $\boldsymbol{C_1}$ & $\boldsymbol{C_2}$ & $\boldsymbol{C_3}$ & $\boldsymbol{C_4}$ & $\boldsymbol{C_5}$ \\
\hline
\arrayrulecolor{black!10}
P1  & 24 & Female & None & \originalbox{orig.} $\rightarrow$ \simplifiedbox{simp.} & \simplifiedbox{simp.} $\rightarrow$ \originalbox{orig.} & \originalbox{orig.} $\rightarrow$ \simplifiedbox{simp.} & \simplifiedbox{simp.} $\rightarrow$ \originalbox{orig.} & \originalbox{orig.} $\rightarrow$ \simplifiedbox{simp.} \\ \hline
P2  & 20 & Male   & None & \originalbox{orig.} $\rightarrow$ \simplifiedbox{simp.} & \simplifiedbox{simp.} $\rightarrow$ \originalbox{orig.} & \originalbox{orig.} $\rightarrow$ \simplifiedbox{simp.} & \simplifiedbox{simp.} $\rightarrow$ \originalbox{orig.} & \originalbox{orig.} $\rightarrow$ \simplifiedbox{simp.} \\ \hline
P3  & 23 & Female & 2 & \originalbox{orig.} $\rightarrow$ \simplifiedbox{simp.} & \simplifiedbox{simp.} $\rightarrow$ \originalbox{orig.} & \originalbox{orig.} $\rightarrow$ \simplifiedbox{simp.} & \simplifiedbox{simp.} $\rightarrow$ \originalbox{orig.} & \originalbox{orig.} $\rightarrow$ \simplifiedbox{simp.} \\ \hline
P4  & 24 & Male   & None & \originalbox{orig.} $\rightarrow$ \simplifiedbox{simp.} & \simplifiedbox{simp.} $\rightarrow$ \originalbox{orig.} & \originalbox{orig.} $\rightarrow$ \simplifiedbox{simp.} & \simplifiedbox{simp.} $\rightarrow$ \originalbox{orig.} & \originalbox{orig.} $\rightarrow$ \simplifiedbox{simp.} \\ \hline
P5  & 19 & Female & None & \originalbox{orig.} $\rightarrow$ \simplifiedbox{simp.} & \simplifiedbox{simp.} $\rightarrow$ \originalbox{orig.} & \originalbox{orig.} $\rightarrow$ \simplifiedbox{simp.} & \simplifiedbox{simp.} $\rightarrow$ \originalbox{orig.} & \originalbox{orig.} $\rightarrow$ \simplifiedbox{simp.} \\ \hline
P6  & 21 & Male   & None & \originalbox{orig.} $\rightarrow$ \simplifiedbox{simp.} & \simplifiedbox{simp.} $\rightarrow$ \originalbox{orig.} & \originalbox{orig.} $\rightarrow$ \simplifiedbox{simp.} & \simplifiedbox{simp.} $\rightarrow$ \originalbox{orig.} & \originalbox{orig.} $\rightarrow$ \simplifiedbox{simp.} \\ \hline
P7  & 20 & Female & None & \originalbox{orig.} $\rightarrow$ \simplifiedbox{simp.} & \simplifiedbox{simp.} $\rightarrow$ \originalbox{orig.} & \originalbox{orig.} $\rightarrow$ \simplifiedbox{simp.} & \simplifiedbox{simp.} $\rightarrow$ \originalbox{orig.} & \originalbox{orig.} $\rightarrow$ \simplifiedbox{simp.} \\ \hline
P8  & 23 & Female & None & \simplifiedbox{simp.} $\rightarrow$ \originalbox{orig.} & \originalbox{orig.} $\rightarrow$ \simplifiedbox{simp.} & \simplifiedbox{simp.} $\rightarrow$ \originalbox{orig.} & \originalbox{orig.} $\rightarrow$ \simplifiedbox{simp.} & \simplifiedbox{simp.} $\rightarrow$ \originalbox{orig.} \\ \hline
P9  & 25 & Male   & None & \originalbox{orig.} $\rightarrow$ \simplifiedbox{simp.} & \simplifiedbox{simp.} $\rightarrow$ \originalbox{orig.} & \originalbox{orig.} $\rightarrow$ \simplifiedbox{simp.} & \simplifiedbox{simp.} $\rightarrow$ \originalbox{orig.} & \originalbox{orig.} $\rightarrow$ \simplifiedbox{simp.} \\ \hline
P10 & 29 & Male   & None & \simplifiedbox{simp.} $\rightarrow$ \originalbox{orig.} & \originalbox{orig.} $\rightarrow$ \simplifiedbox{simp.} & \simplifiedbox{simp.} $\rightarrow$ \originalbox{orig.} & \originalbox{orig.} $\rightarrow$ \simplifiedbox{simp.} & \simplifiedbox{simp.} $\rightarrow$ \originalbox{orig.} \\ \hline
P11 & 30 & Male   & None & \simplifiedbox{simp.} $\rightarrow$ \originalbox{orig.} & \originalbox{orig.} $\rightarrow$ \simplifiedbox{simp.} & \simplifiedbox{simp.} $\rightarrow$ \originalbox{orig.} & \originalbox{orig.} $\rightarrow$ \simplifiedbox{simp.} & \simplifiedbox{simp.} $\rightarrow$ \originalbox{orig.} \\ \hline
P12 & 28 & Female & None & \simplifiedbox{simp.} $\rightarrow$ \originalbox{orig.} & \originalbox{orig.} $\rightarrow$ \simplifiedbox{simp.} & \simplifiedbox{simp.} $\rightarrow$ \originalbox{orig.} & \originalbox{orig.} $\rightarrow$ \simplifiedbox{simp.} & \simplifiedbox{simp.} $\rightarrow$ \originalbox{orig.} \\ \hline
P13 & 29 & Female & None & \simplifiedbox{simp.} $\rightarrow$ \originalbox{orig.} & \originalbox{orig.} $\rightarrow$ \simplifiedbox{simp.} & \simplifiedbox{simp.} $\rightarrow$ \originalbox{orig.} & \originalbox{orig.} $\rightarrow$ \simplifiedbox{simp.} & \simplifiedbox{simp.} $\rightarrow$ \originalbox{orig.} \\ \hline
P14 & 25 & Male   & None & \simplifiedbox{simp.} $\rightarrow$ \originalbox{orig.} & \originalbox{orig.} $\rightarrow$ \simplifiedbox{simp.} & \simplifiedbox{simp.} $\rightarrow$ \originalbox{orig.} & \originalbox{orig.} $\rightarrow$ \simplifiedbox{simp.} & \simplifiedbox{simp.} $\rightarrow$ \originalbox{orig.} \\ \hline
P15 & 20 & Male   & None & \simplifiedbox{simp.} $\rightarrow$ \originalbox{orig.} & \originalbox{orig.} $\rightarrow$ \simplifiedbox{simp.} & \simplifiedbox{simp.} $\rightarrow$ \originalbox{orig.} & \originalbox{orig.} $\rightarrow$ \simplifiedbox{simp.} & \simplifiedbox{simp.} $\rightarrow$ \originalbox{orig.} \\ \hline
P16 & 28 & Female & None & \simplifiedbox{simp.} $\rightarrow$ \originalbox{orig.} & \originalbox{orig.} $\rightarrow$ \simplifiedbox{simp.} & \simplifiedbox{simp.} $\rightarrow$ \originalbox{orig.} & \originalbox{orig.} $\rightarrow$ \simplifiedbox{simp.} & \simplifiedbox{simp.} $\rightarrow$ \originalbox{orig.} \\ \hline
\arrayrulecolor{black}
\hline
\end{tabular}%
}
\end{table*}

\begin{table}[t]
\centering
\renewcommand{\arraystretch}{1.1}
\scriptsize\sffamily
\caption{Participant demographics in expert assessment (C/E=Choreographic Experience, T/E=Teaching Experience).}
\label{tab:participant-demographics}
\setlength{\tabcolsep}{4pt}
\begin{subtable}[t]{0.49\linewidth}
\centering
\begin{tabular}{c c c c c}
\hline
\textbf{Alias} & \textbf{Gender} & \textbf{Genre} & \textbf{C/E (yrs.)} & \textbf{T/E (yrs.)} \\
\hline
\arrayrulecolor{black!10}
P1  & Female & Hip-hop & 8 & 4 \\ \hline
P2  & Female & Hip-hop & 6 & 5 \\ \hline
P3  & Female & Waack & 7 & 3 \\ \hline
P4  & Male   & Contemporary & 4 & 3 \\ \hline
P5  & Female & Contemporary & 6 & 6 \\ \hline
P6  & Female & Lock & 10 & 5 \\ \hline
P7  & Female & Hip-hop & 9 & 7 \\ \hline
P8  & Male   & Hip-hop & 8 & 8 \\ \hline
P9  & Male   & Break & 10 & 8 \\ \hline
P10 & Female & Contemporary & 10 & 8 \\ \hline
\arrayrulecolor{black}
\hline
\end{tabular}
\end{subtable}
\hfill
\begin{subtable}[t]{0.49\linewidth}
\centering
\begin{tabular}{c c c c c}
\hline
\textbf{Alias} & \textbf{Gender} & \textbf{Genre} & \textbf{C/E (yrs.)} & \textbf{T/E (yrs.)} \\
\hline
\arrayrulecolor{black!10}
P11 & Male   & Choreography & 13 & 13 \\ \hline
P12 & Male   & Krump & 13 & 6 \\ \hline
P13 & Male   & Pop & 18 & 13 \\ \hline
P14 & Male   & Choreography, Krump & 3 & 2 \\ \hline
P15 & Female & Ballet & 5 & 5 \\ \hline
P16 & Female & Choreography & 8 & 9 \\ \hline
P17 & Male   & Waack & 10 & 5 \\ \hline
P18 & Male   & Contemporary & 15 & 7 \\ \hline
P19 & Female & Ballet & 13 & 9 \\ \hline
P20 & Male   & Choreo., Krump & 12 & 11 \\ \hline
\arrayrulecolor{black}
\hline
\end{tabular}
\end{subtable}
\end{table}

\begin{table*}[h]
\centering
\small\sffamily
\renewcommand{\arraystretch}{1.25}
\setlength{\tabcolsep}{3pt}
\caption{Participant demographics and learning order per criterion in learner study 2  (D/L=Dance Learning).}
\label{tab:order-by-criterion-version}
\resizebox{\linewidth}{!}{%
\begin{tabular}{c c c c c c c c c c}
\hline
\textbf{Group} & \textbf{Alias} & \textbf{Age} & \textbf{Gender} & \textbf{D/L (mo.)} & $\boldsymbol{C_1}$ & $\boldsymbol{C_2}$ & $\boldsymbol{C_3}$ & $\boldsymbol{C_4}$ & $\boldsymbol{C_5}$ \\
\hline
\arrayrulecolor{black!10}
\multirow{6}{*}{\gtbox{GT}}
 & P1  & 21 & Female & None & \originalbox{orig.} $\rightarrow$ \gtbox{GT} & \gtbox{GT} $\rightarrow$ \originalbox{orig.} & \originalbox{orig.} $\rightarrow$ \gtbox{GT} & \gtbox{GT} $\rightarrow$ \originalbox{orig.} & \originalbox{orig.} $\rightarrow$ \gtbox{GT} \\ \cline{2-10}
 & P2  & 20 & Female & 2 & \gtbox{GT} $\rightarrow$ \originalbox{orig.} & \originalbox{orig.} $\rightarrow$ \gtbox{GT} & \gtbox{GT} $\rightarrow$ \originalbox{orig.} & \originalbox{orig.} $\rightarrow$ \gtbox{GT} & \gtbox{GT} $\rightarrow$ \originalbox{orig.} \\ \cline{2-10}
 & P3  & 20 & Female & None & \originalbox{orig.} $\rightarrow$ \gtbox{GT} & \gtbox{GT} $\rightarrow$ \originalbox{orig.} & \originalbox{orig.} $\rightarrow$ \gtbox{GT} & \gtbox{GT} $\rightarrow$ \originalbox{orig.} & \originalbox{orig.} $\rightarrow$ \gtbox{GT} \\ \cline{2-10}
 & P4  & 22 & Male   & None & \gtbox{GT} $\rightarrow$ \originalbox{orig.} & \originalbox{orig.} $\rightarrow$ \gtbox{GT} & \gtbox{GT} $\rightarrow$ \originalbox{orig.} & \originalbox{orig.} $\rightarrow$ \gtbox{GT} & \gtbox{GT} $\rightarrow$ \originalbox{orig.} \\ \cline{2-10}
 & P5  & 20 & Male   & None & \originalbox{orig.} $\rightarrow$ \gtbox{GT} & \gtbox{GT} $\rightarrow$ \originalbox{orig.} & \originalbox{orig.} $\rightarrow$ \gtbox{GT} & \gtbox{GT} $\rightarrow$ \originalbox{orig.} & \originalbox{orig.} $\rightarrow$ \gtbox{GT} \\ \cline{2-10}
 & P6  & 23 & Male   & None & \gtbox{GT} $\rightarrow$ \originalbox{orig.} & \originalbox{orig.} $\rightarrow$ \gtbox{GT} & \gtbox{GT} $\rightarrow$ \originalbox{orig.} & \originalbox{orig.} $\rightarrow$ \gtbox{GT} & \gtbox{GT} $\rightarrow$ \originalbox{orig.} \\
\arrayrulecolor{black!30}
\hline
\arrayrulecolor{black!10}
\multirow{6}{*}{\rbbox{rule}}
 & P7  & 20 & Female & None & \originalbox{orig.} $\rightarrow$ \rbbox{rule} & \rbbox{rule} $\rightarrow$ \originalbox{orig.} & \originalbox{orig.} $\rightarrow$ \rbbox{rule} & \rbbox{rule} $\rightarrow$ \originalbox{orig.} & \originalbox{orig.} $\rightarrow$ \rbbox{rule} \\ \cline{2-10}
 & P8  & 24 & Male   & 3 & \originalbox{orig.} $\rightarrow$ \rbbox{rule} & \rbbox{rule} $\rightarrow$ \originalbox{orig.} & \originalbox{orig.} $\rightarrow$ \rbbox{rule} & \rbbox{rule} $\rightarrow$ \originalbox{orig.} & \originalbox{orig.} $\rightarrow$ \rbbox{rule} \\ \cline{2-10}
 & P9  & 23 & Male   & None & \originalbox{orig.} $\rightarrow$ \rbbox{rule} & \rbbox{rule} $\rightarrow$ \originalbox{orig.} & \originalbox{orig.} $\rightarrow$ \rbbox{rule} & \rbbox{rule} $\rightarrow$ \originalbox{orig.} & \originalbox{orig.} $\rightarrow$ \rbbox{rule} \\ \cline{2-10}
 & P10 & 22 & Female & 4 & \rbbox{rule} $\rightarrow$ \originalbox{orig.} & \originalbox{orig.} $\rightarrow$ \rbbox{rule} & \rbbox{rule} $\rightarrow$ \originalbox{orig.} & \originalbox{orig.} $\rightarrow$ \rbbox{rule} & \rbbox{rule} $\rightarrow$ \originalbox{orig.} \\ \cline{2-10}
 & P11 & 22 & Female & None & \rbbox{rule} $\rightarrow$ \originalbox{orig.} & \originalbox{orig.} $\rightarrow$ \rbbox{rule} & \rbbox{rule} $\rightarrow$ \originalbox{orig.} & \originalbox{orig.} $\rightarrow$ \rbbox{rule} & \rbbox{rule} $\rightarrow$ \originalbox{orig.} \\ \cline{2-10}
 & P12 & 24 & Male   & None & \rbbox{rule} $\rightarrow$ \originalbox{orig.} & \originalbox{orig.} $\rightarrow$ \rbbox{rule} & \rbbox{rule} $\rightarrow$ \originalbox{orig.} & \originalbox{orig.} $\rightarrow$ \rbbox{rule} & \rbbox{rule} $\rightarrow$ \originalbox{orig.} \\
\arrayrulecolor{black!30}
\hline
\arrayrulecolor{black!10}
\multirow{6}{*}{\lbbox{learning}}
 & P13 & 18 & Female & None & \lbbox{learning} $\rightarrow$ \originalbox{orig.} & \originalbox{orig.} $\rightarrow$ \lbbox{learning} & \lbbox{learning} $\rightarrow$ \originalbox{orig.} & \originalbox{orig.} $\rightarrow$ \lbbox{learning} & \lbbox{learning} $\rightarrow$ \originalbox{orig.} \\ \cline{2-10}
 & P14 & 21 & Female & None & \originalbox{orig.} $\rightarrow$ \lbbox{learning} & \lbbox{learning} $\rightarrow$ \originalbox{orig.} & \originalbox{orig.} $\rightarrow$ \lbbox{learning} & \lbbox{learning} $\rightarrow$ \originalbox{orig.} & \originalbox{orig.} $\rightarrow$ \lbbox{learning} \\ \cline{2-10}
 & P15 & 25 & Male   & None & \lbbox{learning} $\rightarrow$ \originalbox{orig.} & \originalbox{orig.} $\rightarrow$ \lbbox{learning} & \lbbox{learning} $\rightarrow$ \originalbox{orig.} & \originalbox{orig.} $\rightarrow$ \lbbox{learning} & \lbbox{learning} $\rightarrow$ \originalbox{orig.} \\ \cline{2-10}
 & P16 & 22 & Male   & None & \lbbox{learning} $\rightarrow$ \originalbox{orig.} & \originalbox{orig.} $\rightarrow$ \lbbox{learning} & \lbbox{learning} $\rightarrow$ \originalbox{orig.} & \originalbox{orig.} $\rightarrow$ \lbbox{learning} & \lbbox{learning} $\rightarrow$ \originalbox{orig.} \\ \cline{2-10}
 & P17 & 21 & Female & None & \originalbox{orig.} $\rightarrow$ \lbbox{learning} & \lbbox{learning} $\rightarrow$ \originalbox{orig.} & \originalbox{orig.} $\rightarrow$ \lbbox{learning} & \lbbox{learning} $\rightarrow$ \originalbox{orig.} & \originalbox{orig.} $\rightarrow$ \lbbox{learning} \\ \cline{2-10}
 & P18 & 22 & Male   & None & \originalbox{orig.} $\rightarrow$ \lbbox{learning} & \lbbox{learning} $\rightarrow$ \originalbox{orig.} & \originalbox{orig.} $\rightarrow$ \lbbox{learning} & \lbbox{learning} $\rightarrow$ \originalbox{orig.} & \originalbox{orig.} $\rightarrow$ \lbbox{learning} \\
\arrayrulecolor{black}
\hline
\end{tabular}%
}
\end{table*}

% --------------------------------------------------------------------------------------------------

\newpage
\section{Survey Questionnaire for Novice Dance Learners}\label{appendix:survey-choreo-learners}

\subsection{Screening and Experience Background}

\begin{tcolorbox}[colback=gray!10, colframe=black!10, boxrule=0.5pt, left=5pt, right=5pt, top=5pt, bottom=5pt, breakable]
\sffamily
\textbf{(1) Dance learning experience (screening):}

This survey is designed for learners with less than 6 months of total dance learning experience. Is your total dance learning period less than 6 months?

\textit{Options:}
\begin{itemize}[leftmargin=1.5em, label=◦]
    \item Yes
    \item No
\end{itemize}

\vspace{0.1cm}
\textbf{(2) Total duration of choreography learning:}

How long have you been learning dance? Please specify your total experience in months.

\textit{Response format:} Open-ended (numerical, in months)

\vspace{0.1cm}
\rule{10cm}{0.4pt}

\end{tcolorbox}

\subsection{Class Context and Difficulty}

\begin{tcolorbox}[colback=gray!10, colframe=black!10, boxrule=0.5pt, left=5pt, right=5pt, top=5pt, bottom=5pt, breakable]
\sffamily
\textbf{(3) Song used in the class:}

What song was used in the dance class you attended today?

\textit{Response format:} Open-ended

\vspace{0.1cm}
Song title / Artist: \rule{8cm}{0.4pt}

\vspace{0.1cm}
\textbf{(4) Perceived difficulty of today's choreography:}

How would you rate the difficulty level of the choreography you learned today?

\textit{Scale:} 1 = Very easy, 7 = Very difficult

\begin{center}
\begin{tabular}{ccccccc}
◦ 1 & ◦ 2 & ◦ 3 & ◦ 4 & ◦ 5 & ◦ 6 & ◦ 7 \\
\end{tabular}
\end{center}

\end{tcolorbox}

\subsection{Identification of Difficult Movements}

\begin{tcolorbox}[colback=gray!10, colframe=black!10, boxrule=0.5pt, left=5pt, right=5pt, top=5pt, bottom=5pt, breakable]
\sffamily
\textbf{(5) Difficult movements encountered:}

Which movements from today's choreography did you find difficult? Please describe the movements and their context (e.g., timing, body parts involved).

\textit{Response format:} Open-ended

\vspace{0.1cm}
\rule{\textwidth}{0.4pt}
\vspace{0.1cm}
\rule{\textwidth}{0.4pt}

\vspace{0.1cm}
\textbf{(6) Reasons for difficulty:}

Why were these particular movements difficult for you to learn?

\textit{Response format:} Open-ended

\vspace{0.1cm}
\rule{\textwidth}{0.4pt}
\vspace{0.1cm}
\rule{\textwidth}{0.4pt}

\textbf{(7) Learning strategies or know-how:}

Did you or your instructor use any specific strategies or techniques to help you learn these difficult movements?

\textit{Response format:} Open-ended

\vspace{0.1cm}
\rule{\textwidth}{0.4pt}
\vspace{0.1cm}
\rule{\textwidth}{0.4pt}

\vspace{0.1cm}

\textbf{(8) Emotional responses to difficult movements:}

Please describe your feelings, thoughts, or emotions when encountering movements that were difficult to execute.

\textit{Response format:} Open-ended

\vspace{0.1cm}
\rule{\textwidth}{0.4pt}
\vspace{0.1cm}
\rule{\textwidth}{0.4pt}
\end{tcolorbox}

\subsection{Perceived Usefulness of Simplification}

\begin{tcolorbox}[colback=gray!10, colframe=black!10, boxrule=0.5pt, left=5pt, right=5pt, top=5pt, bottom=5pt, breakable]
\sffamily
\textbf{(9) Helpfulness of simplification for learning:}

If difficult movements were presented in a simplified version, how much do you think it would help you learn those movements?

\textit{Scale:} 1 = Not helpful at all, 7 = Extremely helpful

\vspace{0.1cm}
\begin{center}
\begin{tabular}{ccccccc}
◦ 1 & ◦ 2 & ◦ 3 & ◦ 4 & ◦ 5 & ◦ 6 & ◦ 7 \\
\end{tabular}
\end{center}

\vspace{0.1cm}

\textbf{(10) Effect of simplification on learning confidence:}

Do you think learning simplified versions of difficult movements would increase your confidence in eventually mastering the original choreography?

\textit{Scale:} 1 = Would not increase confidence at all, 7 = Would greatly increase confidence

\begin{center}
\begin{tabular}{ccccccc}
◦ 1 & ◦ 2 & ◦ 3 & ◦ 4 & ◦ 5 & ◦ 6 & ◦ 7 \\
\end{tabular}
\end{center}
\end{tcolorbox}

% --------------------------------------------------------------------------------------------------

\section{Configurations for Classifiers}\label{appendix:configs-for-classifiers}

\begin{table}[h]
\centering\sffamily
\renewcommand{\arraystretch}{1.15}
\setlength{\tabcolsep}{3pt}
\caption{Detailed hyperparameter configurations for complexity metric classifiers. All models use 5-fold cross-validation with grid search optimization.}
\label{tab:classifier-configs}
\resizebox{\textwidth}{!}{%
\begin{tabular}{ccccccccccc}
\hline
\textbf{Metric} & \textbf{Algorithm} & \textbf{Estimators} & \textbf{Max Depth} & \textbf{Learning Rate} & \textbf{Subsample} & \textbf{Max Features} & \textbf{Col. Sample} & \textbf{Min Child} & \textbf{Reg. ($\alpha$/$\lambda$)} & \textbf{Early Stop} \\
\hline
\arrayrulecolor{black!10}
$C_1$ & Gradient Boost & 1000 & 12 & 0.02 & 0.8 & sqrt & -- & -- & -- & -- \\ \hline
$C_2$ & XGBoost & 1000 & 8 & 0.02 & 0.8 & -- & 0.8 & 3 & 0.1/1.0 & 50 \\ \hline
$C_3$ & Gradient Boost & 800 & 10 & 0.03 & 0.8 & sqrt & -- & -- & -- & -- \\ \hline
$C_4$ & Gradient Boost & 1000 & 12 & 0.02 & 0.8 & sqrt & -- & -- & -- & -- \\ \hline
$C_5$ & Gradient Boost & 800 & 10 & 0.03 & 0.8 & sqrt & -- & -- & -- & -- \\
\arrayrulecolor{black}
\hline
\end{tabular}%
}
\end{table}

% --------------------------------------------------------------------------------------------------

\section{Pseudocode for Simplification Approaches}\label{appendix:pseudocode}

This appendix provides pseudocode for the rule-based simplification pipeline in \S\ref{subsec:rule-based-model}.

\subsection{Rule-Based Approach}\label{appendix:pseudocode-rule}
\sffamily

\begin{breakablealgorithm}{Rule-Based Dance Motion Simplification (Full Pipeline)}{alg:rule-full}
\label{alg:rule-full}
% \begin{algorithmic}[1]
\Require Joint sequence $\mathbf{j}^p \in \mathbb{R}^{F\times J\times 3}$; thresholds $\varepsilon,\alpha$;
metric thresholds $\{\tau_i\}_{i=1}^5$; minimum segment lengths $\{\ell_i\}_{i=1}^5$;
compression $k\in[0,1]$; slowdown $\lambda\in\mathbb{Z}^+$; target yaw $\psi_{\text{target}}$; direction vector $\vec{v}\in\{-1,1\}^3$
\Ensure Simplified sequence $\mathbf{j}^{p,\textit{gen}}$

\State $\mathbf{C} \gets \Call{GetComplexities}{\mathbf{j}^p}$
\State $\mathbf{j}^{p,\textit{gen}} \gets \mathbf{j}^p$
\State $\mathcal{T} \gets \Call{MotionDetection}{\mathbf{j}^{p,\textit{gen}}, \varepsilon, \alpha}$ \Comment{Alg.~\ref{alg:motion-detect}}

\Statex \Statex \textit{($C_1$: steps/footwork) velocity reduction in lower limbs}
\If{$\mathbf{C}[1] > \tau_1$}
  \State $\mathcal{I} \gets \Call{DetectIntervalsByMetric}{\mathbf{j}^{p,\textit{gen}}, \mathcal{T}, 1, \tau_1, \ell_1}$ \Comment{Alg.~\ref{alg:detect-intervals}}
  \State $\mathbf{j}^{\textit{temp}} \gets \Call{VelocityReduction}{\mathbf{j}^{p,\textit{gen}}, \mathcal{I}, \lambda}$ \Comment{Alg.~\ref{alg:vel-reduce}}
  \State $\mathbf{C}^{\textit{new}} \gets \Call{GetComplexities}{\mathbf{j}^{\textit{temp}}}$
  \If{$\mathbf{C}^{\textit{new}}[1] < \mathbf{C}[1]$}
    \State $\mathbf{j}^{p,\textit{gen}} \gets \mathbf{j}^{\textit{temp}}$;\quad $\mathbf{C}\gets \mathbf{C}^{\textit{new}}$
  \EndIf
\EndIf

\Statex \Statex \textit{($C_2$: movement density) distance compression}
\If{$\mathbf{C}[2] > \tau_2$}
  \State $\mathcal{I} \gets \Call{DetectIntervalsByMetric}{\mathbf{j}^{p,\textit{gen}}, \mathcal{T}, 2, \tau_2, \ell_2}$
  \State $\mathbf{j}^{\textit{temp}} \gets \Call{DistanceCompression}{\mathbf{j}^{p,\textit{gen}}, \mathcal{I}, k}$ \Comment{Alg.~\ref{alg:dist-compress}}
  \State $\mathbf{C}^{\textit{new}} \gets \Call{GetComplexities}{\mathbf{j}^{\textit{temp}}}$
  \If{$\mathbf{C}^{\textit{new}}[2] < \mathbf{C}[2]$}
    \State $\mathbf{j}^{p,\textit{gen}} \gets \mathbf{j}^{\textit{temp}}$;\quad $\mathbf{C}\gets \mathbf{C}^{\textit{new}}$
  \EndIf
\EndIf

\Statex \Statex \textit{($C_3$: orientation change) orientation stabilization}
\If{$\mathbf{C}[3] > \tau_3$}
  \State $\mathcal{I} \gets \Call{DetectIntervalsByMetric}{\mathbf{j}^{p,\textit{gen}}, \mathcal{T}, 3, \tau_3, \ell_3}$
  \State $\mathbf{j}^{\textit{temp}} \gets \Call{OrientationStabilization}{\mathbf{j}^{p,\textit{gen}}, \mathcal{I}, \psi_{\text{target}}}$ \Comment{Alg.~\ref{alg:ori-stab}}
  \State $\mathbf{C}^{\textit{new}} \gets \Call{GetComplexities}{\mathbf{j}^{\textit{temp}}}$
  \If{$\mathbf{C}^{\textit{new}}[3] < \mathbf{C}[3]$}
    \State $\mathbf{j}^{p,\textit{gen}} \gets \mathbf{j}^{\textit{temp}}$;\quad $\mathbf{C}\gets \mathbf{C}^{\textit{new}}$
  \EndIf
\EndIf

\Statex \Statex \textit{($C_4$: multi-limb coordination) distance compression}
\If{$\mathbf{C}[4] > \tau_4$}
  \State $\mathcal{I} \gets \Call{DetectIntervalsByMetric}{\mathbf{j}^{p,\textit{gen}}, \mathcal{T}, 4, \tau_4, \ell_4}$
  \State $\mathbf{j}^{\textit{temp}} \gets \Call{DistanceCompression}{\mathbf{j}^{p,\textit{gen}}, \mathcal{I}, k}$
  \State $\mathbf{C}^{\textit{new}} \gets \Call{GetComplexities}{\mathbf{j}^{\textit{temp}}}$
  \If{$\mathbf{C}^{\textit{new}}[4] < \mathbf{C}[4]$}
    \State $\mathbf{j}^{p,\textit{gen}} \gets \mathbf{j}^{\textit{temp}}$;\quad $\mathbf{C}\gets \mathbf{C}^{\textit{new}}$
  \EndIf
\EndIf

\Statex \Statex \textit{($C_5$: bilateral asymmetry) direction change}
\If{$\mathbf{C}[5] > \tau_5$}
  \State $\mathcal{I} \gets \Call{DetectIntervalsByMetric}{\mathbf{j}^{p,\textit{gen}}, \mathcal{T}, 5, \tau_5, \ell_5}$
  \State $\mathbf{j}^{\textit{temp}} \gets \Call{DirectionalChange}{\mathbf{j}^{p,\textit{gen}}, \mathcal{I}, \vec{v}}$ \Comment{Alg.~\ref{alg:dir-change}}
  \State $\mathbf{C}^{\textit{new}} \gets \Call{GetComplexities}{\mathbf{j}^{\textit{temp}}}$
  \If{$\mathbf{C}^{\textit{new}}[5] < \mathbf{C}[5]$}
    \State $\mathbf{j}^{p,\textit{gen}} \gets \mathbf{j}^{\textit{temp}}$
  \EndIf
\EndIf

\State \Return $\mathbf{j}^{p,\textit{gen}}$
% \end{algorithmic}
\end{breakablealgorithm}

\begin{breakablealgorithm}{Motion Detection (Per-Axis Segmentation + Sweep-Line Merge)}{alg:motion-detect}
\label{alg:motion-detect}
% \begin{algorithmic}[1]
\Require $\mathbf{j}^p \in \mathbb{R}^{F\times J\times 3}$; thresholds $\varepsilon,\alpha$
\Ensure Trend set $\mathcal{T}=\{(j,s,e,\vec{v})\}$, $\vec{v}\in\{-1,0,1\}^3$
\State $\mathcal{T}_{\text{axis}}\gets\emptyset$
\For{$j\gets 1$ \textbf{to} $J$}
  \For{axis $a$ in $\{x,y,z\}$}
    \State Compute $d_t \gets \mathbf{j}^{p}_{t+1}[j]_a-\mathbf{j}^{p}_{t}[j]_a$ for $t=1,\dots,F-1$
    \State Define $\sigma_t \gets \operatorname{sign}_{\varepsilon}(d_t)$ where $\operatorname{sign}_{\varepsilon}(u)=+1$ if $u\ge\varepsilon$, $-1$ if $u\le-\varepsilon$, else $0$
    \State Segment $\sigma_t$ into maximal contiguous runs with constant nonzero value; each run yields an interval $[s,e]$
    \For{each interval $[s,e]$}
      \State Set $v_a \gets \sigma_s$ and store trend $(j,s,e,\vec{v}_{a\text{-only}})$ in $\mathcal{T}_{\text{axis}}$
    \EndFor
  \EndFor
\EndFor
\State Sort $\mathcal{T}_{\text{axis}}$ by $(j,s)$
\State $\mathcal{T}\gets\emptyset$
\For{$j\gets 1$ \textbf{to} $J$}
  \State $\mathcal{A}\gets\emptyset$
  \For{each axis-trend $(j,s_i,e_i,\vec{v}_i)$ in order}
    \State merged $\gets$ false
    \For{each active trend $(j,s_a,e_a,\vec{v}_a)$ in $\mathcal{A}$}
      \State $r \gets \frac{|[s_a,e_a]\cap[s_i,e_i]|}{\min(e_a-s_a,\ e_i-s_i)}$
      \If{$r\ge\alpha$}
        \State $s_a\gets \min(s_a,s_i)$;\quad $e_a\gets \max(e_a,e_i)$
        \State $\vec{v}_a\gets \vec{v}_a+\vec{v}_i$ \Comment{sum axis indicators}
        \State merged $\gets$ true
      \EndIf
    \EndFor
    \If{\textbf{not} merged}
      \State insert $(j,s_i,e_i,\vec{v}_i)$ into $\mathcal{A}$
    \EndIf
    \State remove any $(j,s_a,e_a,\vec{v}_a)$ from $\mathcal{A}$ with $e_a < s_i$
  \EndFor
  \State $\mathcal{T}\gets \mathcal{T}\cup \mathcal{A}$
\EndFor
\State \Return $\mathcal{T}$
% \end{algorithmic}
\end{breakablealgorithm}

\begin{breakablealgorithm}{Interval Detection by Complexity Activation}{alg:detect-intervals}
\label{alg:detect-intervals}
% \begin{algorithmic}[1]
\Require Current sequence $\mathbf{j}^{p,\textit{gen}}$; trends $\mathcal{T}$; metric index $i\in\{1,\dots,5\}$; threshold $\tau_i$; min length $\ell_i$
\Ensure Interval set $\mathcal{I}_i=\{(s,e,\mathcal{J})\}$
\State Compute frame-wise metric activation $m_i[f]$ from \S\ref{subsec:complexity-metrics}
\State $b_i[f]\gets \mathbb{1}[m_i[f]>\tau_i]$
\State Extract maximal contiguous positive runs of $b_i[f]$ with length $\ge \ell_i$ to obtain segments $\mathcal{S}_i$
\State $\mathcal{I}_i\gets\emptyset$
\For{each segment $[s,e]\in\mathcal{S}_i$}
  \State $\mathcal{J}\gets \Call{SelectTargetJoints}{i,\mathcal{T},[s,e]}$
  \State add $(s,e,\mathcal{J})$ to $\mathcal{I}_i$
\EndFor
\State \Return $\mathcal{I}_i$
% \end{algorithmic}
\end{breakablealgorithm}

\begin{breakablealgorithm}{Distance Compression ($C_2$, $C_4$)}{alg:dist-compress}
\label{alg:dist-compress}
% \begin{algorithmic}[1]
\Require $\mathbf{j}^{p,\textit{gen}}$; intervals $\mathcal{I}$; factor $k\in[0,1]$
\Ensure Updated $\mathbf{j}^{p,\textit{gen}}$
\For{each $(s,e,\mathcal{J})\in\mathcal{I}$}
  \State $\mathbf{j}^{p,\textit{src}}\gets \mathbf{j}^{p,\textit{gen}}$
  \For{each joint $j\in\mathcal{J}$}
    \State $\mathbf{j}^{p,\textit{gen}}_{s}[j]\gets \mathbf{j}^{p,\textit{src}}_{s}[j]$
    \For{$t=s+1$ \textbf{to} $e$}
      \State $\Delta \gets \mathbf{j}^{p,\textit{src}}_{t}[j]-\mathbf{j}^{p,\textit{src}}_{t-1}[j]$
      \State $\mathbf{j}^{p,\textit{gen}}_{t}[j]\gets \mathbf{j}^{p,\textit{gen}}_{t-1}[j] + k\cdot \Delta$
    \EndFor
  \EndFor
  \State $\mathbf{j}^{p,\textit{gen}}\gets \Call{JointGroupReattachment}{\mathbf{j}^{p,\textit{src}}, \mathbf{j}^{p,\textit{gen}}, \mathcal{J}, [s,e]}$ \Comment{Alg.~\ref{alg:reattach}}
  \State $\mathbf{j}^{p,\textit{gen}}\gets \Call{TemporalDiscontinuitySmoothing}{\mathbf{j}^{p,\textit{src}}, \mathbf{j}^{p,\textit{gen}}, \mathcal{J}, e}$ \Comment{Alg.~\ref{alg:smooth}}
\EndFor
\State \Return $\mathbf{j}^{p,\textit{gen}}$
% \end{algorithmic}
\end{breakablealgorithm}

\begin{breakablealgorithm}{Velocity Reduction ($C_1$)}{alg:vel-reduce}
\label{alg:vel-reduce}
% \begin{algorithmic}[1]
\Require $\mathbf{j}^{p,\textit{gen}}$; intervals $\mathcal{I}$; slowdown $\lambda\in\mathbb{Z}^+$ ($\lambda>1$)
\Ensure Updated $\mathbf{j}^{p,\textit{gen}}$
\For{each $(s,e,\mathcal{J})\in\mathcal{I}$}
  \State $e_{\text{new}}\gets s+(e-s)\lambda$
  \If{$e_{\text{new}} \ge F$}
    \State \textbf{continue} \Comment{skip if stretched endpoint exceeds sequence length}
  \EndIf
  \State $\mathbf{j}^{p,\textit{src}}\gets \mathbf{j}^{p,\textit{gen}}$
  \For{each joint $j\in\mathcal{J}$}
    \For{$k=0$ \textbf{to} $(e-s-1)$}
      \State $\Delta \gets \mathbf{j}^{p,\textit{src}}_{s+k+1}[j]-\mathbf{j}^{p,\textit{src}}_{s+k}[j]$
      \For{$i=0$ \textbf{to} $(\lambda-1)$}
        \State $f \gets s+k\lambda+i$
        \State $\mathbf{j}^{p,\textit{gen}}_{f}[j]\gets \mathbf{j}^{p,\textit{src}}_{s+k}[j] + \frac{i}{\lambda}\Delta$
      \EndFor
    \EndFor
    \State $\mathbf{j}^{p,\textit{gen}}_{e_{\text{new}}}[j]\gets \mathbf{j}^{p,\textit{src}}_{e}[j]$ \Comment{final keyframe alignment}
  \EndFor
  \State $\mathbf{j}^{p,\textit{gen}}\gets \Call{JointGroupReattachment}{\mathbf{j}^{p,\textit{src}}, \mathbf{j}^{p,\textit{gen}}, \mathcal{J}, [s,e_{\text{new}}]}$
  \State $\mathbf{j}^{p,\textit{gen}}\gets \Call{TemporalDiscontinuitySmoothing}{\mathbf{j}^{p,\textit{src}}, \mathbf{j}^{p,\textit{gen}}, \mathcal{J}, e_{\text{new}}}$
\EndFor
\State \Return $\mathbf{j}^{p,\textit{gen}}$
% \end{algorithmic}
\end{breakablealgorithm}

\begin{breakablealgorithm}{Directional Change ($C_5$)}{alg:dir-change}
\label{alg:dir-change}
% \begin{algorithmic}[1]
\Require $\mathbf{j}^{p,\textit{gen}}$; intervals $\mathcal{I}$; $\vec{v}\in\{-1,1\}^3$
\Ensure Updated $\mathbf{j}^{p,\textit{gen}}$
\For{each $(s,e,\mathcal{J})\in\mathcal{I}$}
  \State $\mathbf{j}^{p,\textit{src}}\gets \mathbf{j}^{p,\textit{gen}}$
  \For{each joint $j\in\mathcal{J}$}
    \State $\mathbf{p}_s \gets \mathbf{j}^{p,\textit{src}}_{s}[j]$
    \State $\mathbf{c}\gets (0,0,0)$
    \For{$t=s+1$ \textbf{to} $e$}
      \State $\Delta \gets \mathbf{j}^{p,\textit{src}}_{t}[j]-\mathbf{j}^{p,\textit{src}}_{t-1}[j]$
      \State $\mathbf{c}\gets \mathbf{c} + (\vec{v}\odot \Delta)$
      \State $\mathbf{j}^{p,\textit{gen}}_{t}[j]\gets \mathbf{p}_s + \mathbf{c}$
    \EndFor
  \EndFor
  \State $\mathbf{j}^{p,\textit{gen}}\gets \Call{JointGroupReattachment}{\mathbf{j}^{p,\textit{src}}, \mathbf{j}^{p,\textit{gen}}, \mathcal{J}, [s,e]}$
  \State $\mathbf{j}^{p,\textit{gen}}\gets \Call{TemporalDiscontinuitySmoothing}{\mathbf{j}^{p,\textit{src}}, \mathbf{j}^{p,\textit{gen}}, \mathcal{J}, e}$
\EndFor
\State \Return $\mathbf{j}^{p,\textit{gen}}$
% \end{algorithmic}
\end{breakablealgorithm}

\begin{breakablealgorithm}{Orientation Stabilization ($C_3$)}{alg:ori-stab}
\label{alg:ori-stab}
% \begin{algorithmic}[1]
\Require $\mathbf{j}^{p,\textit{gen}}$; intervals $\mathcal{I}$; target yaw $\psi_{\text{target}}$
\Ensure Updated $\mathbf{j}^{p,\textit{gen}}$
\For{each $(s,e,\mathcal{J})\in\mathcal{I}$}
  \For{$f=s$ \textbf{to} $e$}
    \State $\mathbf{p}\gets \mathbf{j}^{p,\textit{gen}}_{f}[\text{pelvis}]$
    \State $\mathbf{r}\gets \mathbf{j}^{p,\textit{gen}}_{f}[\text{ref}]$
    \State $\psi_f \gets \Call{YawFromXZ}{\mathbf{p},\mathbf{r}}$
    \State $\theta_f \gets \psi_{\text{target}} - \psi_f$
    \State $\mathbf{R}\gets \Call{RotateYMatrix}{\theta_f}$
    \For{$j\gets 1$ \textbf{to} $J$}
      \State $\mathbf{j}^{p,\textit{gen}}_{f}[j]\gets \mathbf{R}\big(\mathbf{j}^{p,\textit{gen}}_{f}[j]-\mathbf{p}\big)+\mathbf{p}$
    \EndFor
  \EndFor
\EndFor
\State \Return $\mathbf{j}^{p,\textit{gen}}$
% \end{algorithmic}
\end{breakablealgorithm}

\begin{breakablealgorithm}{Joint Group Reattachment via Root Offset}{alg:reattach}
\label{alg:reattach}
% \begin{algorithmic}[1]
\Require Original $\mathbf{j}^{p}$; current $\mathbf{j}^{p,\textit{gen}}$; ordered joint chain $\mathcal{J}$; interval $[s,e]$
\Ensure Updated $\mathbf{j}^{p,\textit{gen}}$
\State $r \gets \mathcal{J}[1]$ \Comment{first joint in the chain}
\For{$f=s$ \textbf{to} $e$}
  \State $\Delta_f \gets \mathbf{j}^{p}_{f}[r] - \mathbf{j}^{p,\textit{gen}}_{f}[r]$
  \For{each $j\in\mathcal{J}$}
    \State $\mathbf{j}^{p,\textit{gen}}_{f}[j] \gets \mathbf{j}^{p,\textit{gen}}_{f}[j] + \Delta_f$
  \EndFor
\EndFor
\State \Return $\mathbf{j}^{p,\textit{gen}}$
% \end{algorithmic}
\end{breakablealgorithm}

\begin{breakablealgorithm}{Temporal Discontinuity Smoothing}{alg:smooth}
\label{alg:smooth}
% \begin{algorithmic}[1]
\Require Original $\mathbf{j}^{p}$; current $\mathbf{j}^{p,\textit{gen}}$; target joints $\mathcal{J}$; boundary frame $e$
\Ensure Updated $\mathbf{j}^{p,\textit{gen}}$
\For{each joint $j\in\mathcal{J}$}
  \State $o \gets \mathbf{j}^{p}_{e}[j]-\mathbf{j}^{p,\textit{gen}}_{e}[j]$ \Comment{endpoint offset $o_e[j]$}
  \For{$f=e+1$ \textbf{to} $F$}
    \State $d \gets \mathbf{j}^{p}_{f}[j]-\mathbf{j}^{p}_{f-1}[j]$ \Comment{$d_f[j]$}
    \State $\mathbf{j}^{p,\textit{gen}}_{f}[j]\gets \mathbf{j}^{p,\textit{gen}}_{f}[j]-o$
    \For{axis $a$ in $\{x,y,z\}$}
      \If{$\Call{Sign}{d_a} = -\Call{Sign}{o_a}$}
        \State $u \gets o_a + d_a$
        \If{$\Call{Sign}{u} \neq \Call{Sign}{o_a}$}
          \State $o_a \gets 0$ \Comment{$\texttt{clip0}$ to prevent overshoot}
        \Else
          \State $o_a \gets u$
        \EndIf
      \EndIf
    \EndFor
  \EndFor
\EndFor
\State \Return $\mathbf{j}^{p,\textit{gen}}$
% \end{algorithmic}
\end{breakablealgorithm}

\subsection{Learning-Based Approach}
\label{appendix:pseudocode-learning}

\begin{breakablealgorithm}{Learning-Based Simplification: Training}{alg:learn-train}
\Require Dataset $\mathcal{D}=\{(\mathbf{x},\mathbf{y},\mathbf{m})\}$, denoiser $\hat{\mathbf{y}}_\theta$, ControlNet encoder $\mathrm{MLP}_\phi$, diffusion schedule $\{\bar{\alpha}_t,\beta_t\}_{t=1}^{T_{\text{diff}}}$, dropout prob. $p_{\text{drop}}$, loss weights
\Ensure Trained parameters $(\theta,\phi)$
\Repeat
  \State Sample minibatch $(\mathbf{x},\mathbf{y},\mathbf{m})\sim\mathcal{D}$ \Comment{$\mathbf{x}$: original, $\mathbf{y}$: simplified}
  \State Sample $t\sim\mathrm{Uniform}\{1,\dots,T_{\text{diff}}\}$, $\boldsymbol{\epsilon}\sim\mathcal{N}(\mathbf{0},\mathbf{I})$
  \State $z_t \gets \sqrt{\bar{\alpha}_t}\mathbf{y} + \sqrt{1-\bar{\alpha}_t}\boldsymbol{\epsilon}$ \Comment{forward diffusion on simplified motion}
  % \Statex \sffamily{Classifier-free conditioning dropout}
  \State $\mathbf{m}_{\text{in}} \gets \begin{cases}
      \boldsymbol{\epsilon}_m \sim \mathcal{N}(\mathbf{0},\mathbf{I}), & \text{with prob. } p_{\text{drop}}\\
      \mathbf{m}, & \text{otherwise}
  \end{cases}$ \Comment{Classifier-free conditioning dropout}
  \State $\mathbf{x}_{\text{in}} \gets \begin{cases}
      \boldsymbol{\epsilon}_x \sim \mathcal{N}(\mathbf{0},\mathbf{I}), & \text{with prob. } p_{\text{drop}}\\
      \mathbf{x}, & \text{otherwise}
  \end{cases}$
  \State $\mathbf{h}_x \gets \mathrm{MLP}_\phi(\mathbf{x}_{\text{in}})$; inject $\mathbf{h}_x$ into denoiser via residual paths
  \State $\hat{\mathbf{y}} \gets \hat{\mathbf{y}}_\theta(z_t, t, \mathbf{m}_{\text{in}}, \mathbf{x}_{\text{in}})$
  % \Statex \sffamily{Losses (Eqs.~\eqref{eq:L_simple}, \eqref{eq:L_vlb}, \eqref{eq:L_hybrid}, \eqref{eq:L_joint}, \eqref{eq:L_va}, \eqref{eq:L_body}, \eqref{eq:L_aux}, \eqref{eq:L_complexity}, \eqref{eq:L})}
  \State $\mathcal{L}_{\text{simple}} \gets \|\mathbf{y}-\hat{\mathbf{y}}\|_2^2$
  \State $\mathcal{L}_{\text{vlb}} \gets \mathbb{E}_{t\sim p_t}\!\left[\left\lceil L_t/p_t\right\rceil\right],\ \ p_t\propto \sqrt{\mathbb{E}[L_t^2]}$
  \State $\mathcal{L}_{\text{hybrid}} \gets \mathcal{L}_{\text{simple}} + \lambda\,\mathcal{L}_{\text{vlb}}$
  \State $\mathcal{L}_{\text{joint}} \gets \frac{1}{N}\sum_{i=1}^{N}\|FK(\mathbf{y}^{(i)})-FK(\hat{\mathbf{y}}^{(i)})\|_2^2$
  \State $\mathcal{L}_{\text{vel.,acc.}} \gets \frac{1}{N}\sum_{i=1}^{N}\|\mathbf{y}'^{(i)}-\hat{\mathbf{y}}'^{(i)}\|_2^2+\|\mathbf{y}''^{(i)}-\hat{\mathbf{y}}''^{(i)}\|_2^2$
  \State $\mathcal{L}_{\text{body}} \gets \frac{1}{N-1}\sum_{i=1}^{N-1}\| (FK(\hat{\mathbf{y}}^{(i+1)})-FK(\hat{\mathbf{y}}^{(i)}))\cdot \hat{\mathbf{c}}^{(i)}\|_2^2$
  \State $\mathcal{L}_{\text{auxiliary}} \gets \lambda_{\text{joint}}\mathcal{L}_{\text{joint}}+\lambda_{\text{vel.,acc.}}\mathcal{L}_{\text{vel.,acc.}}+\lambda_{\text{body}}\mathcal{L}_{\text{body}}$
  \State $\mathcal{L}_{\text{complexity}} \gets \sum_{i=1}^{5}\lambda_{C_i}\mathcal{L}_{C_i}(\hat{\mathbf{y}})$
  \State $\mathcal{L} \gets \mathcal{L}_{\text{hybrid}}+\mathcal{L}_{\text{auxiliary}}+\mathcal{L}_{\text{complexity}}$
  \State Update $(\theta,\phi)$ by gradient descent on $\mathcal{L}$
\Until{convergence}
\end{breakablealgorithm}

\begin{breakablealgorithm}{Learning-Based Simplification: Inference with Compositional Guidance}{alg:learn-infer}
\Require Trained model, input motion $\mathbf{x}$, music $\mathbf{m}$, steps $T_{\text{diff}}$, guidance parameters $(w,\chi,\beta_{\mathbf{m}},\beta_{\mathbf{x}})$
\Ensure Generated simplified motion $\hat{\mathbf{y}}_{0}$
\State $\hat{z}_{T_{\text{diff}}}\sim\mathcal{N}(\mathbf{0},\mathbf{I})$
\For{$t=T_{\text{diff}}$ down to $1$}
  \State $\hat{\mathbf{y}}_{\emptyset}\gets \hat{\mathbf{y}}_\theta(\hat{z}_t,t,\emptyset,\emptyset)$
  \State $\hat{\mathbf{y}}_{\mathbf{m},\mathbf{x}}\gets \hat{\mathbf{y}}_\theta(\hat{z}_t,t,\mathbf{m},\mathbf{x})$
  \State $\hat{\mathbf{y}}_{\mathbf{m}}\gets \hat{\mathbf{y}}_\theta(\hat{z}_t,t,\mathbf{m},\emptyset)$
  \State $\hat{\mathbf{y}}_{\mathbf{x}}\gets \hat{\mathbf{y}}_\theta(\hat{z}_t,t,\emptyset,\mathbf{x})$
  \State $\tilde{\mathbf{y}}_t \gets \hat{\mathbf{y}}_{\emptyset}
  + w\!\left[
    \chi\big(\hat{\mathbf{y}}_{\mathbf{m},\mathbf{x}}-\hat{\mathbf{y}}_{\emptyset}\big)
    +(1-\chi)\!\left(
      \beta_{\mathbf{m}}\big(\hat{\mathbf{y}}_{\mathbf{m}}-\hat{\mathbf{y}}_{\emptyset}\big)
      +\beta_{\mathbf{x}}\big(\hat{\mathbf{y}}_{\mathbf{x}}-\hat{\mathbf{y}}_{\emptyset}\big)
    \right)\right]$
  \State $\hat{z}_{t-1}\sim q\!\left(\tilde{\mathbf{y}}_t,\ t-1\right)$ \Comment{reverse diffusion transition}
\EndFor
\State $\hat{\mathbf{y}}_{0}\gets \Call{DecodeOrProject}{\hat{z}_{0}}$
\State Apply boundary smoothing (translation/rotation) across generated slices
\State \Return $\hat{\mathbf{y}}_{0}$
\end{breakablealgorithm}

\end{document}